\documentclass[reprint,preprintnumbers,aps,prd,amsmath,amssymb,nobibnotes,nofootinbib,onecolumn,unsortedaddress]{revtex4-2}

\usepackage[T1]{fontenc}
\usepackage{lmodern}
\usepackage{graphicx}
\usepackage{xcolor}
\usepackage{multirow}
\usepackage{xspace}
\usepackage{todonotes}

\usepackage[hidelinks]{hyperref}

\newcommand{\acronymsw}[1]{\textmd{\textsc{#1}}}
\newcommand{\Pepper}{\acronymsw{Pepper}\xspace}
\newcommand{\Chili}{\acronymsw{Chili}\xspace}
\newcommand{\Vegas}{\acronymsw{Vegas}\xspace}
\newcommand{\Sherpa}{\acronymsw{Sherpa}\xspace}
\newcommand{\NNPDF}{\acronymsw{NNPDF}\xspace}
\newcommand{\Kokkos}{\acronymsw{Kokkos}\xspace}
\newcommand{\Pyper}{\acronymsw{Pyper}\xspace}
\newcommand{\madgraph}{\acronymsw{MadGraph5\_aMC@NLO}\xspace}

\newcommand{\zjjjj}{$pp \to e^+e^- + 4j$\xspace}
\newcommand{\zjjjjj}{$pp \to e^+e^- + 5j$\xspace}
\newcommand{\ttjjjj}{$pp \to t \bar t + 4j$\xspace}
\newcommand{\jjjj}{$pp \to 4j$\xspace}
\newcommand{\jjjjj}{$pp \to 5j$\xspace}

\newcommand{\pepvegas}{\Pepper{}+\Vegas}
\newcommand{\pepflow}{\Pepper{}+Flow\xspace}

\begin{document}

\preprint{MBI-ML-26-04}
\preprint{MCNET-26-23}
\preprint{MSUHEP-26-012}

\title{Efficient Event Generation for High-Multiplicity LHC Processes:\\
An End-to-End GPU Workflow with Normalizing Flows}

\author{Enrico Bothmann}
\affiliation{CERN, 1217 Meyrin, Switzerland}

\author{Joshua Isaacson}
\affiliation{Department of Physics and Astronomy, Michigan State University, East Lansing, MI 48824, USA}

\author{Claudius Krause}
\affiliation{Marietta Blau Institute for Particle Physics (MBI), Austrian Academy of Sciences (ÖAW), 1010 Vienna, Austria}

\author{Carla J. López-Zurita}
\affiliation{ETH Zürich, 8092 Zürich, Switzerland}
\affiliation{CERN, 1217 Meyrin, Switzerland}

\author{Maximilian Spannring}
\affiliation{Marietta Blau Institute for Particle Physics (MBI), Austrian Academy of Sciences (ÖAW), 1010 Vienna, Austria}
\affiliation{Atominstitut, Technische Universität Wien, 1020 Vienna, Austria}

\author{Daohan Wang}
\email[Corresponding author: ]{daohan.wang@oeaw.ac.at}
\affiliation{Marietta Blau Institute for Particle Physics (MBI), Austrian Academy of Sciences (ÖAW), 1010 Vienna, Austria}

\date{\today}

\begin{abstract}
Producing very large unweighted event samples for high-multiplicity processes is limited by expensive matrix-element evaluations and low unweighting efficiencies. We present the first end-to-end GPU-resident event-generation workflow that integrates normalizing-flow proposals with the parton-level event generator \Pepper. Helicity-conditioned coupling flows are trained using online updates supplemented by sample replay and deployed across all subprocesses of complete proton--proton collision processes with many final-state jets. In this workflow, a Python-based control layer and \Pepper exchange flow-generated phase-space points and the corresponding target-density evaluations directly in device memory. The control layer performs flow sampling, proposal-density evaluation, and unweighting, while \Pepper evaluates the matrix elements, PDFs, and phase-space factors defining the target density and writes the accepted events in standard formats. We compare subprocess-specific flows, with one flow per partonic subprocess, to grouped conditional flows that share parameters among subprocesses with related parton content. The workflow is benchmarked for \zjjjj, \zjjjjj, \ttjjjj, \jjjj, and \jjjjj production. On four H100 GPUs, we generate \(10^9\) unweighted events for each benchmark process. Including the cost of flow training, the workflow achieves end-to-end speedups of up to two orders of magnitude over standalone \Pepper event generation and turns a multi-week task into a sub-day computation. It thereby makes billion-event production more practical and offers a pathway to alleviating the Monte Carlo statistics bottleneck in high-multiplicity collider physics.
\end{abstract}

\maketitle
\tableofcontents

\section{Introduction}
\label{sec:introduction}

The LHC has entered Long Shutdown 3, during which crucial upgrades for the high-luminosity phase (HL-LHC, starting around 2030) are being installed. With an anticipated increase of the integrated luminosity by roughly an order of magnitude~\cite{Apollinari:2015wtw}, many analyses are expected to become increasingly limited by the statistical uncertainty of available Monte Carlo (MC) simulation samples~\cite{HEPSoftwareFoundation:2017ggl,HSFPhysicsEventGeneratorWG:2020gxw,HSFPhysicsEventGeneratorWG:2021xti}.
The computational significance of event generation is illustrated by recent ATLAS and CMS offline CPU-usage projections. Their HL-LHC baseline computing models attribute \(17\%\) and \(9\%\) of their respective total CPU demands in a representative Run-4 year to event generation~\cite{CERN-LHCC-2022-005,CMS:2026CDR}.
To address the growing need for simulated events with finite computing resources, more efficient ways to generate events have been explored, as reviewed in Ref.~\cite{Campbell:2022qmc}.

This challenge is particularly severe for processes with many
final-state jets,
such as $V + \text{jets}$, $t \bar t + \text{jets}$ and pure QCD multi-jet production,
which have large cross sections~\cite{ATL-PHYS-PUB-2024-011,CMS-XS-XJETS-SUMMARY},
are used for many Standard Model measurements,
and form an important set of backgrounds for a broad range of LHC measurements and searches.
As the jet multiplicity increases, the number of contributing partonic subprocesses and helicity configurations grows rapidly, and the matrix-element evaluation becomes increasingly expensive. At the same time, the growing dimensionality and complexity of the phase-space integrand make it increasingly difficult for sampling proposals to match the target density, resulting in broader event-weight distributions and lower unweighting efficiencies.

The cost of producing unweighted events is therefore amplified twice: each proposed phase-space point requires an expensive matrix-element evaluation, and many such points may be
rejected before one accepted event is obtained~\cite{Hoche:2019flt,Bothmann:2023siu}.
These effects are visible in realistic \Sherpa\ benchmarks. For a typical ATLAS production setup, PDF evaluation and matrix-element generation together can account for about \(80\%\) of the total runtime~\cite{Bothmann:2022thx}, while unweighting efficiencies of orders \(\mathcal O(10^{-4})\)--\(\mathcal O(10^{-5})\) have been reported for setups with 4 or more final-state jets~\cite{Bothmann:2023siu}. Consequently, producing large samples of unweighted events for such high-multiplicity processes can require substantial computing resources.

A natural way to improve the unweighting efficiency is to construct adaptive importance samplers that follow the structure of the integrand more closely. This idea has a long history in collider event generation, including factorized adaptive grids such as VEGAS~\cite{Lepage:1977sw,Ohl:1998jn,Lepage:1980dq,Lepage:2020tgj}, cellular partitioning as in FOAM~\cite{Jadach:2002kn}, and multi-channel methods that combine process-specific phase-space mappings and adapt their relative weights~\cite{Kleiss:1994qy}. These methods are highly effective when the dominant singular structures can be captured by suitable coordinates or channels, but complex correlations and overlapping peak structures can make the proposal difficult to optimize. Machine-learning approaches have also been explored for adaptive phase-space sampling, including boosted decision trees and neural networks trained as importance samplers~\cite{Bendavid:2017zhk,Klimek:2018mza}. A broader overview of machine-learning approaches to LHC event generation is given in Ref.~\cite{Butter:2022rso}. Normalizing flows provide a particularly useful realization of this idea~\cite{papamakarios2021normalizing,Durkan:2019}: they transform a simple base density into a flexible proposal through a sequence of invertible mappings, while retaining efficient sampling and exact density evaluation. The proposal density can therefore be used to compute corrected event weights, so imperfect learning affects the sampling efficiency rather than the correctness of the generated distribution. Early collider applications demonstrated that flow-based proposals can capture non-trivial phase-space structures and improve MC integration and event-generation efficiency~\cite{Bothmann:2020ywa,Gao:2020zvv,Gao:2020vdv,Stienen:2020gns}.
MadNIS subsequently combined learned multi-channel weights with normalizing-flow proposals for automated phase-space integration~\cite{Heimel:2022wyj}. MadNIS Reloaded implemented and refined this framework within \madgraph~\cite{Alwall:2014hca} for single partonic channels~\cite{Heimel:2023ngj}, while MadNIS-Lite introduced differentiable phase-space mappings augmented by small, interpretable learnable flow elements~\cite{Heimel:2024wph}. Most recently, MadSpace provided modular, GPU-ready phase-space construction, adaptive and neural importance sampling, and event unweighting~\cite{Heimel:2026hgp}, while MadNIS at next-to-leading order (NLO) extended
neural multi-channel integration to real and virtual next-to-leading-order corrections~\cite{DeCrescenzo:2026tsp}.
As a complementary approach to importance-sampling applications, Normalizing flows have been used as trainable control variates~\cite{Heimel:2026cxh}.

While learned phase-space proposals can improve the sampling efficiency, the matrix-element evaluation itself remains a major component of the event-generation cost, especially in high-multiplicity processes. This has motivated several efforts
to study algorithmic improvements,
and to move matrix-element evaluation and complete MC workflows to GPUs and other hardware accelerators~\cite{Bothmann:2021nch,Carrazza:2021gpx,Valassi:2023yud,Bothmann:2023gew,Hageboeck:2023blb,Hagebock:2025jyk,Valassi:2025gmq,Gu:2026zbf},
building on earlier exploratory work~\cite{Giele:2010ks,Hagiwara:2010ujr,Hagiwara:2010oca,Hagiwara:2013oka}.
The parton-level event generator framework \Pepper~\cite{Bothmann:2021nch,Bothmann:2023gew},
including its phase-space sampler \Chili~\cite{Bothmann:2023siu},
are built from scratch to provide end-to-end GPU-based event generation.
This makes \Pepper a natural framework for testing how far Machine-Learning based proposals, themselves typically trained on GPU, can improve fully GPU-accelerated event-generation workflows.
Recently, helicity-conditioned continuous normalizing flows trained with flow matching have been applied to many-jet phase-space sampling with
\Pepper~\cite{Bothmann:2025lwg,Bothmann:2026dar}. For representative Drell--Yan and $t \bar t + \text{jets}$ partonic channels, these models achieve higher unweighting efficiencies and better scaling with increasing multiplicity than coupling-layer flows, albeit at a larger evaluation cost. 
Using RegFlow, a sizable fraction of the efficiency gains of the continuous models can be transferred to faster coupling-layer flows, yielding substantial gains in parton-level event-generation wall time~\cite{Bothmann:2026dar}.

The present work is technically distinct from earlier representative-channel studies and develops a production-scale framework
with coupling-layer flow based sampling for complete many-jet processes in \Pepper, with the following characteristics:

\begin{itemize}
    \item We integrate trained normalizing-flow proposals directly into the event-generation loop. A new Python layer controls flow sampling, proposal-density evaluation, subprocess allocation, and accept-reject unweighting, while \Pepper evaluates the matrix elements, PDFs, and phase-space factors defining the target density, and writes out the generated parton-level events.
    To reduce round-trip latency, all operations are performed on the GPU, with no memory transfers required to the CPU or between devices.

    \item We use helicity-conditioned rational-quadratic coupling-layer flows whose fast sampling and exact density evaluation make them well suited to high-throughput event production, where the proposal must be sampled and evaluated for every generated phase-space point. The flows are initialized from the existing \Vegas grids and trained with online samples from the evolving proposal together with replay-buffer samples from previous training rounds. This allows already evaluated phase-space points to continue improving the proposal without additional matrix-element evaluations.

    \item We develop two strategies for scaling learned proposals to the full subprocess collections of many-jet processes. In the subprocess-specific setup, each subprocess is assigned an independent flow. In the grouped setup, subprocesses with related initial- and final-state parton content share a conditional flow and are distinguished through trainable process embeddings.
\end{itemize}

We provide end-to-end leading-order many-jet runtime measurements for this framework, including both the cost of training the learned proposal and the subsequent generation of unweighted parton-level events. We benchmark \zjjjj, \zjjjjj, \ttjjjj, \jjjj, and \jjjjj production, and also report on the generation-only performance using trained checkpoints.

The paper is organized as follows.  Section~\ref{sec:event-generation} reviews the many-jet event-generation problem and the conventional \Pepper workflow. Section~\ref{sec:flow-optimized-sampling} introduces the helicity-conditioned normalizing-flow proposal and its training strategy for a single subprocess. In Sec.~\ref{sec:subprocess-specific-flows} we extend this construction to full many-jet processes by training one independent flow for each subprocess, while Sec.~\ref{sec:grouped-flows} introduces grouped conditional flows with shared parameters and process embeddings.  Section~\ref{sec:results} presents the runtime benchmarks for \zjjjj production on four H100 GPUs, comparing subprocess-specific and grouped flows in both training-plus-generation and
generation-only settings.  Additional training diagnostics, four-H100 benchmarks for the remaining processes, and single-RTX results are collected in the appendices.  We conclude in Sec.~\ref{sec:conclusion}.

\section{Many-Jet Event Generation with Pepper}
\label{sec:event-generation}

\Pepper~\cite{Bothmann:2021nch,Bothmann:2023gew} is a parton-level event
generator developed to address the computing challenges of event generation
for the HL-LHC. It focuses on the computationally most demanding
process classes in typical LHC simulation campaigns, such as vector-boson,
top-quark-pair, and pure multijet production in association with many
additional jets.
Currently, this is ready for production at LO, with the extension to NLO ongoing.
\Pepper is written in C++ and uses the
\textsc{Kokkos} performance-portability framework~\cite{CarterEdwards20143202,Trott:2021arv}, such that a single
implementation runs efficiently on conventional CPUs as well as on GPUs from
different vendors. Its event-generation workflow is data-parallel throughout:
phase-space points are proposed, evaluated, and unweighted in large batches,
which maps naturally onto the massively parallel execution model of modern
GPUs and, as we will see below, onto the batched training and sampling of
normalizing-flow proposals.

For each process, \Pepper decomposes the calculation into partonic
subprocesses and evaluates the corresponding tree-level matrix elements using
recursive Berends--Giele-type techniques~\cite{BERENDS1988759,Berends:1988yn,Berends:1990ax},
with helicity degrees of freedom
treated by MC sampling. Phase-space points are generated with the
\Chili phase-space sampler~\cite{Bothmann:2023siu}, which combines a small
number of generic mappings tailored to \(t\)-channel-dominated hadron-collider
kinematics with a \Vegas-style adaptive remapping of the uniform input random
numbers. Together with PDF and running-coupling evaluations, this defines the
event weight used for unweighting, and accepted events are written in
standardized output formats suitable for downstream particle-level simulation.
In this paper, \Pepper serves both as the baseline event generator, defining
the conventional workflow that our flow-based approach is benchmarked against,
and as the provider of the target density that the normalizing-flow proposals
are trained on.

The remainder of this section describes this conventional workflow in more
detail. Section~\ref{sec:subprocesses-helicities} discusses the subprocess
decomposition and the helicity-amplitude treatment.
Sections~\ref{sec:phase-space-unweighting} and~\ref{sec:chili} review
phase-space sampling and unweighting in general and the \Chili mappings with
their \Vegas optimization in particular, and
Sec.~\ref{sec:Pepper-workflow} summarizes the resulting end-to-end \Pepper
event-generation workflow.

\subsection{Processes and Helicity Amplitudes}
\label{sec:subprocesses-helicities}

At leading order, the cross section of a \(2\to n\)
process at a hadronic collider is given by the factorization formula
\begin{equation}
    \sigma
    =
    \sum_{ab}
    \int \mathrm{d}x_a \, \mathrm{d}x_b \,
    f_a(x_a,\mu_F^2)\, f_b(x_b,\mu_F^2)
    \int \mathrm{d}\Phi_n \,
    \frac{1}{2\hat{s}}
    \sum_{h}
    \overline{\bigl|\mathcal{M}^{ab}_h(\Phi_n;\mu_R^2)\bigr|^2}
    \;\Theta(\Phi_n) \,,
\end{equation}
where the outer sum runs over the partonic channels \(ab\) contributing to
the process, \(f_a\) and \(f_b\) are the parton distribution functions
evaluated at the momentum fractions \(x_{a,b}\) and the factorization scale
\(\mu_F\), with $\alpha_s$ evaluated at the renormalization scale $\mu_R$, \(\mathrm{d}\Phi_n\) is the \(n\)-particle phase-space measure,
and \(\hat{s}=x_a x_b s\) relates the partonic center-of-mass energy $\hat{s}$ with the hadronic one $s$.
The matrix element \(\mathcal{M}^{ab}_h\) is evaluated for a fixed assignment \(h\) of
the external-particle helicities, the bar denotes the color sum together
with the average over initial-state colors and helicities, and
\(\Theta(\Phi_n)\) implements the phase-space cuts listed in
Sec.~\ref{sec:event-generation-parameters}.

For many-jet final states, the number of contributing partonic channels
grows rapidly with the jet multiplicity, since each additional jet can be a
gluon or any of the massless (anti)quark flavors.
However, many of these channels can be related because their matrix elements
are invariant under flavor changes;
they only differ through their PDF factors of the initial-state flavors.
\Pepper exploits this redundancy by organizing the partonic channels into
flavor-process groups. Each group is represented by a base channel, for
which the matrix element is evaluated, while the remaining mapped channels
of the group are accounted for through their PDF weights. The
flavor-process group is the unit that \Pepper samples, optimizes, and
unweights independently, and we therefore refer to a flavor-process group
simply as a ``subprocess'' throughout this paper. Even after this grouping,
complete many-jet processes retain a large number of subprocesses;
\zjjjj, for example, decomposes into \(134\) subprocesses.

The helicity sum in the factorization formula runs over all configurations
of the external-particle helicities. Their number grows exponentially with
the number of external legs, but a sizable fraction of the configurations
vanishes identically, for example due to helicity conservation along
massless quark lines. \Pepper determines the set of non-vanishing, or
active, helicity configurations \(\mathcal{H}_{\rm act}\) of each subprocess
during initialization, and estimates the helicity sum stochastically: for
every phase-space point, a single configuration is drawn from a discrete
proposal distribution \(\pi(h)\) over the active configurations, and the
summand is corrected by the corresponding probability,
\begin{equation}
    \sum_{h\in\mathcal{H}_{\rm act}}
    \overline{\bigl|\mathcal{M}_h\bigr|^2}
    =
    \biggl\langle
        \frac{\overline{\bigl|\mathcal{M}_h\bigr|^2}}{\pi(h)}
    \biggr\rangle_{h\sim\pi} .
\end{equation}
In the conventional \Pepper workflow, \(\pi(h)\) is adjusted over the
\(N_{\rm act}\) active configurations during the initial optimization
phase with a multi-channel approach~\cite{Kleiss:1994qy}.
The helicity choice during event generation is then encoded in
one additional uniform random number that \Pepper consumes along with the
phase-space random numbers. Sampling instead of summing helicities reduces
the cost per phase-space point from \(N_{\rm act}\) amplitude evaluations to
a single one, at the price of an additional source of event-weight
fluctuations, as quantified for pure gluon amplitudes in~\cite[Sec.~4.2]{Bothmann:2021nch}.
The helicity label thereby becomes a discrete component of the
sampling problem, on the same footing as the continuous phase-space
variables. The flow-based proposals of
Sec.~\ref{sec:flow-optimized-sampling} exploit precisely this structure, by
combining a learned, non-uniform helicity proposal with phase-space
proposals conditioned on the sampled helicity.
This helicity conditioning had first been proposed in~\cite{Bothmann:2025lwg}.

For a given subprocess and helicity configuration, the color-summed squared
amplitude is computed using Berends--Giele
recursion
after color decomposition using the minimal color basis proposed in~\cite{Melia:2013bta,Melia:2013xok,Johansson:2015oia,Melia:2015ika}.
The recursion constructs
amplitudes from recursively defined off-shell currents rather than from
individual Feynman diagrams, avoiding the factorial growth of the
diagrammatic expansion, and the color sum is performed exactly, without
stochastic color sampling. The recursion consists of regular,
arithmetically dense operations that are applied identically to every phase-space
point, which is what makes the batched, portable \Pepper implementation
particularly efficient for batched lock-step evaluation on GPUs.

\subsection{Phase-space Sampling and Unweighting}
\label{sec:phase-space-unweighting}
To produce events distributed according to the differential cross section, one needs to be able to sample from the probability density defined by the differential rate normalized by the total rate.
Evaluating the total cross section requires the calculation of high-dimensional integrals.
This is traditionally handled through the use of Monte-Carlo methods in which the integral is approximated as
\begin{equation}
\sigma \approx \frac{1}{N}\sum_{i=1}^N f(x^{(i)}) = \frac{1}{N} \sum_{i=1}^N \sum_{ab} f_a\left(x^{(i)}_a, \mu_F^2\right) f_b\left(x^{(i)}_b,\mu_F^2\right)\frac{1}{2\hat{s}}\sum_h
    \overline{\bigl|\mathcal{M}^{ab}_h(\Phi^{(i)}_n;\mu_R^2)\bigr|^2}
    \;\Theta(\Phi^{(i)}_n) \,,
\end{equation}
where the superscript $(i)$ labels a random sample from some base distribution.
Each Monte-Carlo sample obtained in this manner is traditionally referred to as a weighted event.
These weighted events are not ideal for experimentalists to use in the analysis pipeline
due to having to propagate many potentially small weight events through expensive detector simulation and reconstruction steps.
Therefore, an accept-reject step is performed to produce unweighted events, i.e.\ events with a uniform weight.
The probability to accept an event in the unweighting is given by
\begin{equation}
P(x^{(i)}) = \frac{f(x^{(i)})}{\max(\{f(x^{(i)})\})} \equiv \frac{f(x^{(i)})}{f_{\max}} \,.
\end{equation}
The unweighting efficiency $\varepsilon_\text{unw}$ can be defined as the average of the accept-reject probability: $\varepsilon_\text{unw} = \langle f(x^{(i)}\rangle/f_{\max}$.
After the unweighting step, the events are distributed as a Bernoulli variable ($a_i$) with the value of either 0 for rejected events or 1 for accepted events.
The estimate of the total cross section is then given as
\begin{equation}
\sigma \approx \frac{f_{\max}}{N} \sum_{i=1}^N a_i\,.
\end{equation}
However, the metric we care about most is the computational cost to generate the required Monte-Carlo sample for a given experiment.
This is dominated by the detector simulation, thus unweighted events are ideal to reduce the need for simulations on negligible weight events.
Thus, maximizing the unweighting efficiency is of upmost importance.
This is maximized when the relative variance on the cross section is minimized:
\begin{equation}
\frac{\textrm{Var}(\sigma)}{\sigma^2} = \frac{f_{\max}^2\varepsilon_{\rm unw}(1-\varepsilon_{\rm unw})}{N\sigma^2}=\frac{1-\varepsilon_{\rm unw}}{N\varepsilon_{\rm unw}}=\frac{1-\varepsilon_{\rm unw}}{N_{\rm eff}}\,,
\end{equation}
where $N_{\rm eff}=N\varepsilon_{\rm unw}$ is the effective sample size.

The unweighting efficiency is determined by how we sample the random numbers used for the phase space. Ideally, we would sample directly from the cross section distribution. However, in practice that is not possible.
Therefore, samples are drawn from a distribution $g(x)$ such that $w =f(x)/g(x) \approx 1$ for all values of $x$.
After this change of variables, the unweighting probability is given as
\begin{equation}
P(x^{(i)}) = \frac{w(x^{(i)})}{\max(\{w(x^{(i)}))\})}\,.
\end{equation}
One major downside to this approach occurs if $g(y)\rightarrow 0$ for some value $y$. In this region of parameter space, the maximum weight approaches infinity and the unweighting efficiency goes to zero.

The solution to this is to modify the unweighting probability to use a new definition of the maximum. In this case, we define the variable $w_{\rm max, eff}(\varepsilon)$ to be the value such that the weights above it contribute at most a fraction of $\varepsilon$ to the total weight of all events. 
This defines the partial-unweighting scheme, in which the weight of an event is given as
\begin{equation}
a_i = \Theta\left(\frac{w_i}{w_{\rm max, eff}(\varepsilon)}-R\right)\times \max\left(\frac{w_i}{w_{\rm max, eff}(\varepsilon)}, 1\right)\,,
\end{equation}
with $R$ a random number between $[0,1]$. 
With that in place, the goal is to find a function $g(x)$ that closely approximates the differential cross section, while still being able to analytically integrate it to obtain the required cumulative distribution function to draw random samples according to $g(x)$. This is traditionally handled through the \Vegas algorithm, discussed in the following section.

\subsection{\Chili Phase-Space Mappings and \Vegas Optimization}
\label{sec:chili}

The general phase space for a $2\rightarrow n$ scattering process, with incoming particles labeled by $a$ and $b$ and outgoing particles labeled by $1\ldots n$, can be expressed as
\begin{equation}
{\rm d}\Phi_n\left(a,b;1,\ldots,n\right) = \left[\prod_{i=1}^{n}\frac{{\rm d}^3\vec{p}_i}{(2\pi)^32E_i}\right]\left(2\pi\right)^4\delta^{(4)}\left(p_a+p_b-\sum_{i=1}^{n}p_i\right)\,.
\end{equation}
The full differential phase space can be reduced to lower-multiplicity phase-space elements by introducing an intermediate psuedo-particle $\pi$ of virtuality $s_\pi=p^2_\pi$, as described in Ref.~\cite{Byckling:1969luw}:
\begin{equation}
{\rm d}\Phi_n\left(a,b;1,\ldots,n\right) = {\rm d}\Phi_{n-m+1}(a,b;\pi,m+1,\ldots,n)\frac{{\rm d}s_\pi}{2\pi}{\rm d}\Phi_m(\pi;1,\ldots,m)\,.
\end{equation}
Given a set of basic building blocks for the phase space, the above can be iterated until all that remains are those building blocks. In the \Chili\cite{Bothmann:2023siu} approach, these building blocks consist of a single $t$-channel production process and a fixed number of $s$-channel decays. All $s$-channel decay chains can be iterated until only two-particle decays remain with differential phase-space elements ${\rm d}\Phi_2$.

The $t$-channel building block is built from leveraging variables traditionally used in experimental cuts and the mapping of the initial state momenta effected by the parton distribution functions (PDF).
First, consider the single-particle differential phase-space element
\begin{equation}
\frac{{\rm d}^3\vec{p}_i}{(2\pi)^32E_i} = \frac{1}{16\pi^2}{\rm d}p_{i,T}^2{\rm d}y_i\frac{{\rm d}\phi_i}{2\pi}\,,
\end{equation}
where $p_{i,T}$, $y_i$, and $\phi_i$ are the transverse momentum, rapidity, and azimuthal angle of momentum $i$ in the laboratory frame, respectively. The choice of representing the phase-space element with these variables is ideal for high efficiency sampling, since many experimental analyses at hadron colliders require cuts on the transverse momentum and rapidity of jets and other particles.
The remaining component for the $t$-channel requires the evaluation of the momentum-conserving delta function and the convolution over the PDFs,
\begin{align}
{\rm d}x_a{\rm d}x_b{\rm d}\Phi_n\left(a,b;1,\ldots,n\right) &= \frac{{\rm d}P_+{\rm d}P_-}{s}\left[\prod_{i=1}^{n-1}\frac{1}{16\pi^2}{\rm d}p_{i,\perp}^2{\rm d}y_i\frac{{\rm d}\phi_i}{2\pi}\right]\nonumber\\
&\times\frac{{\rm d}^4p_n}{(2\pi)^3}\delta\left(p_n^2-s_n\right)\Theta(E_n)\left(2\pi\right)^4\delta^{(4)}\left(p_a+p_b-\sum_{i=1}^{n-1}p_i-p_n\right)\,,
\end{align}
where $s$ is the hadronic center-of-mass energy, and $P_\pm=P_0\pm P_z$ is defined using $P = \sum_{i=1}^{n-1} p_i$. The evaluation of the delta functions can be easily accomplished through a change of variables from $P_+$ and $P_-$ to $s_n$ and $y_n$. The result is the final form of the $t$-channel building block:
\begin{equation}
{\rm d}x_a{\rm d}x_b{\rm d}\Phi_n\left(a,b;1,\ldots,n\right)=\frac{2\pi}{s}\left[\prod_{i=1}^{n-1}\frac{1}{16\pi^2}{\rm d}p_{i,\perp}^2{\rm d}y_i\frac{{\rm d}\phi_i}{2\pi}\right] {\rm d}y_n\,.
\end{equation}
This building block is particularly efficient for the production of electroweak vector bosons ($W$, $Z$, and $\gamma$) in association with any number of jets.

In addition to the $t$-channel building block, the $s$-channel building block is required to efficiently handle the decay of resonances,
e.g., for the $Z$ boson resonance in $pp\rightarrow e^+e^-+nj$.
The factorization given above can be repeated until only iterative two-body decays remain in addition to the $t$-channel block.
This decay can be expressed in the center-of-mass frame of the combined momentum $p_1 + p_2$, and is given as
\begin{equation}
{\rm d}\Phi_2\left(\left\{1, 2\right\};1,2\right)=\frac{1}{16\pi^2}\frac{\sqrt{(p_1p_2)^2-p_1^2p_2^2}}{(p_1+p_2)^2}{\rm d}\cos\theta_1^{\{1,2\}}{\rm d}\phi_1^{\{1,2\}}\,.
\end{equation}
In the original \Chili phase-space integrator, the number of $s$-channel terms could be controlled by the user. In \Pepper, this is fixed to the ``\Chili{}(basic)'' mode, in which there is exactly one $s$-channel term.
For additional details on \Chili see Ref.~\cite{Bothmann:2023siu}.

The phase-space components can be combined in different ways to produce a set of sampling ``channels''. This approach leads to the introduction of multiple integration channels and the multichannel optimization approach that is standard in particle physics~\cite{Kleiss:1994qy}.
However, \Chili{}(basic) only consists of a single channel, so the details of multichannel optimization is not relevant here.

While the phase space mapping handles the dominant structures of the integral, an importance sampler is used to further refine the results.
Importance sampling attempts to minimize the standard deviation of an estimate of the exact integral, given as
\begin{equation}
\sigma_I^2 = \frac{1}{N}\left(\int_0^1 {\rm d}y\ J^2(y)f^2(x(y))-I^2\right)\,,
\end{equation}
where $N$ is the number of samples for the estimate, $I$ is the exact integral value, $f$ is the function to be integrated, $J(y)$ is the Jacobian transformation, and $x(y)$ is the mapping from random variables $y$ to $x$. It can be shown~\cite{Lepage:1977sw} that the variance is minimized if
\begin{equation}
J(y(x)) = \frac{\int_a^b {\rm d}x\ \left|f(x)\right|}{\left| f(x) \right|}\,.
\end{equation}
An optimal choice of $J(y(x))$ will result in the product $J(y)f(x(y))$ to have no peaks.
Traditionally in particle physics, this has been carried out via the \Vegas algorithm~\cite{Lepage:1977sw,Lepage:1980dq,Lepage:2020tgj}.
This algorithm approximates the multi-dimensional integrand by a product of one-dimensional functions.
The one-dimensional functions are then approximated by a piecewise constant function with $M$ nodes given as
\begin{align}
    x_0 & = a \nonumber \\
    x_1 & = x_0 + \Delta x_0 \\
    \cdots \nonumber \\
    x_M & = x_{M-1} + \Delta x_{M-1}=b\,, \nonumber
\end{align}
with $a$ and $b$ being the phase-space boundaries in the corresponding dimension.
The points in $y$ are defined such that $x(y=i/M) = x_i$, with linear interpolation in-between.
With this, the Jacobian would be given as $J(y) = J_i = M \Delta x_i$.
The variance on the integral estimate is thus minimized when 
\begin{equation}
\frac{J_i^2}{\Delta x_i}\int_{x_i}^{x_{i+1}}{\rm d}x\ f^2(x) = {\rm constant}\,
\end{equation}
for all i. This is accomplished by adjusting the values for $\Delta x_i$, which results in small regions where $|f(x)|$ is large.
Since this is only an estimate, this procedure is iterated until the grid converges.
At this point, the \Vegas integrator has achieved its optimal performance.
The major drawbacks of this approach are the approximation that the function to be integrated is separable and the assumption that it can be approximated precisely with a piecewise constant function with a finite number of bins.

\subsection{The \Pepper Event-Generation Workflow}
\label{sec:Pepper-workflow}

The components described in the previous subsections are combined into the
following standalone end-to-end workflow, which we refer to as ``\pepvegas'' or simply ``\Pepper'' event
generation throughout this paper. A run is steered by an input card that
specifies the process, the physics parameters, and the phase-space cuts of
Sec.~\ref{sec:event-generation-parameters}. During initialization, \Pepper
enumerates the contributing subprocesses and determines their active
helicity configurations (Sec.~\ref{sec:subprocesses-helicities}), and
constructs the \Chili phase-space mappings. In the subsequent optimization stage, one \Vegas grid per subprocess is adapted as described in Sec.~\ref{sec:chili}. This stage also yields estimates of the subprocess cross sections, which determine how a requested unweighted-event sample is distributed across the subprocesses, together with the maximum-weight estimates that enter the unweighting step of Sec.~\ref{sec:phase-space-unweighting}.

Event generation then proceeds in large batches of events that traverse the full
pipeline together, with each event of the batch being mapped to one thread on the GPU.
For each batch, \Pepper draws uniform random
numbers---the phase-space random-number vector together with the helicity
random number introduced in Sec.~\ref{sec:subprocesses-helicities}---and
maps them through the \Vegas-refined \Chili parameterization to partonic
momenta and initial-state momentum fractions. The phase-space cuts are
applied, and the PDFs, running couplings, and the color-summed squared
matrix element for the sampled helicity configuration are evaluated for the
points that pass. Together with the phase-space and flux factors, this
defines the raw event weight of each point. The batch is then unweighted by
an accept--reject step against the maximum weight, following the prescription with
tail tolerance \(\varepsilon\) described in
Sec.~\ref{sec:phase-space-unweighting}.
Finally, the accepted events are written
to the standard \Pepper HDF5 output, based on the LHEH5 event
format~\cite{Hoche:2019flt,Bothmann:2023ozs}, for downstream particle-level
simulation.

All stages of this pipeline operate on contiguous batches of event data and
are implemented in a single, performance-portable code base, so that the
identical workflow runs on CPUs and on GPUs, with a batch size
that can be configured by the user to saturate the target device.
Larger production runs can be parallelized over
multiple devices and nodes using MPI, with each rank generating an
independent stream of batches; this setup has been demonstrated to scale to
the size of current leadership-class systems~\cite{Bothmann:2023gew}. The
\pepvegas workflow on identical GPU hardware defines the baseline
against which the flow-based benchmarks of Sec.~\ref{sec:results} are
measured.

A structural property of this workflow that is essential for the remainder
of this paper is that the proposal stage communicates with the physics
evaluation exclusively through random numbers: once the process and
its parameters are fixed, the phase-space construction, the weight
evaluation, and the event record are deterministic functions of the supplied
random-number vector. The \Vegas proposal can therefore be easily
exchanged for any external sampler that provides random numbers
together with a tractable proposal density.

To exploit this property, \Pepper can alternatively be operated from Python
through its \texttt{pybind11}-based~\cite{pybind11} \Pyper bindings, which
have been developed in the context of this work and are used here for the
first time. A \Pyper session is configured from the same input card as a
standalone run and performs the identical initialization and optimization
stages, including the determination of the active helicity configurations,
the \Vegas adaptation, and the unweighting setup. Instead of entering the
internal generation loop, however, the session disables \Pepper{}'s own
generation of proposal random numbers, and the driving Python process
supplies them externally. For each batch, the caller passes to the
\texttt{process\_batch} routine the memory addresses of an input array,
holding the phase-space random numbers of all events of the batch together
with one helicity random number per helicity block, and of an output array,
which receives the corresponding raw event weights. Both arrays are wrapped
as unmanaged \Kokkos views on the compute device, such that GPU-resident
NumPy or PyTorch buffers are accessed in place and the event weights are
written directly into the caller's memory, avoiding transfers between host
and device. The subprocess to be evaluated can be selected through its
flavor-process-group index before each call, otherwise it is selected
by \Pepper. The event output remains under
\Pepper{}'s control: the standard HDF5 event file is opened when the session
is created, filled with the processed batches and the output weights
(potentially modified by the caller to account for a Jacobian),
and finalized when the session is closed, so that
externally driven runs produce the same type of event samples as the
standalone workflow. When the supplied random numbers are drawn uniformly,
and \Pepper{}’s own \Vegas optimization is not explicitly disabled,
\Pyper-driven generation reproduces standalone \Pepper generation.
Supplying the random numbers from a learned, non-uniform proposal
instead---and correcting the event weights by the corresponding proposal
density---is the basis of the flow-based workflow developed in the following
sections.

\section{Flow-Optimized Phase-Space Sampling}
\label{sec:flow-optimized-sampling}

The previous section described the conventional \pepvegas event-generation workflow, in which phase-space points are generated from a \Vegas-optimized proposal and then unweighted according to their event weights.  The efficiency of this procedure is controlled by the fluctuations of the corrected weights: large weight tails increase the effective maximum weight and reduce the unweighting efficiency. The goal of this section is to replace the fixed phase-space proposal by a trainable conditional normalizing-flow density \(g_\theta(z\mid h)\), together with an adaptive helicity proposal \(\pi(h)\).

Normalizing flows are a class of generative models that transform a simple base probability distribution (here a uniform distribution $p_0$) into a more complex one, through a series of invertible and differentiable transformations $T$, offering exact density evaluation as well as efficient sampling~\cite{papamakarios2021normalizing}. For the conditional flow used here,
\(u,z\in[0,1]^D\), and we write
\begin{equation}
    z=T_\theta(u;h),
    \qquad
    u\sim p_0.
\end{equation}
The resulting density is obtained from the change-of-variables formula,
\begin{equation}
    g_\theta(z\mid h)
    =
    p_0\!\left(T_\theta^{-1}(z;h)\right)
    \left|
        \det
        \frac{\partial T_\theta^{-1}(z;h)}
             {\partial z}
    \right|.
\end{equation}

To keep the density evaluation tractable, \(T_\theta\) is constructed as a composition of coupling layers, each of which has a block-triangular Jacobian. Each layer partitions the \(D\) coordinates into two approximately equal subsets: one subset is passed to the conditioner, which parametrizes the invertible transformation of the other subset. The log-Jacobian determinant is
therefore obtained by summing the scalar contributions from the approximately \(D/2\) transformed coordinates. Its evaluation cost is
\(\mathcal{O}(D/2)=\mathcal{O}(D)\) per layer, compared with
\(\mathcal{O}(D^3)\) for the determinant of a general dense \(D\times D\) Jacobian. The specific helicity-conditioned rational-quadratic spline coupling architecture used here is described in Sec.~\ref{sec:helicity-conditioned-flows}.

The flow is used as a replacement of \Pepper’s internal \Vegas
to generate the random numbers \(z\in[0,1]^D\) and helicity label \(h\).
\Pepper transforms them the random numbers into external four-momenta, evaluates the squared the matrix element, phase-space factors and cuts, PDFs and couplings, and returns a raw, potentially signed event weight \(p_h(z)\). We define the corresponding non-negative target density
used for proposal training and unweighting as
\begin{equation}
    \rho_h(z)
    =
    |p_h(z)|.
\end{equation}
The sign of \(p_h(z)\) is retained when the accepted events are written.

Throughout the training studies below, we monitor the quality of the learned proposal using the corrected event weights
\begin{equation}
    w_i
    =
    \frac{\rho_{h_i}(z_i)}
         {\pi(h_i)\,g_\theta(z_i\mid h_i)} .
\end{equation}
From these weights we report two diagnostics. The first is the batch-level
effective unweighting efficiency,
\begin{equation}
    A_{\varepsilon}
    =
    \frac{\langle w\rangle}
         {w_{\max,\mathrm{eff}}(\varepsilon)} .
\end{equation}
Here \(w_{\max,\mathrm{eff}}(\varepsilon)\) is determined from the same batch by choosing a threshold such that the weights above it contribute at most a fraction \(\varepsilon\) of the total batch weight. For the training
diagnostics shown below, we use \(\varepsilon=10^{-2}\). The second diagnostic is the effective sample size ratio,
\begin{equation}
    \frac{N_{\rm eff}}{N}
    =
    \frac{\left(\sum_i w_i\right)^2}
         {N\sum_i w_i^2}.
\end{equation}
Both diagnostics increase as the variance of the corrected weights decreases and therefore provide measures of sampling efficiency during training. 

In this section we describe the construction and training of the flow proposal for a single \Pepper subprocess, using its definition as a flavor-process group introduced in Sec.~\ref{sec:event-generation}. The proposal is conditioned on the sampled helicity configuration, initialized using the existing \Vegas grid, and trained with an objective designed to reduce event-weight fluctuations. The extension from a single subprocess to the full multi-subprocess event generator is described in the following section.

\subsection{Helicity-Conditioned Coupling Flows}
\label{sec:helicity-conditioned-flows}

For a fixed subprocess, the flow models a helicity-conditioned proposal density
\begin{equation}
    g_\theta(z\mid h),
    \qquad z\in[0,1]^D ,
\end{equation}
where \(z\) denotes the \(D\)-dimensional unit-hypercube random-number vector passed to \Pepper, and \(h\) is the sampled helicity label. Each subprocess is assigned its own trainable helicity-embedding matrix,
\begin{equation}
    E_{\rm hel}\in\mathbb{R}^{64\times64}.
\end{equation}
Given a discrete helicity label \(h\), the corresponding conditioning vector is obtained by selecting the \(h\)-th row,
\begin{equation}
    c_{\rm hel}(h)=E_{\rm hel}[h,:]\in\mathbb{R}^{64}.
\end{equation}
The label is therefore treated as a categorical index. The \(64\) rows accommodate the maximum number of helicity labels considered in the implementation. For a given subprocess, only the rows corresponding to its active helicities are accessed during training and generation. Using the same embedding-matrix shape for all subprocesses also allows their model parameters to be stacked for vectorized parallel training, as described in Sec.~\ref{sec:subprocess-specific-flows}.  In every coupling layer, \(c_{\rm hel}(h)\) is concatenated with the coordinates left unchanged by the coupling mask and passed to the coupling network that predicts the spline parameters. The embedding matrix is optimized jointly with the remaining flow parameters through the same training objective.

The base distribution factorizes over dimensions as independent uniform
distributions on \([-1,1]\),
\begin{equation}
    p_0(u)
    =
    \prod_{i=1}^{D}
    \frac{1}{2}\,
    \mathbf{1}_{[-1,1]}(u_i).
\end{equation}
Equivalently, \(p_0(u)=2^{-D}\) inside the box \([-1,1]^D\) and vanishes
outside. In the implementation this is a context-independent box-uniform
distribution, so its log probability does not depend on the helicity context. A small boundary tolerance is used to avoid numerical failures for points that lie exactly on the edge of the box.

Since \Pepper uses random numbers in the unit hypercube, we include an explicit affine transform between the \Pepper domain and the internal flow domain. For density evaluation, the input is mapped as
\begin{equation}
    u = 2z - 1,
    \qquad z\in[0,1]^D,\quad u\in[-1,1]^D ,
\end{equation}
with log-Jacobian
\begin{equation}
    \log \left|
    \det \frac{\partial u}{\partial z}
    \right|
    =
    D\log 2 .
\end{equation}
For sampling, the inverse map
\begin{equation}
    z = \frac{u+1}{2}
\end{equation}
is applied after the coupling transformations. This makes the learned density \(g_\theta(z\mid h)\) a normalized proposal on the same unit hypercube used by \Pepper.

The main transform consists of eight rational-quadratic coupling layers based on neural spline flows~\cite{Durkan:2019}. The flow implementation is built on the
\texttt{nflows} package~\cite{nflows:2020}. The coupling masks alternate between even and odd dimensions, so that all
random-number coordinates are transformed across successive layers. No additional coordinate permutations are applied between the coupling layers. Each layer uses $K=10$ bins, a tail bound of \(1\), and minimum bin width, bin height, and derivative
\begin{equation}
    w_{\min}=h_{\min}=d_{\min}=10^{-4}.
\end{equation}
The spline parameters are predicted by a conditional multilayer perceptron (MLP). In each coupling block, the input coordinates are split into unchanged
coordinates \(z_{\rm id}\) and transformed coordinates \(z_{\rm tr}\). The coupling network takes the concatenated vector
\begin{equation}
    [z_{\rm id}, c_{\rm hel}(h)]
\end{equation}
as input and predicts the spline parameters used to transform \(z_{\rm tr}\).
The network is a two-hidden-layer MLP,
\begin{equation}
    \mathrm{Linear}
    \to
    \mathrm{SiLU}
    \to
    \mathrm{Linear}
    \to
    \mathrm{SiLU}
    \to
    \mathrm{Linear},
\end{equation}
where SiLU stands for Sigmoid Linear Unit, and the hidden layer width is \(256\). Its final layer outputs the unconstrained widths, heights, and derivatives of the rational-quadratic spline for the transformed coordinates.

\subsection{\Vegas-Informed Initialization}
\label{sec:Vegas-initialization}

The flow is initialized using information from the existing \Vegas grid. The first six coupling layers are initialized to the identity transformation: in each coupling MLP, the weights of the final linear layer are set to zero, and the output bias is chosen to give uniform spline widths, uniform spline heights, and unit derivatives. Consequently, these layers leave the input unchanged at initialization.

The last two coupling layers are initialized from the \Vegas grid, following the \Vegas-to-spline initialization strategy used in MadNIS~\cite{Heimel:2023ngj}. For each transformed dimension, the \Vegas bin edges define an initial set of bin widths. Since the flow uses only \(K=10\) spline bins, the \Vegas bins are merged down to this lower resolution. At each step, the procedure merges the neighboring pair with the smallest absolute difference between their width-to-height ratios, restricting the candidates, whenever possible, to pairs whose combined width and height are both smaller than \(2/K\). If no pair satisfies these bounds, all neighboring pairs are considered. This preserves the coarse structure of the \Vegas map while reducing it to the spline resolution used by the coupling transform. The resulting bin widths, heights, and derivatives are then projected to satisfy the minimum-width, minimum-height, and minimum-derivative constraints, and converted to the unconstrained spline parameters predicted by the coupling network. To retain the derivative information from the \Vegas map, we use a minor extension of the rational-quadratic spline implementation that allows the spline boundary derivatives to be initialized directly. The resulting flow therefore starts from a proposal that approximately reproduces the \Vegas-optimized phase-space map used by \Pepper, while all parameters remain trainable during the subsequent optimization.

\begin{figure*}[t]
    \centering
    \begin{tabular}{cc}
        \includegraphics[width=0.48\textwidth]{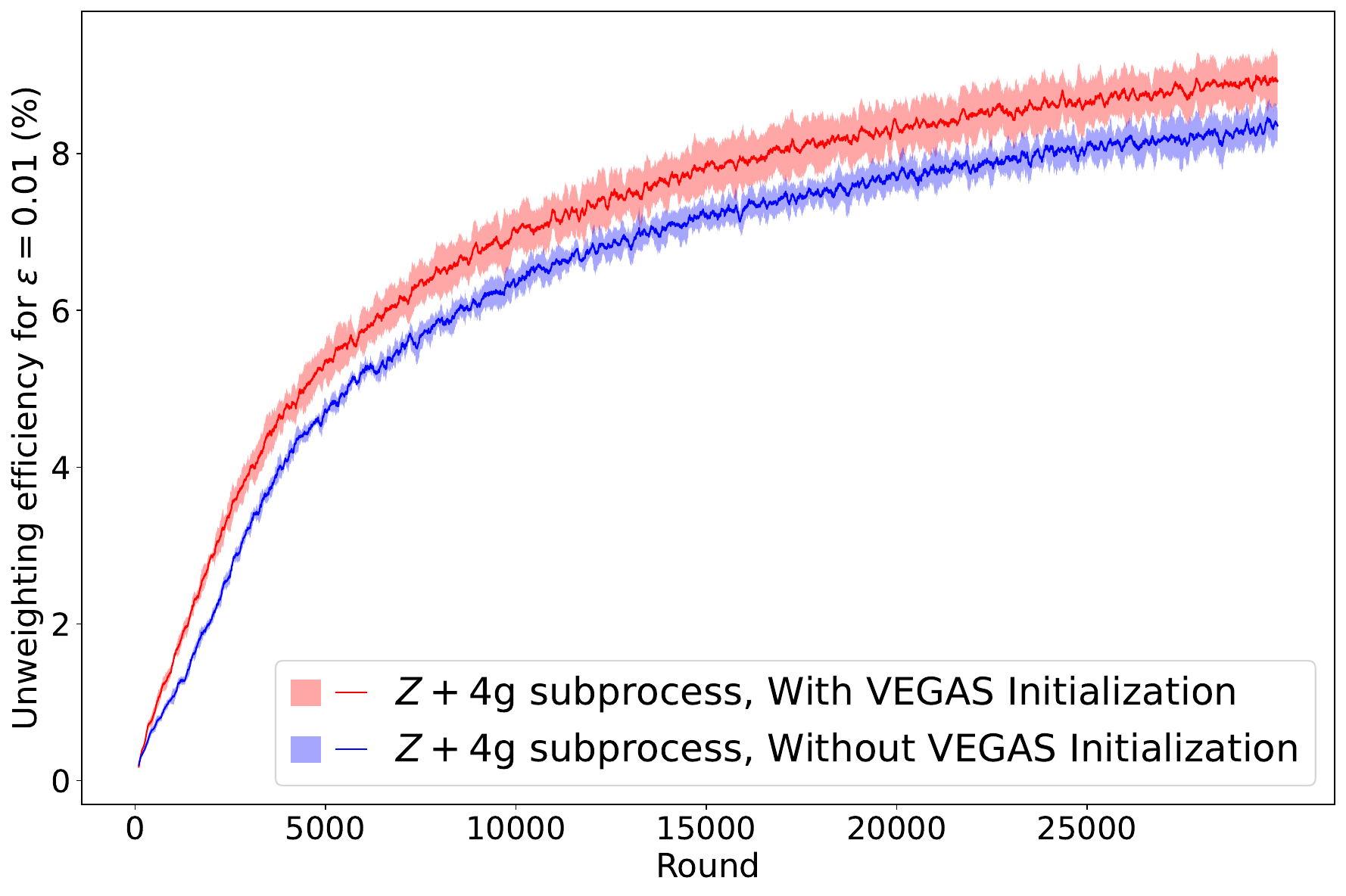} &
        \includegraphics[width=0.48\textwidth]{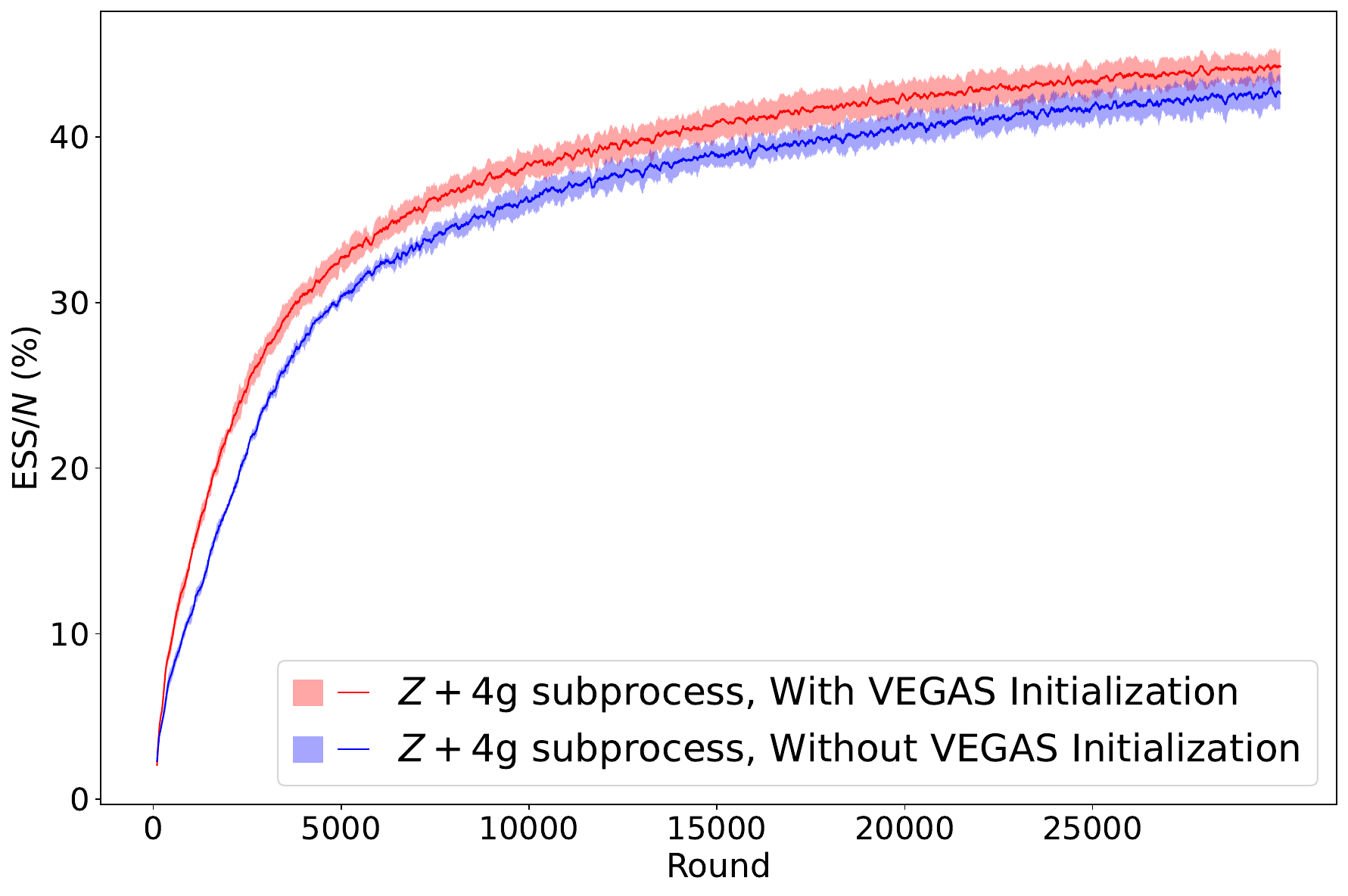} \\
        \includegraphics[width=0.48\textwidth]{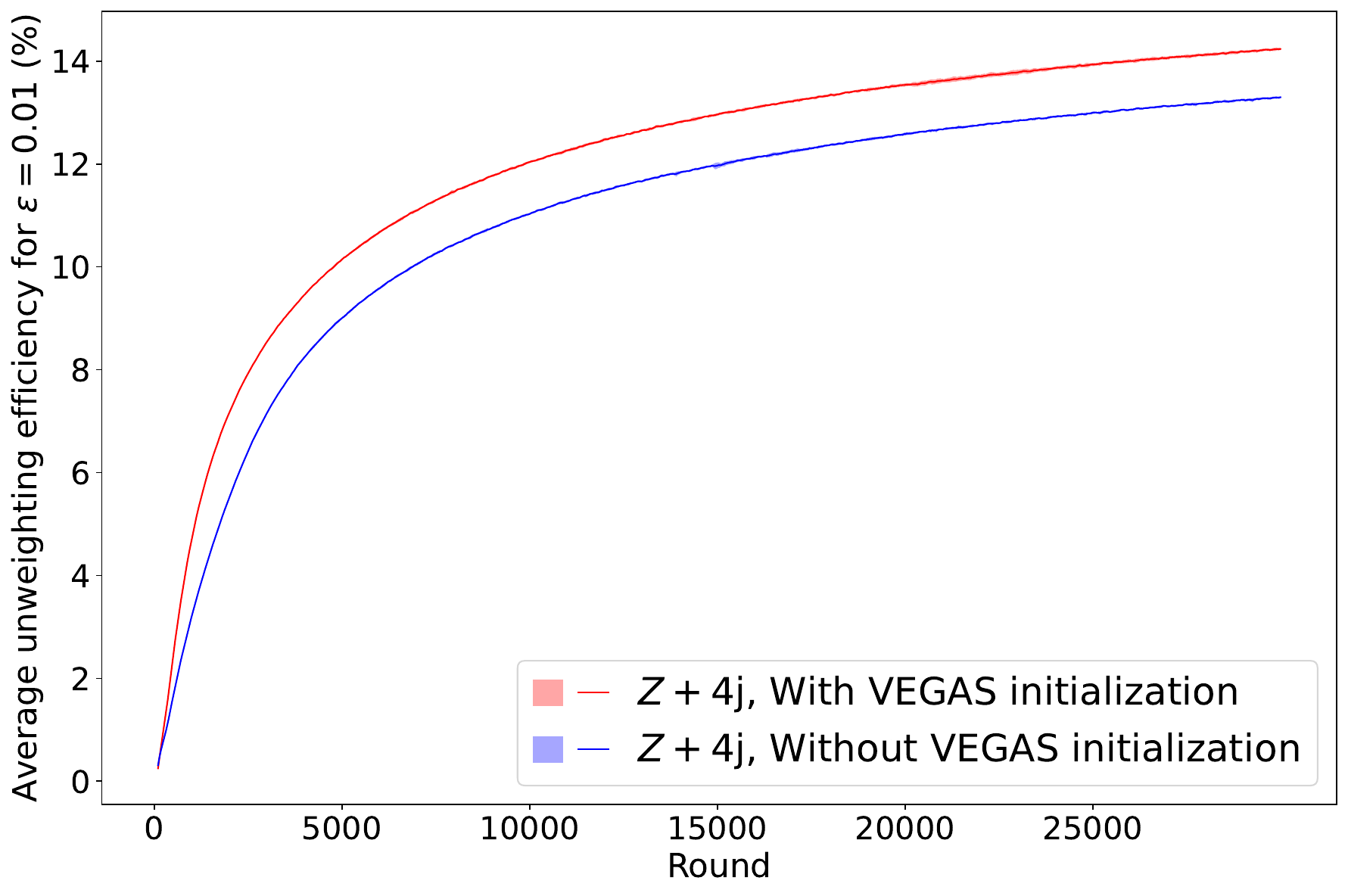} &
        \includegraphics[width=0.48\textwidth]{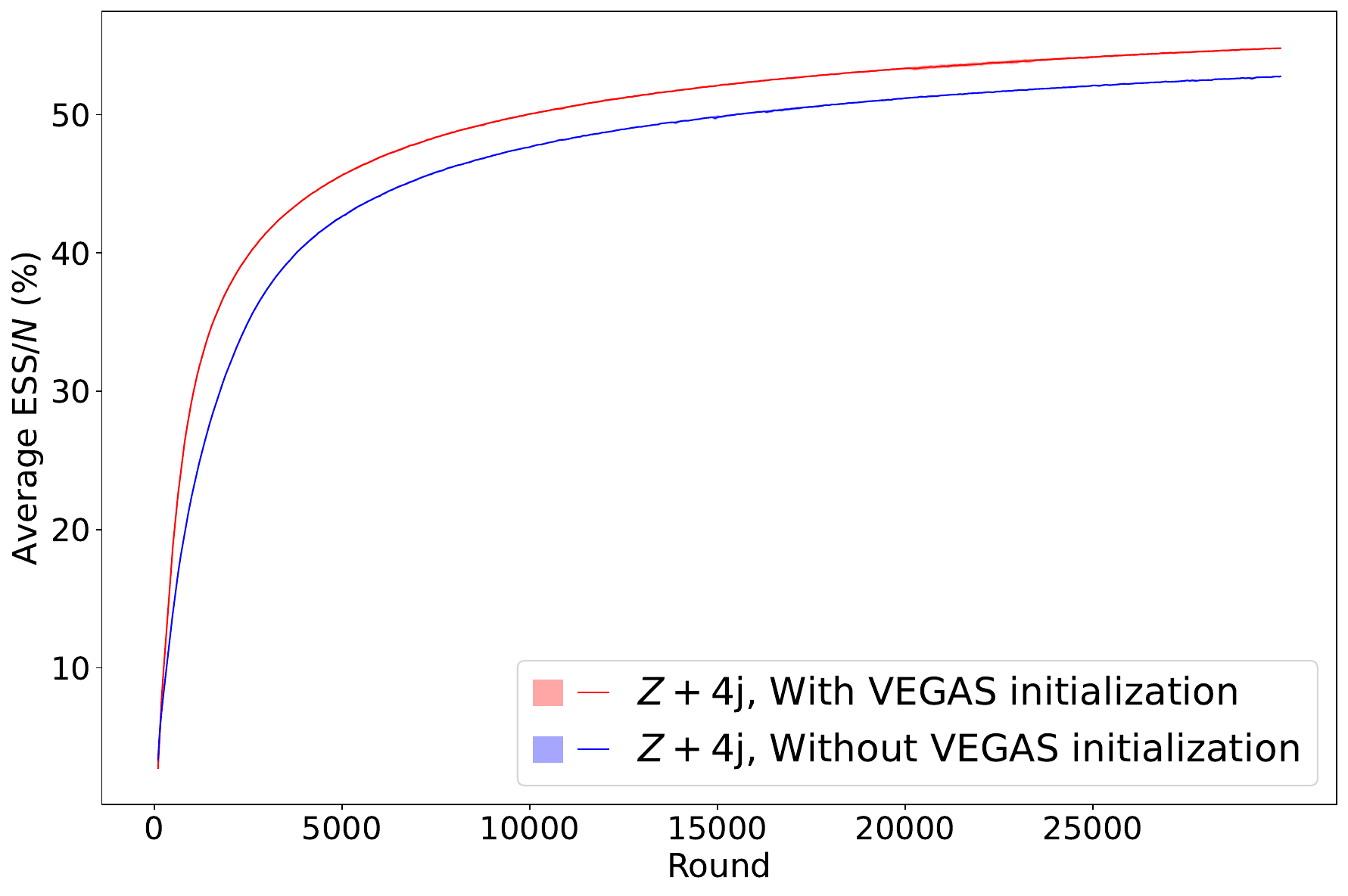}
    \end{tabular}
    \caption{
    Effect of \Vegas-informed initialization on the flow training.
    The top row shows a representative $Z+4g$ subprocess, while the bottom row shows the average over all $134$ subprocesses in $Z+4j$ production.
    The left column shows the unweighting efficiency, and the right column shows the effective sample size normalized by the number of samples, $\mathrm{ESS}/N$, as functions of the training round. For both the \Vegas-initialized and non-initialized trainings, the solid curves show the mean over five independent runs and the shaded bands show the corresponding standard deviation. Across both the individual $Z+4g$ subprocess and the full $Z+4j$ average, \Vegas-informed initialization yields higher mean unweighting efficiency and higher mean $\mathrm{ESS}/N$ throughout training.}
    \label{fig:Vegas-initialization}
\end{figure*}

Figure~\ref{fig:Vegas-initialization} motivates the use of this
\Vegas-informed initialization by comparing trainings with and without it. Here one training round denotes one online batch per subprocess. For the representative \(Z+4g\) subprocess, the \Vegas-initialized flow achieves a higher mean unweighting efficiency and a higher mean \(\mathrm{ESS}/N\) over the full training range. The same improvement is observed for the complete \zjjjj process, where the curves are averaged over all \(134\) subprocesses. The shaded bands show the run-to-run variation over five independent trainings; they are not uniformly smaller for the \Vegas-initialized setup. The main effect of the initialization is therefore not a reduction of training fluctuations, but a systematic upward shift in sampling efficiency. This indicates that the \Vegas grid provides a useful starting proposal that the flow can further refine during training.

\subsection{Training Objective}
\label{sec:training-objective}

In the loss definitions below, we allow the training samples to be drawn from a known proposal density rather than necessarily from the current flow. We denote
by
\begin{equation}
    q_{\rm prop}(z\mid h)
\end{equation}
the conditional phase-space density under which a given sample was generated.
Together with the helicity proposal probability in effect when the sample was generated, denoted by \(\pi_{\rm prop}(h)\), the joint proposal density is
\begin{equation}
    r(h,z)=\pi_{\rm prop}(h)\,q_{\rm prop}(z\mid h).
\end{equation}
The helicity proposal is updated during training; \(\pi(h)\) denotes its
current value, whereas \(\pi_{\rm prop}(h)\) denotes the value used when the corresponding sample was generated. This notation lets us write the loss for samples drawn from any known proposal density. The practical origin of \(q_{\rm prop}\), including the case of reused training samples, is described after the loss definitions.

For a batch of samples \(x_i=(h_i,z_i)\), we use the shorthand
\begin{equation}
    p_i=p_{h_i}(z_i),
    \qquad
    \rho_i=\rho_{h_i}(z_i)=|p_i|,
    \qquad
    q_{{\rm prop},i}=q_{\rm prop}(z_i\mid h_i),
    \qquad
    g_{\theta,i}=g_\theta(z_i\mid h_i),
    \qquad
    \pi_{{\rm prop},i}=\pi_{\rm prop}(h_i).
\end{equation}

We compare two objectives for training the flow: a forward Kullback–Leibler (KL) loss and a variance-oriented loss, both of which
were previously discussed for online and buffered neural importance
sampling in Ref.~\cite{Heimel:2022wyj}. We focus on these two objectives to provide a controlled comparison between a standard density-matching loss and a loss that directly targets corrected-weight fluctuations and is therefore more closely aligned with efficient importance sampling and event unweighting. Both objectives are estimated on each batch using self-normalized importance sampling, with the importance weights normalized within the batch.

\paragraph{Forward KL loss.}

The first objective is the forward KL divergence from the normalized
non-negative target density to the joint flow proposal. Denoting the normalized target by
\begin{equation}
    \widetilde\rho(x)
    =
    \frac{\rho(x)}
         {\int dx'\,\rho(x')},
\end{equation}
and recalling that
\begin{equation}
    q_\theta(h,z)
    =
    \pi(h)\,g_\theta(z\mid h),
\end{equation}
the complete forward KL divergence is
\begin{equation}
    D_{\rm KL}
    \bigl(
        \widetilde\rho
        \,\Vert\,
        q_\theta
    \bigr)
    =
    \int dx\,
    \widetilde\rho(x)
    \left[
        \log\rho(x)
        -
        \log\!\left(\int dx'\,\rho(x')\right)
        -
        \log\pi(h)
        -
        \log g_\theta(z\mid h)
    \right].
\end{equation}
During each flow-parameter update, the helicity proposal \(\pi(h)\) is treated
as fixed, while the normalization of the target density is also independent of
the flow parameters. Dropping these terms, while retaining
\(\log\rho(x)\) for the importance-sampling representation below, gives the
equivalent objective
\begin{equation}
    \mathcal{L}_{\rm KL}(\theta)
    =
    \int dx\,
    \widetilde\rho(x)
    \left[
        \log\rho(x)
        -
        \log g_\theta(z\mid h)
    \right].
\end{equation}

The training samples are not drawn from \(\widetilde\rho(x)\), but from the
proposal density
\begin{equation}
    r(x)
    =
    r(h,z)
    =
    \pi_{\rm prop}(h)\,q_{\rm prop}(z\mid h).
\end{equation}
We therefore estimate the target expectation using self-normalized importance sampling. For the forward KL loss, the unnormalized importance weights are
\begin{equation}
    \omega_i^{\rm KL}
    =
    \frac{\rho_i}
         {\pi_{{\rm prop},i}\,q_{{\rm prop},i}},
\end{equation}
To remove the unknown overall normalization of \(\rho\) and make the update invariant under its rescaling, we normalize these weights within each batch:
\begin{equation}
    \widehat\omega_i^{\rm KL}
    =
    \frac{\omega_i^{\rm KL}}
         {\sum_{j=1}^{N}\omega_j^{\rm KL}}.
\end{equation}

This gives the batch loss
\begin{equation}
    \mathcal{L}_{\rm KL}
    =
    \sum_{i=1}^{N}
    \widehat\omega_i^{\rm KL}
    \left(
        \log\rho_i
        -
        \log g_{\theta,i}
    \right).
\end{equation}
Since the term involving \(\log\rho_i\) is independent of the flow parameters, it does not contribute to the gradient. In practice, the normalized importance weights are detached from the computational graph, and we optimize
\begin{equation}
    \mathcal{L}_{\rm KL}
    =
    -\sum_{i=1}^{N}
    \operatorname{stopgrad}
    \left(
        \widehat\omega_i^{\rm KL}
    \right)
    \log g_{\theta,i}.
\end{equation}
This objective encourages the flow to place probability mass in regions where the non-negative target density \(\rho_h(z)\) is large.

\paragraph{Variance loss.}

The second objective is designed to reduce the variance of the corrected
weight magnitudes obtained when sampling from the joint flow proposal. Using
\(x=(h,z)\), \(\rho(x)\equiv\rho_h(z)\), and
\begin{equation}
    q_\theta(x)
    \equiv
    q_\theta(h,z)
    =
    \pi(h)\,g_\theta(z\mid h),
\end{equation}
the target-to-proposal weight magnitude is
\begin{equation}
    w_\theta(x)
    =
    \frac{\rho(x)}
         {q_\theta(x)}
    =
    \frac{\rho_h(z)}
         {\pi(h)\,g_\theta(z\mid h)}.
\end{equation}
The corresponding weight variance is
\begin{equation}
    \operatorname{Var}_{x\sim q_\theta}
    \left[
        w_\theta(x)
    \right]
    =
    \left\langle
        \left(
            \frac{\rho(x)}
                 {q_\theta(x)}
        \right)^2
    \right\rangle_{x\sim q_\theta}
    -
    \left\langle
        \frac{\rho(x)}
             {q_\theta(x)}
    \right\rangle_{x\sim q_\theta}^{2}.
\end{equation}

Writing the expectation values explicitly, the variance becomes
\begin{equation}
    \operatorname{Var}_{x\sim q_\theta}
    \left[
        w_\theta(x)
    \right]
    =
    \int dx\,
    \frac{\rho(x)^2}
         {q_\theta(x)}
    -
    \left[
        \int dx\,\rho(x)
    \right]^2.
\end{equation}
The second term is the square of the normalization of the non-negative target density and does not depend on the flow parameters. Thus, minimizing the weight variance with respect to \(\theta\) is equivalent to minimizing
\begin{equation}
    \mathcal{L}_{\rm var}(\theta)
    =
    \int dx\,
    \frac{\rho(x)^2}
         {q_\theta(x)}
    =
    \sum_h
    \int dz\,
    \frac{\rho_h(z)^2}
         {\pi(h)\,g_\theta(z\mid h)}.
\end{equation}
This loss penalizes regions where \(\rho(x)\) is large while the joint proposal \(q_\theta(x)\) is small, and is therefore directly connected to improving the unweighting efficiency.

Since the training samples are drawn from
\begin{equation}
    r(x)
    =
    r(h,z)
    =
    \pi_{\rm prop}(h)\,q_{\rm prop}(z\mid h),
\end{equation}
the variance-loss gradient can be written as
\begin{equation}
    \nabla_\theta\mathcal{L}_{\rm var}
    =
    -
    \int dx\,
    \frac{\rho(x)^2}{q_\theta(x)}
    \nabla_\theta\log q_\theta(x)
    =
    -
    \int dx\,
    r(x)\,
    \frac{\rho(x)^2}
         {r(x)\,q_\theta(x)}
    \nabla_\theta\log q_\theta(x).
\end{equation}
Using \(r(x)=\pi_{\rm prop}(h)q_{\rm prop}(z\mid h)\) and
\(q_\theta(x)=\pi(h)g_\theta(z\mid h)\), we estimate this gradient using
self-normalized importance sampling. The corresponding unnormalized weights are
\begin{equation}
    \omega_i^{\rm var}
    =
    \frac{\rho_i^2}
         {\pi_{{\rm prop},i}\,
          \pi(h_i)\,
          q_{{\rm prop},i}\,
          g_{\theta,i}},
\end{equation}
and the corresponding normalized weights
\begin{equation}
    \widehat\omega_i^{\rm var}
    =
    \frac{\omega_i^{\rm var}}
         {\sum_{j=1}^{N}\omega_j^{\rm var}}.
\end{equation}
As in the forward KL loss, the normalized importance weights are detached from the computational graph. Since \(\pi(h)\) is treated as fixed during each flow-parameter update,
\begin{equation}
    \nabla_\theta\log q_\theta(h,z)
    =
    \nabla_\theta\log g_\theta(z\mid h),
\end{equation}
and we optimize
\begin{equation}
    \mathcal{L}_{\rm var}
    =
    -\sum_{i=1}^{N}
    \operatorname{stopgrad}
    \left(
        \widehat\omega_i^{\rm var}
    \right)
    \log g_{\theta,i}.
\end{equation}
This is the self-normalized importance-sampling estimate of the variance-loss gradient.

Compared to the forward KL objective, this loss gives more weight to the tail of the event-weight distribution. When the sample-generating proposal is close to the current joint proposal,
\begin{equation}
    q_{\rm prop}(z_i\mid h_i)\simeq g_\theta(z_i\mid h_i),
    \qquad
    \pi_{{\rm prop},i}\simeq\pi(h_i),
\end{equation}
the KL weights scale as
\begin{equation}
    \omega_i^{\rm KL}
    \propto
    \frac{\rho_i}
         {\pi(h_i)\,g_{\theta,i}}
    =
    w_{\theta,i},
\end{equation}
whereas the variance weights scale as
\begin{equation}
    \omega_i^{\rm var}
    \propto
    \left[
        \frac{\rho_i}
             {\pi(h_i)\,g_{\theta,i}}
    \right]^2
    =
    w_{\theta,i}^2.
\end{equation}
In this limit, the KL loss therefore penalizes large corrected-weight
magnitudes linearly, while the variance loss penalizes them quadratically.

\begin{figure*}[t]
    \centering
    \begin{tabular}{cc}
        \includegraphics[width=0.48\textwidth]{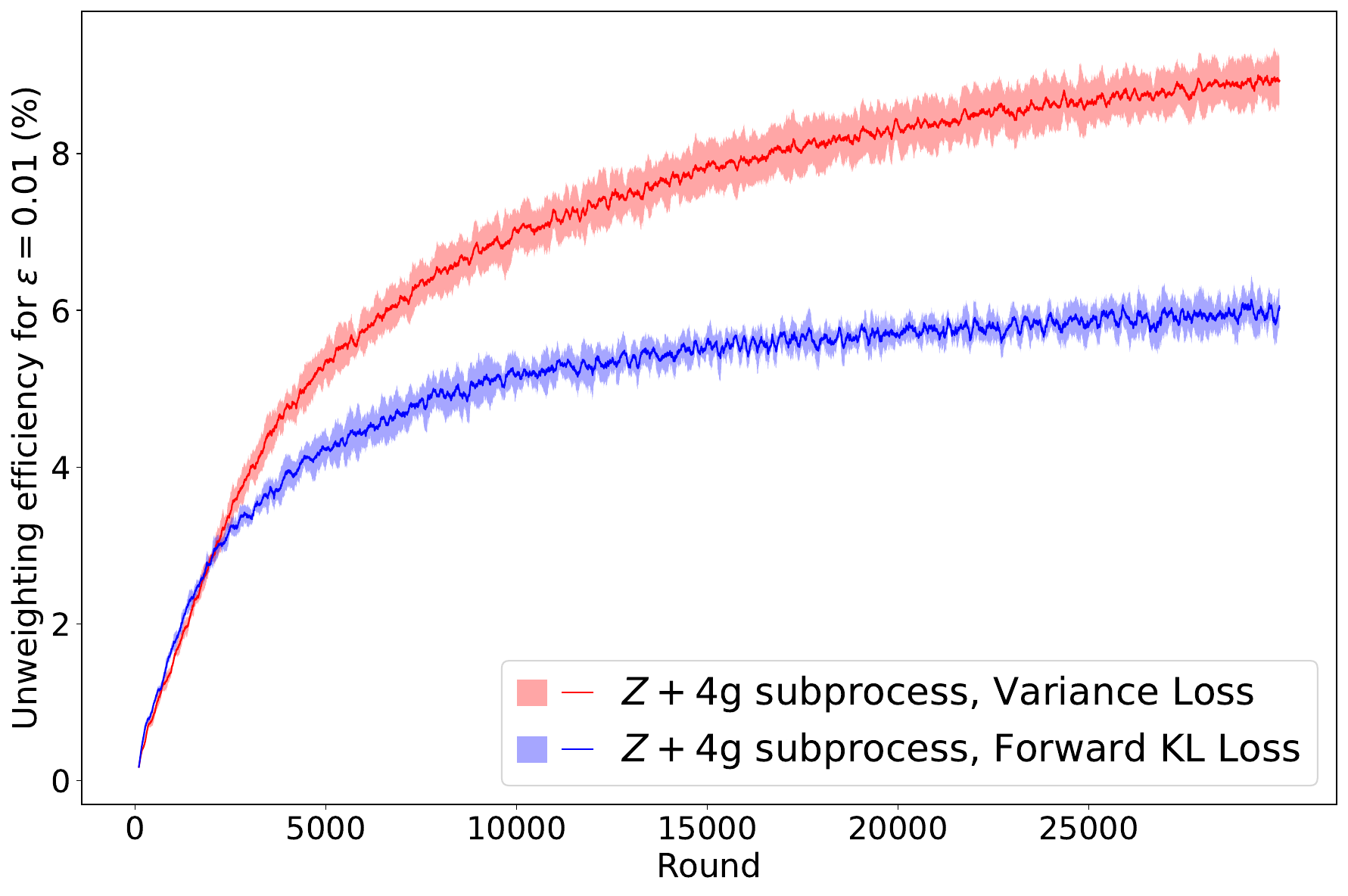} &
        \includegraphics[width=0.48\textwidth]{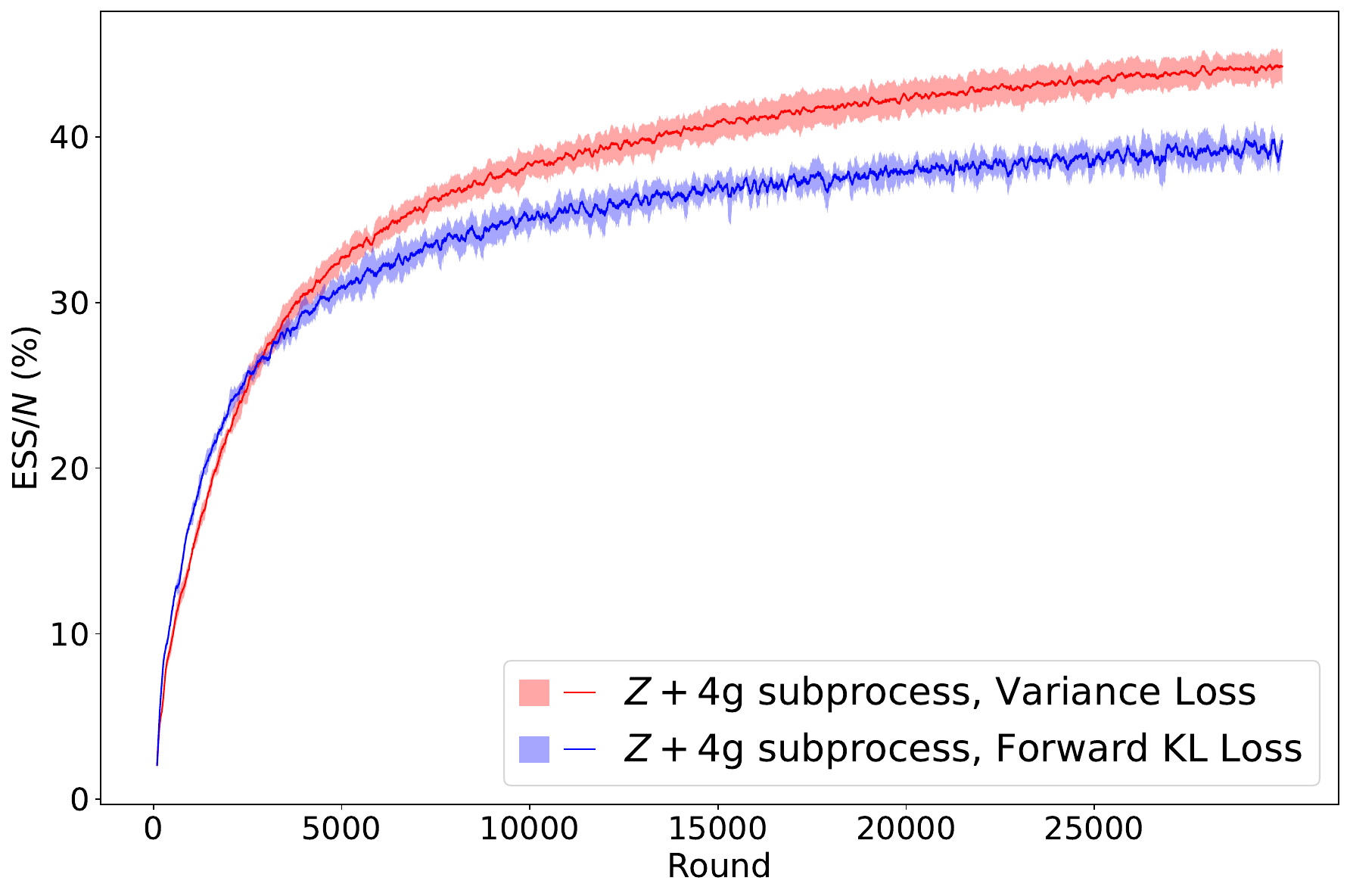} \\
        \includegraphics[width=0.48\textwidth]{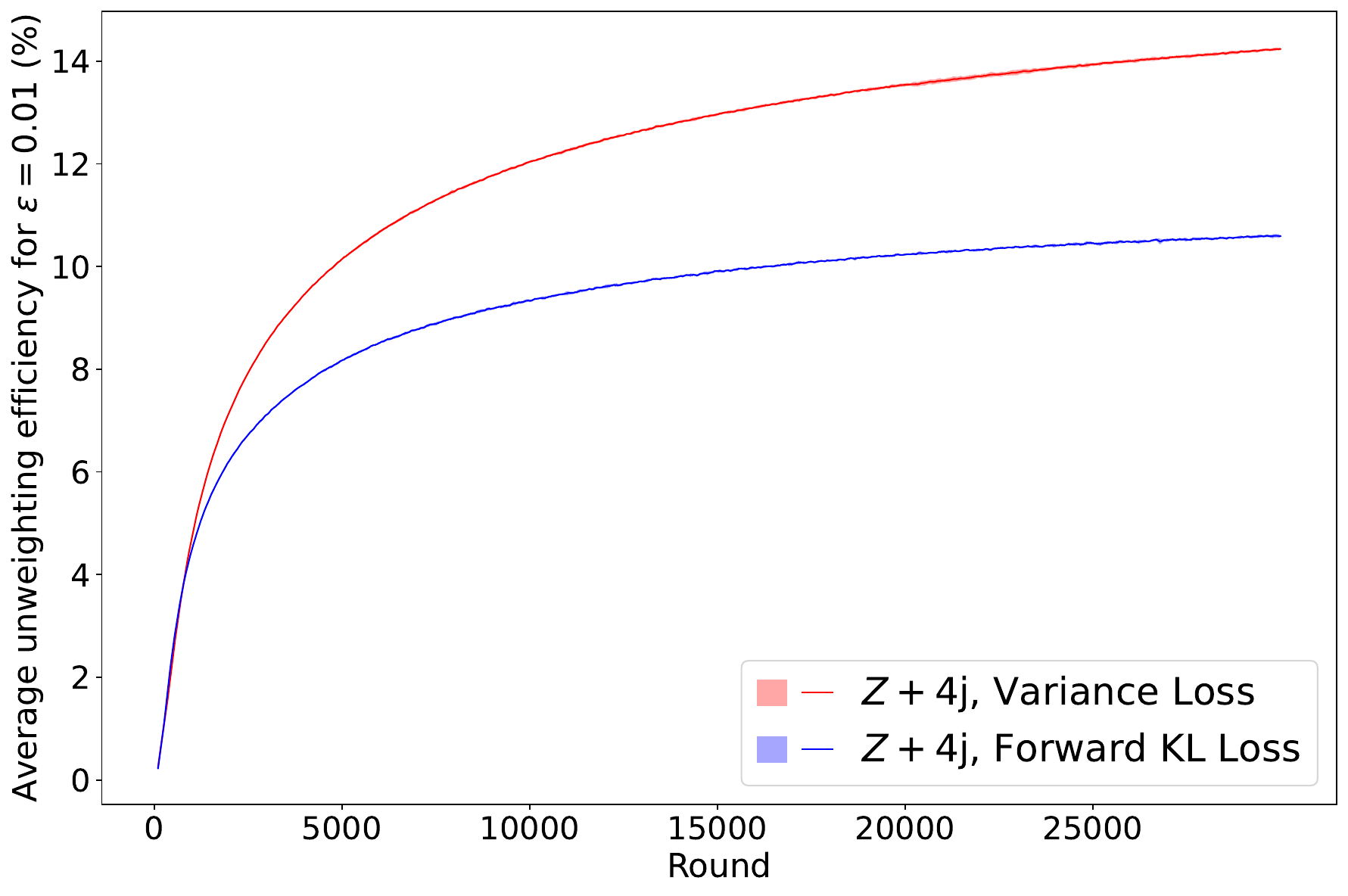} &
        \includegraphics[width=0.48\textwidth]{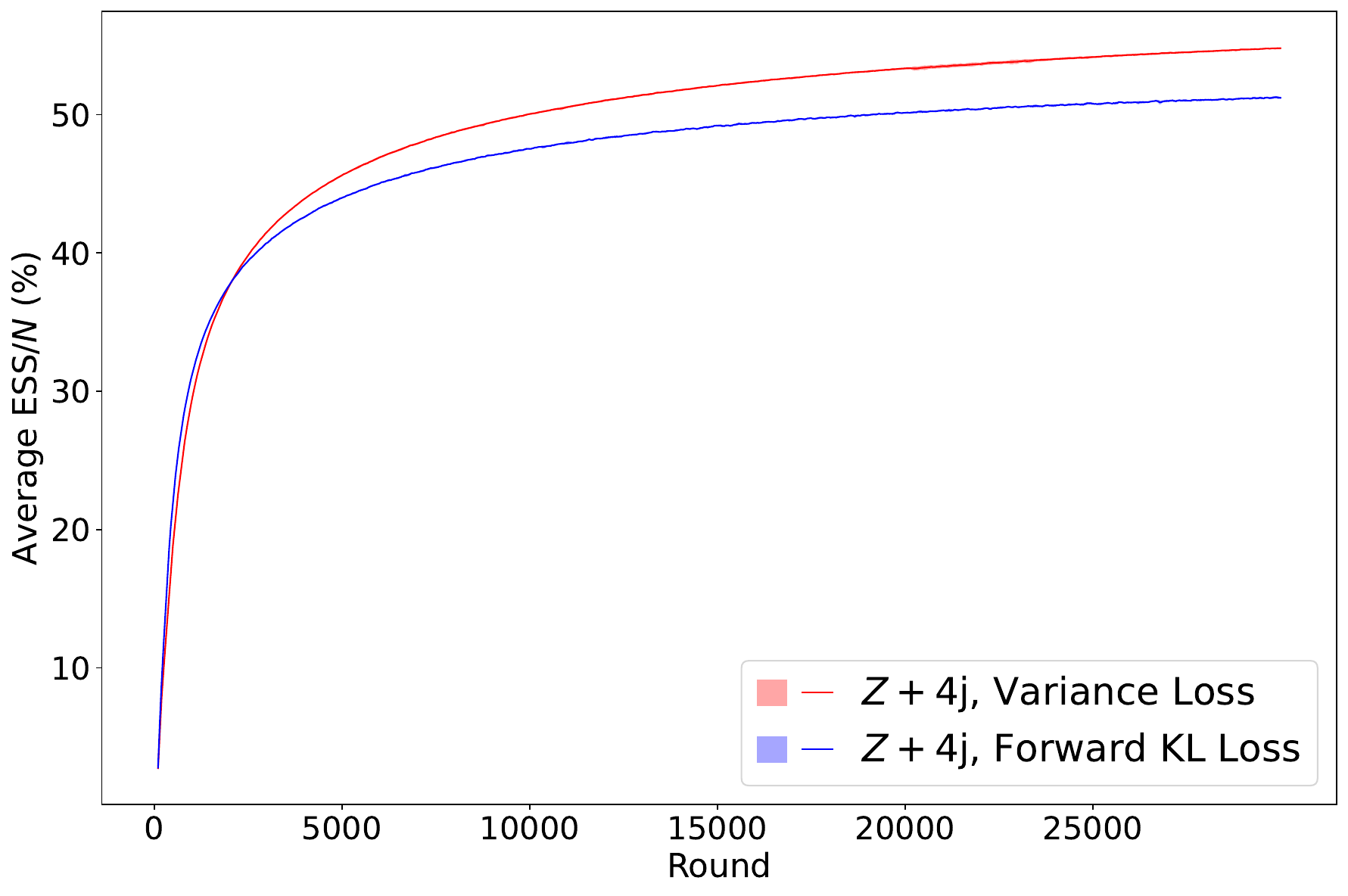}
    \end{tabular}
    \caption{
    Comparison of the variance loss and the forward KL loss for training the flow proposal.
    The top row shows a representative $Z+4g$ subprocess, while the bottom row shows the average over the $134$ partonic subprocesses contributing to $Z+4j$ production.
    The left column shows the unweighting efficiency for $\varepsilon=10^{-2}$, and the right column shows the effective sample size normalized by the number of samples, $\mathrm{ESS}/N$, as functions of the training round.
    For each objective, the solid curves denote the mean over five independent trainings, and the shaded bands denote the corresponding standard deviation.
    The variance loss gives substantially higher unweighting efficiency and $\mathrm{ESS}/N$ than the forward KL loss over most of the training trajectory, both for the individual subprocess and for the full $Z+4j$ average.
    }
    \label{fig:variance-vs-forward-KL}
\end{figure*}

Figure~\ref{fig:variance-vs-forward-KL} confirms this expectation. Here one training round denotes one online batch per subprocess. For the representative $Z+4g$ subprocess, the variance loss leads to a substantially higher unweighting efficiency than the forward KL loss after the early training stage, and the gap continues to grow over the full optimization. The same behavior is seen in $\mathrm{ESS}/N$: although the two objectives are comparable during the first few thousand rounds, the variance loss gives a clearly larger effective sample size throughout the later training. Averaging over all $134$ subprocesses in $Z+4j$ shows the same pattern. The variance loss therefore provides a more effective objective for optimizing the proposal distribution for unweighted event generation, where the relevant target is not likelihood maximization itself but the reduction of weight fluctuations and the corresponding improvement in sampling efficiency.

\subsection{Practical Training Procedure}
\label{sec:training-procedure}

In all production trainings we use the variance loss described above. For each training point, the loss requires the non-negative target value
\(\rho_i=|p_i|\), the current conditional-flow density \(g_{\theta,i}\), the current helicity proposal probability \(\pi(h_i)\), and the sample-generating proposal density
\(\pi_{{\rm prop},i}q_{{\rm prop},i}\). \Pepper unweighting is disabled during training, so each \Pepper call returns the raw, potentially signed event weights \(p_i\) rather than already unweighted events. Their magnitudes \(\rho_i\) are used to construct the importance weights in the flow objective.

To specify the helicity proposal \(\pi(h_i)\) used during training, let \(\mathcal{H}_{\rm act}\) denote the set of active
helicity configurations for each subprocess and define
\begin{equation}
    N_{\rm act}
    =
    |\mathcal{H}_{\rm act}|.
\end{equation}
The active helicity configurations are indexed consecutively as
\begin{equation}
    h=0,\ldots,N_{\rm act}-1.
\end{equation}
The helicity embedding table contains \(64\) entries, corresponding to the
maximum value of \(N_{\rm act}\) among the subprocesses considered. For a
given subprocess, only the first \(N_{\rm act}\) embedding entries are used.

The initial helicity configuration proposal, denoted by \(\pi_0(h)\), is uniform over the active helicity configurations,
\begin{equation}
    \pi_0(h)
    =
    \frac{1}{N_{\rm act}},
    \qquad
    h=0,\ldots,N_{\rm act}-1.
\end{equation}
At the beginning of each training round, helicity configurations are sampled from the
current proposal,
\begin{equation}
    h_i\sim\pi(h),
\end{equation}
and phase-space random numbers are sampled from the conditional flow,
\begin{equation}
    z_i\sim g_\theta(z\mid h_i).
\end{equation}
The subprocess-dependent helicity random number used by \Pepper is divided into \(N_{\rm act}\) equal intervals. The discrete local helicity label is
mapped to the corresponding interval according to
\begin{equation}
    r_{h,i}
    =
    \frac{h_i+u_i}{N_{\rm act}},
    \qquad
    u_i\sim U(0,1).
\end{equation}
The full random-number input to \Pepper is then formed from the phase-space random numbers \(z_i\) and the helicity random number \(r_{h,i}\). \Pepper evaluates the corresponding raw event weights,
\begin{equation}
    p_i=p_{h_i}(z_i),
    \qquad
    \rho_i=|p_i|,
\end{equation}
with unweighting disabled.

The flow model, including its parameters, helicity embeddings, sampling, and density evaluation, operates in \texttt{float32}. The phase-space coordinates generated by the flow are cast to \texttt{float64} before being passed to \Pepper, as it is its default precision required for momentum conservation; this conversion changes their data type but does not increase their underlying numerical resolution. The raw \Pepper weights and the subsequent corrected-weight calculations are handled in \texttt{float64}. 

Compared with an otherwise identical all-\texttt{float64} setup for
\(Z+4j\) production, this mixed-precision setup reduces the training time per round from \(5.21\,\mathrm{s}\) to \(3.91\,\mathrm{s}\), corresponding to a \(25\%\) reduction in wall time. It also reduces the device-memory footprint sufficiently to train all subprocesses of each process considered here concurrently. Because the present study is restricted to leading-order event generation, we expect this precision setup to be adequate for the benchmarks considered here. Its suitability for higher-order calculations would require dedicated validation.

For freshly generated online samples, \(q_{{\rm prop},i}\) and \(\pi_{{\rm prop},i}\) are given by the current conditional-flow density and
helicity-configuration proposal, respectively, evaluated at the time of generation. These values are detached and stored before the flow and helicity-configurations proposals are updated. The variance loss is then evaluated using the non-negative target values \(\rho_i\), the current model log-density \(\log g_{\theta,i}\), the current helicity-configuration proposal probabilities \(\pi(h_i)\), and the stored sample-generating factors \(q_{{\rm prop},i}\) and \(\pi_{{\rm prop},i}\). The helicity-configuration probabilities and importance weights are treated as fixed during the flow-parameter update, and gradients are propagated only through the current model log-density. The subprocess losses are summed, and the flow and helicity-embedding parameters are updated with the AdamW optimizer~\cite{adamw:2019} using learning rate \(10^{-3}\).

After each online batch, the helicity-configuration proposal is updated using the same batch. For each configuration, we estimate its fixed-helicity contribution by dividing out the conditional flow density,
\begin{equation}
    s_h
    =
    \left\langle
    \frac{\rho_h(z)}
         {g_\theta(z\mid h)}
    \right\rangle_{z\sim g_\theta(z\mid h)} .
\end{equation}
The next helicity-configuration proposal is set proportional to this estimate,
\begin{equation}
    \pi_{\rm new}(h)
    =
    \frac{\max(s_h,\delta)}
         {\sum_{h'}\max(s_{h'},\delta)} ,
\end{equation}
with a small numerical floor \(\delta\) to avoid assigning exactly zero
probability to any active helicity configuration. This adaptively increases the sampling
probability for helicity configurations with larger estimated contributions.

The unweighting efficiency and \(\mathrm{ESS}/N\) diagnostics defined above are evaluated on each online batch and saved for validation and for comparing different training runs.

\paragraph{Replay-buffer training.}

The notation with \(q_{\rm prop}\) also allows us to reuse samples generated by earlier versions of the flow~\cite{Heimel:2022wyj}. In addition to online training on newly generated batches, we use a replay buffer to reuse recent \Pepper evaluations. This is useful because the \Pepper matrix-element evaluation accounts for a substantial fraction of the online-update time. In a representative
\zjjjj timing measurement over \(100\) online updates, \Pepper evaluation required \(96.2\,\mathrm{s}\) out of a total of \(211.9\,\mathrm{s}\), corresponding to approximately \(45\%\). For stored samples, this evaluation is reused, while only the current flow density needs to be recomputed. After each online batch, the tuples
\begin{equation}
    \left\{
    z_i,\,
    h_i,\,
    \rho_i,\,
    q_{{\rm prop},i},\,
    \pi_{{\rm prop},i}
    \right\}
\end{equation}
are stored in subprocess replay buffers, each of which retains the most recent \(50\) batches.

After the online optimizer step, we perform \(k_{\rm buff}=4\) replay updates. For each replay update, one stored batch is sampled from each subprocess buffer. The current flow density \(g_\theta(z_i\mid h_i)\) is recomputed for the stored points, and the current helicity-configuration probability \(\pi(h_i)\) is used in the joint model proposal \(q_\theta(h_i,z_i)\), while the original sample-generating
factors \(q_{{\rm prop},i}\) and \(\pi_{{\rm prop},i}\) remain fixed. The same variance loss is then evaluated with the off-policy importance correction described above. Thus samples generated by earlier versions of the flow remain usable because their original
sampling density is stored explicitly.

Each training round therefore consists of one online update using newly generated \Pepper weights, followed by four replay updates using recent stored batches. Checkpoints are saved every \(50\) training rounds for each subprocess, including the flow parameters, helicity embedding, current helicity-configuration proposal probabilities, and update counter. The unweighting efficiency and \(\mathrm{ESS}/N\) histories are saved separately for validation and for comparing different training runs.

\subsection{\Pepper{}+Flow Event Generation}
\label{sec:python-direct-generation}

After training, the selected flow checkpoints are used in a
workflow controlled by our new Python layer to generate unweighted events. We refer to this end-to-end workflow
as ``\Pepper{}+Flow'' event generation throughout this paper. For each subprocess, we load the trained flow, the helicity embedding, and the final helicity-configuration proposal probabilities. The flow then replaces the \Vegas proposal used in the conventional \pepvegas workflow: helicity configurations are sampled from the learned discrete proposal \(\pi(h)\), and phase-space random numbers are sampled from the conditional flow,
\begin{equation}
    h\sim \pi(h),
    \qquad
    z\sim g_\theta(z\mid h).
\end{equation}
The sampled helicity-configuration label is converted to the continuous helicity-configuration random number expected by \Pepper, and the full random-number vector is passed to \Pepper for a matrix-element evaluation.

For each generated point, \Pepper returns the raw, potentially signed event
weight \(p_h(z)\), while the corresponding non-negative target value is
\(\rho_h(z)=|p_h(z)|\). The flow-corrected weight magnitude is
\begin{equation}
    w(z,h)
    =
    \frac{\rho_h(z)}
         {\pi(h)\,g_\theta(z\mid h)},
\end{equation}
with the same normalization convention as in training. Before event generation, the effective maximum weight is determined independently for each subprocess. To ensure a like-for-like comparison with \pepvegas{}, this preliminary maximum-finding pass uses, for each subprocess, exactly the same number of raw trial points as used by \pepvegas for its corresponding maximum-weight determination. From these weights and for a chosen tail tolerance \(\varepsilon\), a single subprocess-specific effective maximum \(w_{\max,\mathrm{eff}}^{(i)}(\varepsilon)\) is obtained, where \(i\) labels the subprocess.

Once determined, \(w_{\max,\mathrm{eff}}^{(i)}(\varepsilon)\) is held fixed and used for every subsequent production batch of subprocess \(i\). The unweighting is therefore global across all production batches of a given subprocess. The runtime of this dedicated maximum-finding pass is included in both the reported \pepflow training-plus-generation and the \pepflow generation-only benchmarks reported below. An event belonging to subprocess \(i\) is accepted with probability
\begin{equation}
    P_{\rm acc}(z,h\mid i)
    =
    \min\left(
        1,\,
        \frac{w(z,h)}
             {w_{\max,\mathrm{eff}}^{(i)}(\varepsilon)}
    \right).
\end{equation}
Accepted events satisfying
\(w(z,h)\leq w_{\max,\mathrm{eff}}^{(i)}(\varepsilon)\) are assigned the
common output-weight magnitude
\(w_{\max,\mathrm{eff}}^{(i)}(\varepsilon)\), multiplied by
\(\operatorname{sign}[p_h(z)]\) to retain the sign of the raw \Pepper
weight. Overweight events satisfying
\(w(z,h)>w_{\max,\mathrm{eff}}^{(i)}(\varepsilon)\) are accepted with unit probability and retain their original corrected-weight magnitude \(w(z,h)\), again multiplied by \(\operatorname{sign}[p_h(z)]\). The overweight contributions are therefore preserved rather than truncated. More precisely, \(w_{\max,\mathrm{eff}}^{(i)}(\varepsilon)\) is chosen as
the smallest threshold satisfying
\begin{equation}
    \frac{
        \displaystyle
        \sum_{j:\,
        w_j^{(i)}>w_{\max,\mathrm{eff}}^{(i)}(\varepsilon)}
        w_j^{(i)}
    }{
        \displaystyle\sum_j w_j^{(i)}
    }
    \leq \varepsilon ,
\end{equation}
where \(j\) indexes the trial events used in the maximum-finding pass.

The Python generation script also determines how the requested unweighted-event sample is allocated across subprocesses. It reads the per-subprocess cross-section estimates obtained during the preceding \Pepper integration from the cached results. The subprocesses are then generated sequentially, with the total number of requested events split between the available GPUs. For each subprocess, event batches are sampled, evaluated, and unweighted until its assigned number of accepted events is reached. If a batch does not yield enough accepted events, another batch is processed; accepted events beyond the remaining target in the final batch are discarded.

To write the resulting events in the standard \Pepper HDF5 format, based on the LHEH5 format specified in Refs.~\cite{Hoche:2019flt,Bothmann:2023ozs} and implemented using the HDF5 library~\cite{hdf5} through HighFive~\cite{highfive}, we use the \Pepper event writer exposed through the \Pyper interface (Sec.~\ref{sec:Pepper-workflow}). The Python code controls the sampling and unweighting decision, while \Pepper remains responsible for writing the accepted events, including their momenta, weights, scales, and flavor information, to the output file. Concretely, the Python code opens the \Pepper output writer, evaluates each batch using externally supplied random numbers, and constructs the final batch weights after unweighting. Rejected events are assigned zero weight, while the weights of the accepted events are passed back to \Pepper. \Pepper then writes the nonzero-weight events using its existing event writer. This procedure is repeated batch by batch until the requested number of accepted events has been generated, after which the writer is finalized.

The \pepvegas and \pepflow benchmarks use the same cached integration results, physics settings, value of \(\varepsilon\), per-subprocess raw-trial budgets for the effective-maximum determination, unweighting and overweight-event prescription, and \Pepper HDF5 writer. At the level of Monte Carlo sampling, the two workflows therefore differ only in the proposal density: \pepflow uses the learned helicity-conditioned proposal \(\pi(h)g_\theta(z\mid h)\) in place of the conventional proposal used by \pepvegas. Although the two proposals generally produce different corrected-weight distributions and hence different numerical effective maxima, these maxima are determined independently using matched per-subprocess trial statistics and the same prescription.

The learned proposal density is included explicitly in the corrected event weights, so changing the proposal affects sampling efficiency rather than the target distribution. Although \pepflow performs the accept--reject decision in the Python layer whereas \pepvegas performs it natively, the underlying unweighting prescription is the same. For identical physics settings and the same value of \(\varepsilon\), event samples produced by the two workflows should therefore agree up to statistical fluctuations.

Together, these steps make the learned proposal part of a complete
event-production workflow rather than a standalone phase-space sampler. The construction used here is technically distinct from the earlier \Pepper studies based on continuous normalizing flows trained with flow matching and on RegFlow~\cite{Bothmann:2025lwg,Bothmann:2026dar}: neither continuous-flow nor RegFlow models are used in the present workflow. Instead, we train helicity-conditioned rational-quadratic coupling flows directly and deploy them as subprocess-specific or group-shared conditional proposals. The operational unit is the complete physical process: the Python layer coordinates sampling, subprocess allocation, and unweighting across the full subprocess collection, while \Pepper writes standard event samples. The resulting benchmark therefore measures the production of a complete event sample rather than the performance of a representative partonic channel.

\section{Subprocess-Specific Flow Generation}
\label{sec:subprocess-specific-flows}

The previous section described how to construct and train a helicity-conditioned flow proposal for a single partonic subprocess. We now apply this construction to full many-jet processes. Unless stated otherwise, the benchmark processes studied in this work are trained and generated on four NVIDIA H100 SXM5 GPUs, with \(94\,\mathrm{GB}\) of memory per GPU. Note that for all five processes considered in this paper, this four-GPU configuration provides sufficient device memory to distribute all subprocess-specific flows across the GPUs and train them concurrently, including the \(168\) flows required for \zjjjjj production. As a complementary single-GPU study, we also test the subprocess-specific workflow on a single NVIDIA RTX 4000 Ada Generation GPU; these results are collected in Appendix~\ref{app:rtx-results}.

Although our benchmarks use at most four GPUs, the workflow is not intrinsically limited to this configuration. During training, the independent subprocess-specific flows require no gradient synchronization and can be distributed over additional devices, up to assigning one GPU to each flow. Event generation can be parallelized beyond this granularity: a trained flow can be replicated across multiple GPUs, with each device generating independent batches for the same subprocess.

We use \zjjjj production as the main example. The full \zjjjj process contains \(134\) partonic subprocesses, each with its own phase-space structure, active helicity configurations, and event weight distribution. We therefore assign an independent conditional flow to each subprocess,
\begin{equation}
    g_{\theta_a}(z\mid h),
\end{equation}
where \(a\) labels the subprocess. The flow parameters \(\theta_a\), helicity embedding, helicity-configuration proposal, replay buffer, and training diagnostics are all subprocess-specific.

This one-flow-per-subprocess strategy allows each proposal to specialize to the corresponding partonic channel. This is useful because different subprocesses can have different singularity patterns and different optimization difficulty. In practice, the flows are distributed across GPUs for parallel training, with flow-density evaluations vectorized over the models assigned to each GPU.

\subsection{Subprocess-Specific Flows and Parallel Training}
\label{sec:parallel-training}

The subprocess list is kept in \Pepper subprocess-index order and divided into four contiguous subsets whose sizes differ by at most one. Each subset is assigned to a separate worker running on one H100 GPU. Before each \Pepper call, the corresponding subprocess index is used to select the appropriate matrix element.

Within each worker, the \(N_{\rm local}\) subprocess models are trained together, where \(N_{\rm local}\) denotes the number of subprocesses assigned to that worker. We wrap each flow and its helicity embedding into a single module and stack the corresponding parameters and buffers into tensors of shape \((N_{\rm local},\ldots)\).

The log-density evaluation is vectorized with a functional call. For one
subprocess, we define
\begin{equation}
    \ell_a
    =
    \log g_{\theta_a}(z_a\mid h_a).
\end{equation}
The same function is then mapped over the local subprocess dimension using
\texttt{vmap},
\begin{equation}
    \{\ell_a\}_{a=1}^{N_{\rm local}}
    =
    \mathrm{vmap}
    \left[
        \log g_{\theta_a}(z_a\mid h_a)
    \right].
\end{equation}
In practice, the batch of phase-space random numbers has shape
\begin{equation}
    (N_{\rm local},\,N_{\rm batch},\,D),
\end{equation}
and the helicity batch has shape
\begin{equation}
    (N_{\rm local},\,N_{\rm batch}).
\end{equation}
The vectorized call returns the model log-density for all local subprocesses in one operation.

In each training round, for every local subprocess \(a\), we sample helicity configurations from its current proposal \(\pi_a(h)\), sample phase-space random numbers from its flow, and call \Pepper with the corresponding subprocess index to obtain the raw event weights. After all local subprocess batches have been generated, the flow log-densities are evaluated in vectorized form over the subprocess dimension. The variance loss is constructed separately for each subprocess, and the losses are summed before the optimizer step,
\begin{equation}
    \mathcal{L}_{\rm worker}
    =
    \sum_{a=1}^{N_{\rm local}}
    \mathcal{L}_a .
\end{equation}
Thus each subprocess keeps its own model, helicity-configuration proposal, replay buffer, and diagnostic histories, while the neural-network density evaluation and gradient computation are batched over subprocesses whenever possible.

The replay-buffer updates use the same stacked representation. For each replay step, one stored batch is sampled uniformly from each local subprocess buffer. Once the buffer is full, the most recently added batch is excluded from the replay draw. The current flow log-density and current helicity-configuration probability enter the current joint model proposal, while the stored proposal density and helicity-configuration probability
are used for the off-policy importance correction. The replay losses are summed over subprocesses and used for an additional optimizer step. As described in Sec.~\ref{sec:training-objective}, each training round consists of one online update followed by four replay updates. After each update, the stacked parameters are synchronized back to the individual subprocess modules used for sampling.

We also use PyTorch compilation~\cite{Ansel:2024pytorch2} for the most frequently called flow operations. The vectorized log-probability function, including the functional model call and the \texttt{vmap} over subprocesses, is compiled with \texttt{torch.compile}. The sampling wrapper for each subprocess is also compiled. These compilation steps do not change the training objective or the generated distribution, but accelerate the repeated flow-density and sampling operations by reducing dispatch overhead and enabling graph-level optimizations. For \zjjjj production, compilation reduces the average wall time per training round from \(14.65\,\mathrm{s}\) to \(3.91\,\mathrm{s}\), corresponding to a \(73\%\) reduction. The averages are measured over \(1000\) training rounds and include
the initial compilation overhead. The \Pepper matrix-element evaluation is not part of the PyTorch-compiled graph, but is performed by the compiled C++ \Pepper code through \texttt{process\_batch}, the \Pyper batch-evaluation interface for externally supplied random numbers (Sec.~\ref{sec:Pepper-workflow}).

\section{Grouped Conditional Flow Generation}
\label{sec:grouped-flows}

\subsection{Grouping by Initial- and Final-State Parton Content}
\label{sec:grouping-parton-content}

As an alternative to assigning a completely independent flow to every subprocess, we investigate whether subprocesses with related partonic structures can benefit from a common conditional flow with shared neural-network weights. Subprocesses within the same parton-content group are expected to exhibit
similar kinematic features, which independent flows would otherwise have to learn separately. A conditional flow shared within the group allows training information from several subprocesses to shape a common set of transformation layers, while retaining the flexibility to model subprocess-dependent differences. This construction therefore tests whether parameter sharing provides a useful inductive bias, particularly for processes with many subprocesses, and yields a more compact parameterization whose number of independent flow models scales with the number of groups rather than with the number of subprocesses. Within each group, the flow is conditioned on both the helicity configuration and the subprocess index.

The grouping is based only on the partonic content of the subprocess.  Each external parton is first mapped to one of three categories,
\begin{equation}
    g,\qquad q,\qquad \bar q ,
\end{equation}
where \(q\in\{u,d,s,c,b\}\) and \(\bar q\in\{\bar u,\bar d,\bar s,\bar c,\bar b\}\). For each subprocess we then classify the initial state into one of the categories
\begin{equation}
    gg,\quad qg,\quad \bar q g,\quad qq,\quad q\bar q,\quad \bar q\bar q .
\end{equation}
Since the total number of final-state partons is fixed for each process, the final-state parton content is classified by \(n_q^{\rm final}\), the number
of final-state quarks and antiquarks; the number of final-state gluons is then fixed automatically.

The group label for a subprocess is therefore defined as
\begin{equation}
    G =
    \left(
        C_{\rm init},
        n_q^{\rm final}
    \right),
\end{equation}
where \(C_{\rm init}\) is the initial-state category.  Subprocesses with the same value of \(G\) are assigned to the same group.  Applying this rule to the \zjjjj subprocess list gives \(13\) groups in total.  Each group contains subprocesses with the same initial-state parton category and the same number of final-state quarks and gluons, while the individual quark flavors \(u,d,s,c,b\) may differ within the group.

\subsection{Shared Flows with Process Embeddings}
\label{sec:shared-flows-process-embeddings}

\paragraph{Grouped-flow architecture.}

For each group \(G\), we build a single conditional flow shared by all subprocesses in that group. The density modeled by the grouped flow is
\begin{equation}
    g_{\theta_G}(z\mid h,a),
\end{equation}
where \(h\) is the helicity configuration and \(a\) labels the subprocess, as in Sec.~\ref{sec:subprocess-specific-flows}. Here \(a\) denotes the local index within group \(G\). Thus, unlike the subprocess-specific setup where each subprocess has its own flow parameters, the grouped setup uses one set of flow parameters per group and distinguishes subprocesses through an additional conditioning variable.

For a group \(G\) with \(n_G\) subprocesses, the subprocess embedding
table has \(n_G\) entries, each of dimension \(d_{\rm proc}\),
\begin{equation}
    c_{\rm proc}(a)\in\mathbb{R}^{d_{\rm proc}}.
\end{equation}
Each group also has its own helicity embedding table, with one entry of dimension \(d_{\rm hel}\) for each possible helicity-configuration label,
\begin{equation}
    c_{\rm hel}(h)\in\mathbb{R}^{d_{\rm hel}}.
\end{equation}
Only the active helicity-configuration labels for the corresponding
subprocess are sampled during training and generation. The context vector passed to the coupling networks is the concatenation of these two embeddings,
\begin{equation}
    c(h,a)=\left[c_{\rm proc}(a),c_{\rm hel}(h)\right],
    \qquad
    d_{\rm context}=d_{\rm proc}+d_{\rm hel}.
\end{equation}
The same context vector is provided to every coupling layer of the shared group flow.

For the \zjjjj grouped-flow setup, the largest parton-content group contains
\(36\) subprocesses, and the largest helicity space contains \(64\) helicity
configurations. We therefore choose \(d_{\rm proc}=36\) and
\(d_{\rm hel}=64\), giving \(d_{\rm context}=100\).
For the other benchmark processes, the subprocess- and helicity-embedding dimensions are chosen analogously according to the largest subprocess group and helicity-configuration space of the respective process.

The flow architecture itself is the same as in the subprocess-specific model.  Each grouped flow uses a uniform base distribution on \([-1,1]^D\), followed by the unit-interval transform that maps between the \Pepper random-number domain \(z\in[0,1]^D\) and the internal flow domain. The main transform consists of eight rational-quadratic coupling layers with alternating masks and \(K=10\) spline bins.  The only architectural change is that the coupling MLPs receive the enlarged context vector \(c(h,a)\), rather than only the helicity embedding.

\paragraph{Process and helicity embeddings.}

The helicity embedding is shared by all subprocesses within a group.  It is implemented as an trainable embedding table with \(d_{\rm hel}\) possible helicity-configuration labels and embedding dimension \(d_{\rm hel}\), where \(d_{\rm hel}=64\) for the \zjjjj setup. Each label \(h\) selects one row
\(c_{\rm hel}(h)\) of this table. The table is randomly initialized and optimized jointly with the grouped flow, without any normalization or
projection constraint. Since each subprocess may have fewer active helicity configurations, only its active helicity-configuration labels are used during training and generation.

The subprocess embedding is also group specific. For a group with \(n_G\) subprocesses, the embedding table has \(n_G\) rows, one for each subprocess in the group. The association between a subprocess and its embedding is fixed by this row index: subprocess \(a\) always uses the \(a\)-th row of the table, although the entries of that row are trainable. We initialize the subprocess embedding as a one-hot vector,
\begin{equation}
    c_{\rm proc}^{(0)}(a)_j =
    \begin{cases}
        1, & j=a,\\
        0, & j\neq a,
    \end{cases}
\end{equation}
with \(a=1,\ldots,n_G\) and \(j=1,\ldots,d_{\rm proc}\). Equivalently, the \(n_G\times d_{\rm proc}\) embedding table is initialized as
\begin{equation}
    E_{\rm proc}^{(0)}
    =
    \left[
        I_{n_G}
        \,\middle|\,
        0_{n_G\times(d_{\rm proc}-n_G)}
    \right].
\end{equation}
This gives every subprocess a distinct context vector and enables the subprocess-specific \Vegas corrections to be inserted through the context-to-output weights, as described below. After initialization, all embedding components are optimized jointly with the flow. No normalization or projection is applied, so the embeddings need not remain one-hot and neither their row nor column sums are constrained during training.

\paragraph{\Vegas initialization for grouped flows.}

The \Vegas-informed initialization is modified to support multiple subprocesses sharing the same flow. The first six coupling layers are initialized to identity transformations, as in the subprocess-specific model.

For the last two coupling layers, the initialization uses the \Vegas grids of all subprocesses in the group.  We start from the identity spline bias, $b_{\rm id}$,
which corresponds to uniform spline widths, uniform spline heights, and unit derivatives.  For each subprocess \(a\) in the group, we construct the \Vegas spline bias, $b_{\rm \Vegas}^{(a)}$,from its own \Vegas grid, using the same bin-merging and projection procedure as before.  The subprocess-specific \Vegas correction is then
\begin{equation}
    \Delta b^{(a)}
    =
    b_{\rm \Vegas}^{(a)} - b_{\rm id}.
\end{equation}

To implement the subprocess-dependent initialization, we introduce, in parallel with each nonlinear coupling MLP, a bias-free linear branch that maps the context directly to the raw spline parameters. The output of this branch is added to the output of the MLP. Denoting the nonlinear network in coupling layer \(\ell\) by
\(f_{\phi_\ell}(x_\ell,c)\), its output is
\begin{equation}
    s(x,c)
    =
    f_\phi(x,c)
    +
    W_{\rm ctx}c,
    \qquad
    W_{\rm ctx}
    =
    \left[
        W_{\rm proc},
        W_{\rm hel}
    \right],
\end{equation}
where \(x\) denotes the coupling-layer input and
\(c=[c_{\rm proc},c_{\rm hel}]\). The weights of the final linear layer of
\(f_\phi\) are initialized to zero, and its bias is set to
\(b_{\rm id}\), so that \(f_\phi(x,c)=b_{\rm id}\) at initialization.
The matrix \(W_{\rm ctx}\) is initially set to zero. In the last two coupling layers, its \(a\)-th subprocess column is then set to
\(\Delta b^{(a)}\), while the helicity-related columns remain zero.

Since \(c_{\rm proc}^{(0)}(a)\) is one-hot, it selects the corresponding
column of \(W_{\rm proc}\), giving
\begin{equation}
    s(x,c)
    =
    b_{\rm id}
    +
    W_{\rm proc}c_{\rm proc}^{(0)}(a)
    =
    b_{\rm id}
    +
    \Delta b^{(a)}
    =
    b_{\rm \Vegas}^{(a)}.
\end{equation}
Thus, the same conditional flow reproduces the subprocess-specific
\Vegas initialization selected by \(a\). During training, \(f_\phi\), \(W_{\rm ctx}\), and both embedding tables are optimized jointly,
so their relative contributions are determined by the training objective. The direct linear term depends only on the context and therefore acts as a context-dependent baseline; input-dependent phase-space correlations are learned through the nonlinear network \(f_\phi(x,c)\).

\subsection{Parallel Training across Subprocess Groups}
\label{sec:grouped-flow-training}

The grouped-flow setup is parallelized at the level of subprocess groups.  The \(13\) groups in the \zjjjj benchmark are distributed over four H100 GPUs, with the assignment chosen to approximately balance the number of subprocesses on
each GPU.  Each group is assigned to exactly one GPU, and different GPUs train disjoint sets of groups.  Since different groups have independent flow parameters, process embeddings, and helicity embeddings, no gradient synchronization between GPUs is required.

On each H100 GPU, all groups assigned to that worker are trained concurrently in the same training loop.  In each training round, the worker samples helicity configurations and phase-space points for all of its local subprocesses.  The subprocesses are then collected according to their group, and the corresponding shared group flow
is evaluated with the local subprocess index and helicity-configuration label as conditioning variables.  Thus all local groups are advanced together: their losses are formed in the same round, combined into a single worker objective, and updated by the
same optimizer step.

The training sample budget is kept the same as in the subprocess-specific setup. Although several subprocesses share one flow in the grouped setup, each subprocess is still sampled with the same number of events per training round as in the one-flow-per-subprocess training. Thus the grouped and subprocess-specific comparisons use the same per-subprocess training statistics; only the proposal parameterization changes, from independent subprocess-specific flows to group-shared conditional flows.

At the implementation level, the grouped-flow training also differs from the subprocess-specific training described in
Sec.~\ref{sec:parallel-training}. In the subprocess-specific setup, all flows on a worker have the same architecture and tensor shapes, so their parameters can be stacked and the log-density evaluations can be vectorized with \texttt{vmap}.  In the grouped-flow setup, however, different groups contain
different numbers of subprocesses.  Consequently, the process-embedding tables and the group-level batch tensors have group-dependent shapes.  The grouped models therefore cannot be stacked into a single uniform tensor structure for a single \texttt{vmap} call.  Instead, the implementation batches computations within each group and then combines the group objectives at the worker level.

For a group \(G\), the online objective is obtained by averaging the losses of the subprocesses in that group,
\begin{equation}
    \mathcal{L}_{G}
    =
    \frac{1}{|\mathcal{P}_G|}
    \sum_{a\in\mathcal{P}_G}
    \mathcal{L}_{G,a},
\end{equation}
where \(\mathcal{P}_G\) denotes the set of subprocesses assigned to group \(G\).
The objective optimized on a worker is the sum over the groups assigned to that worker,
\begin{equation}
    \mathcal{L}_{\mathcal{W}}
    =
    \sum_{G\in\mathcal{W}}
    \mathcal{L}_{G}.
\end{equation}
A single AdamW optimizer is used on each worker to update all local group flows
and their process and helicity embeddings.

The replay-buffer updates follow the same structure.  Each subprocess has its own replay buffer, while replay entries are regrouped by their associated shared flow before the model density is evaluated.  Replay losses are averaged
within each group and summed over the groups on the worker, exactly as for the online update.  Checkpoints are also saved separately for each group, including the shared flow parameters, the process and helicity embeddings, the local subprocess map, and the current helicity-configuration proposal probabilities.

\section{Flow-Based Event-Generation Benchmarks}
\label{sec:results}

\subsection{Benchmark Processes and Event-Generation Parameters}
\label{sec:event-generation-parameters}

Using the implementation of \Pepper 1.10.0,
we study five proton--proton benchmark processes at parton level for a center-of-mass energy of \(\sqrt{s}=13\,\mathrm{TeV}\):
\begin{equation}
  \text{\zjjjj},\qquad \text{\zjjjjj},\qquad \text{\ttjjjj},\qquad \text{\jjjj}, \qquad\text{and}\qquad \text{\jjjjj}.
\end{equation}
We use \Pepper's default PDF choice,
the \NNPDF3.0 NLO PDF set with \(\alpha_s(m_Z)=0.118\)~\cite{NNPDF:2014otw}, and the running strong coupling is evaluated
from the same set through \textsc{LHAPDF}6~\cite{Buckley:2014ana}. The electroweak input parameters are
\begin{equation}
\sin^2\theta_W=0.23155 \qquad \text{and} \qquad \alpha=1/128.80.
\end{equation}
The top quark mass is set to $m_t=173.21\,\mathrm{GeV}$, while
the remaining quarks are treated as massless.
The massive vector boson masses $m$ and decay widths $\Gamma$
are set as follows
\begin{equation}
m_Z=91.1876\,\mathrm{GeV}, \qquad m_W=80.379\,\mathrm{GeV}, \qquad
\Gamma_Z=2.4952\,\mathrm{GeV}, \qquad \text{and} \qquad \Gamma_W=2.085\,\mathrm{GeV}.
\end{equation}

we use
The renormalization and factorization scales are chosen process by process. For lepton-pair production, \(pp\to e^+e^-+nj\) with \(n=4,5\), we choose
\begin{equation}
    \mu_R^2=\mu_F^2=\frac{H_T'^2}{2},
    \qquad
    H_T'=
    \sqrt{m_{e^+e^-}^2+p_{T,e^+e^-}^2}
    +\sum_j p_{T,j},
\end{equation}
where the sum runs over the final-state partons accompanying the
lepton pair~\cite{Bern:2013gka}. For pure multijet production, \(pp\to nj\) with
\(n=4,5\), and for top-pair production,
\(pp\to t\bar t+4j\), we use
\begin{equation}
    \mu_R^2=\mu_F^2=\frac{H_T^2}{2},
    \qquad
    H_T=\sum_{i\in\mathrm{final\ state}}p_{T,i},
\end{equation}
where the sum includes the top quarks for \(t\bar t+4j\)
production~\cite{Berger:2009ep}. The same parton-level jet cuts are applied to all five benchmark processes. All massless final-state partons are required to satisfy
\begin{equation}
    p_{T,j} > 20\,\mathrm{GeV}, \qquad |y_j| < 5,
    \qquad \Delta R_{ij} > 0.4 ,
\end{equation}
where \(p_{T,j}\) and \(y_j\) are the transverse momentum and rapidity of parton \(j\), and \(\Delta R_{ij}=\sqrt{\Delta y_{ij}^2 + \Delta \phi_{ij}^2}\) is the separation between two massless final-state partons in terms of the rapidity $y$ and the azimuthal angle $\phi$. For the two lepton-pair production processes we additionally require
the dilepton invariant mass to to be close to the $Z$-boson resonance,
\begin{equation}
    66\,\mathrm{GeV} \leq m_{e^+e^-} \leq 116\,\mathrm{GeV}.
\end{equation}
No additional phase-space cuts are imposed on the top quarks in
\ttjjjj production.

In the main text we focus on the results for \zjjjj,
which is our primary benchmark for comparing
subprocess-specific flows and grouped flows
to the baseline \pepvegas results.
The results for \zjjjjj, \ttjjjj, \jjjj, and \jjjjj are collected in Appendices~\ref{app:additional-h100-subprocess-specific-results} and~\ref{app:additional-h100-grouped-flow-results}.

\subsection{Computational Setup and Benchmark Procedure}
\label{sec:computational-setup}

Our main setup consists of four H100 GPUs that are installed in a single dual-socket compute node equipped with two AMD EPYC 9654 96-Core processors.
Results for a single RTX GPU with subprocess-specific flows are shown separately in Appendix~\ref{app:rtx-results}.

All runtime benchmarks are reported for four target samples of unweighted events,
\begin{equation}
    N_{\rm evt}=10^6,\;10^7,\;10^8,\;10^9,
\end{equation}
and for two unweighting thresholds,
\begin{equation}
\varepsilon=10^{-3},10^{-2}.
\end{equation}

Throughout this paper, all \pepflow configurations use a per-subprocess batch size of \(32768\) phase-space points for training and \(131072\) raw trial points for event generation, irrespective of the process, flow architecture, or hardware setup.
With these batch sizes, the device memory is close to being fully utilized for the process with the highest complexity in our study, \zjjjjj, and the throughputs are maximized.

For \zjjjj, both the subprocess-specific and grouped-flow models are trained for \(30000\) rounds.  We define one training round by the generation of one such fresh batch for each of the \(134\) partonic subprocesses. Thus, both setups use the same online sample budget of \(134\times32768\) newly generated, \Pepper-evaluated samples per round. Replay updates reuse buffered batches and therefore require no additional matrix-element evaluations.

The \pepvegas reference is run on the same four-H100 node, using four ranks with one rank assigned to each GPU and a batch size of \(524288\) events per GPU. For the single-RTX benchmarks, \pepvegas uses a batch size of \(131072\). These values were selected because they provide the highest \Pepper throughput on the corresponding GPUs. Before timing, the \Vegas grids are optimized separately for each process
using process-dependent phase-space optimization settings. These settings are chosen such that all scheduled grid updates are performed without iterations being skipped because of insufficient sampling statistics. We verify convergence by checking that both the cross-section estimate and the unweighting efficiency are stable over the final optimization iterations. For each process, the \pepvegas and \pepflow workflows use the same \Pepper cache containing the optimized \Vegas grids. The construction of this shared cache is therefore treated as a common preprocessing step and its runtime is excluded from all reported benchmarks.

The \pepvegas runtime for \(10^6\) unweighted events is measured directly, with the preceding \Vegas optimization excluded from the timing. The runtimes for \(10^7\), \(10^8\), and \(10^9\) events are obtained by linear rescaling in \(N_{\rm evt}\). In contrast, all \pepflow event-generation runtimes reported for the four-H100 benchmarks are measured by directly generating the requested event samples.

As described in Sec.~\ref{sec:training-objective}, checkpoints are saved every \(50\) training rounds in all \pepflow runs, regardless of the hardware setup or flow architecture. The checkpoint selections below are therefore restricted to this discrete set of saved rounds. We report two \pepflow timing prescriptions. In the first, the quoted total runtime includes both training and event generation. To select a checkpoint for a target sample size \(N_{\rm evt}\), we scan the saved checkpoints and measure the time \(T_{\rm gen}(c;10^6)\) required to generate \(10^6\) unweighted events with each checkpoint \(c\). For checkpoint selection only, we estimate the
total runtime as
\begin{equation}
    \widehat{T}_{\rm total}(c;N_{\rm evt})
    =
    T_{\rm train}(c)
    +
    \frac{N_{\rm evt}}{10^6}
    T_{\rm gen}(c;10^6),
\end{equation}
where \(T_{\rm train}(c)\) is the wall time required to train up to
checkpoint \(c\). The checkpoint that minimizes this estimate is selected separately for each target sample size. After the checkpoint has been selected, we directly generate the requested number of events and report the resulting measured generation time together with the training time. Thus, the linear rescaling is used only for checkpoint selection and not for the reported \pepflow runtime.

In the second prescription, we isolate the event-generation cost after training. For this generation-only benchmark, we use the checkpoint at round \(30000\), which gives the highest average training unweighting efficiency among the saved checkpoints for both the subprocess-specific and grouped-flow models. We directly generate \(10^6\), \(10^7\), \(10^8\), and \(10^9\) unweighted events from this fixed checkpoint and measure the corresponding generation times. These two prescriptions separate the end-to-end cost of obtaining a usable trained generator from the throughput of the final trained proposal. For commonly used processes with fixed generation settings, the trained flow checkpoints could be produced once at a central site and distributed to users. In such a deployment model, the training cost is amortized across multiple users and production runs, and the generation-only benchmarks represent the runtime incurred by an individual user.

\subsection{Training Diagnostics and Checkpoint Selection for \zjjjj Flow Proposals}
\label{sec:results-training-diagnostics}

\begin{figure*}[t]
    \centering
    \begin{tabular}{cc}
        \includegraphics[width=0.48\textwidth]{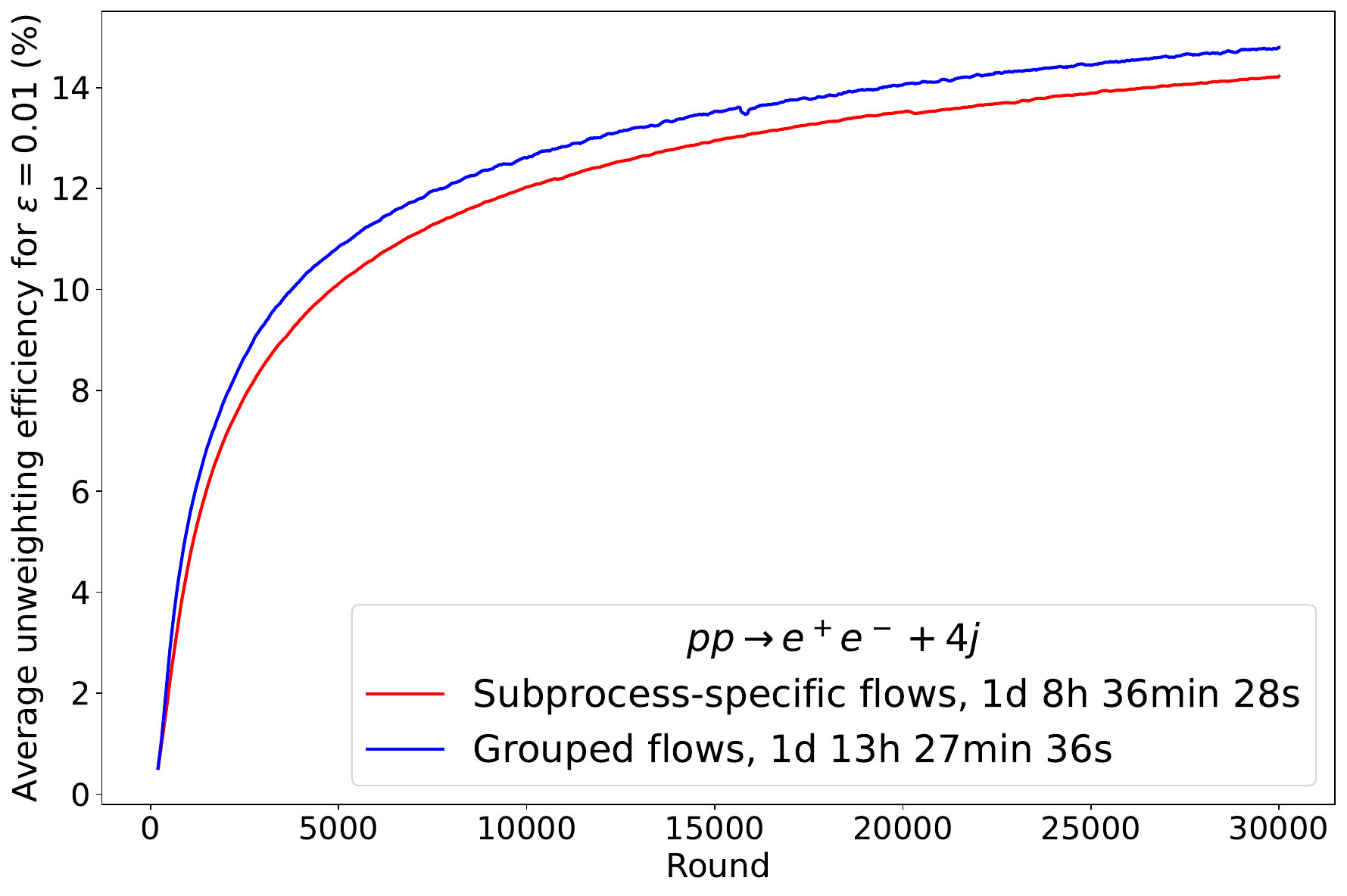} &
        \includegraphics[width=0.48\textwidth]{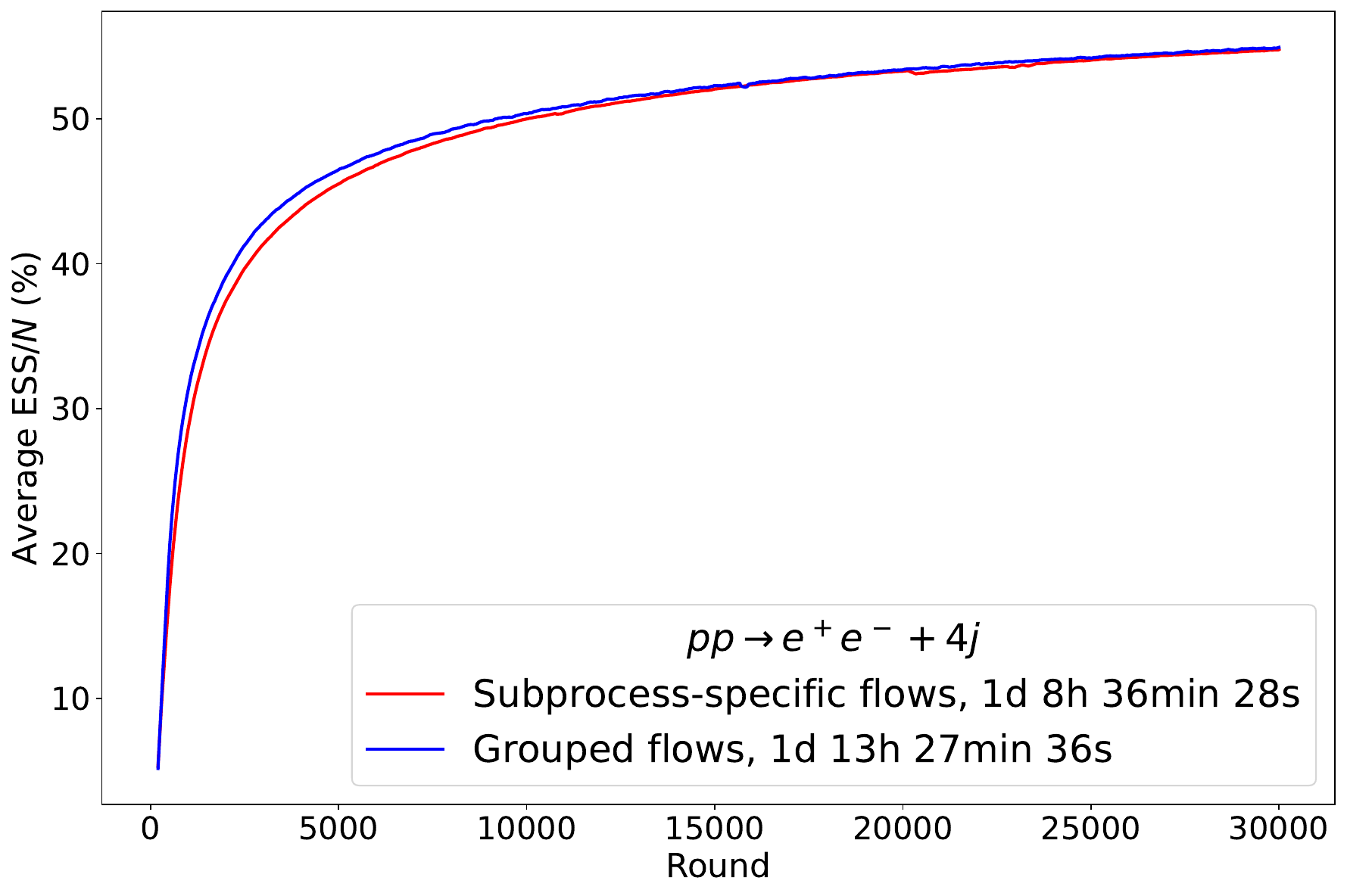}
    \end{tabular}
    \caption{
    Training diagnostics for subprocess-specific flows (red line) and grouped flows (blue line) in \zjjjj production on four H100 GPUs. The left panel shows the average unweighting efficiency for \(\varepsilon=10^{-2}\), and the right panel shows the average effective sample size normalized by the number of samples, \(\mathrm{ESS}/N\), as functions of the training round.
    Both quantities are averaged over the \(134\) subprocesses.
    The grouped-flow setup uses shared conditional flows with process embeddings, while the subprocess-specific setup trains one flow for each subprocess. The grouped flows achieve a slightly higher average unweighting efficiency and a comparable, slightly higher \(\mathrm{ESS}/N\) throughout training, although
    the total training wall time is longer.
    }
    \label{fig:z4j-grouped-training-diagnostics}
\end{figure*}

Figure~\ref{fig:z4j-grouped-training-diagnostics} compares the training diagnostics of subprocess-specific and grouped flows for \zjjjj production. Both setups show a steady improvement in unweighting efficiency and \(\mathrm{ESS}/N\) throughout training, indicating that the learned proposal continues to reduce event-weight fluctuations. The grouped-flow model reaches a slightly higher average unweighting efficiency over the \(134\) subprocesses and gives a comparable, mildly improved \(\mathrm{ESS}/N\). This suggests that, for a process with many subprocesses, sharing a conditional flow within parton-content groups can be beneficial: the shared parameters receive training information from several kinematically related subprocesses, while the process embedding retains enough flexibility to represent subprocess-dependent corrections. In this sense, the grouped flow acts as a regularized conditional proposal rather than simply a smaller collection of models. Corresponding training diagnostics for the remaining benchmark processes are provided in Appendix~\ref{app:additional-training-diagnostics}.

The tradeoff is that the grouped-flow training has a longer wall time. The training sample budget is the same as in the subprocess-specific setup, so the total number of \Pepper matrix-element evaluations is not reduced. In addition, the grouped implementation cannot stack all local models into a single uniform \texttt{vmap} call, because different groups have different numbers of subprocesses and therefore group-dependent batch and embedding shapes. The longer training time therefore reflects the less uniform group structure and the reduced batching efficiency, even though the resulting proposal quality is slightly improved.

\begin{table*}[t]
    \centering
    \caption{Checkpoint rounds selected for the \pepflow ``total'' runtime benchmarks in \zjjjj production on \(4\times\) H100 GPUs.
    For each flow setup, unweighting threshold~$\varepsilon$, and target event sample,
    the selected checkpoint minimizes the estimated sum of training and
    generation times defined in the text. The generation-only benchmarks
    instead use the final checkpoint at round \(30000\).}
    \label{tab:z4j-selected-checkpoints}

    \renewcommand{\arraystretch}{1.35}
    \begin{tabular*}{0.7\textwidth}
        {@{\extracolsep{\fill}} l c c c c c @{}}
        \hline
        \multirow{2}{*}{Flow setup}
        & \multirow{2}{*}{\(\varepsilon\)}
        & \multicolumn{4}{c}{Target sample size \(N_{\rm evt}\)} \\
        \cline{3-6}
        & & \(10^6\) & \(10^7\) & \(10^8\) & \(10^9\) \\
        \hline
        \multirow{2}{*}{Subprocess-specific flow}
        & \(10^{-3}\) & \(100\) & \(400\) & \(900\) & \(3600\) \\
        & \(10^{-2}\) & \(50\) & \(200\) & \(600\) & \(1500\) \\
        \hline
        \multirow{2}{*}{Grouped flow}
        & \(10^{-3}\) & \(100\) & \(250\) & \(750\) & \(3150\) \\
        & \(10^{-2}\) & \(50\) & \(150\) & \(450\) & \(1250\) \\
        \hline
    \end{tabular*}
\end{table*}

The combined evolution of proposal quality and training cost is reflected in the checkpoint selections summarized in
Table~\ref{tab:z4j-selected-checkpoints}. The selected checkpoint is neither fixed nor always the first or final saved checkpoint, but changes systematically with the target event sample. For both flow setups and both unweighting thresholds, larger samples favor later checkpoints because the additional training cost can be compensated by the improved generation efficiency over more events. At a fixed sample size, the tighter threshold, \(\varepsilon=10^{-3}\), also consistently favors a later checkpoint than \(\varepsilon=10^{-2}\), since the generation time is more sensitive to improvements in the proposal at the tighter threshold.

For \(N_{\rm evt}\geq10^7\), the grouped-flow checkpoints are selected at somewhat earlier rounds than their subprocess-specific counterparts. Under the total-runtime objective, additional training is worthwhile only when the resulting reduction in generation time exceeds the extra training cost. Because grouped-flow rounds are more expensive while the grouped proposal already achieves comparable or slightly better average sampling quality, this balance is reached at a smaller round number. The earlier selected checkpoint therefore represents a different cost-optimal stopping point, rather than faster convergence in wall time. All selected checkpoints occur well before the final training round \(30000\), demonstrating that maximizing the final proposal quality is not the same as minimizing the end-to-end runtime.

\subsection{Subprocess-Specific Flow Results on Four H100 GPUs}
\label{sec:results-h100-subprocess-specific}

\begin{figure*}[t]
    \centering
    \begin{tabular}{cc}
        \includegraphics[width=0.48\textwidth]{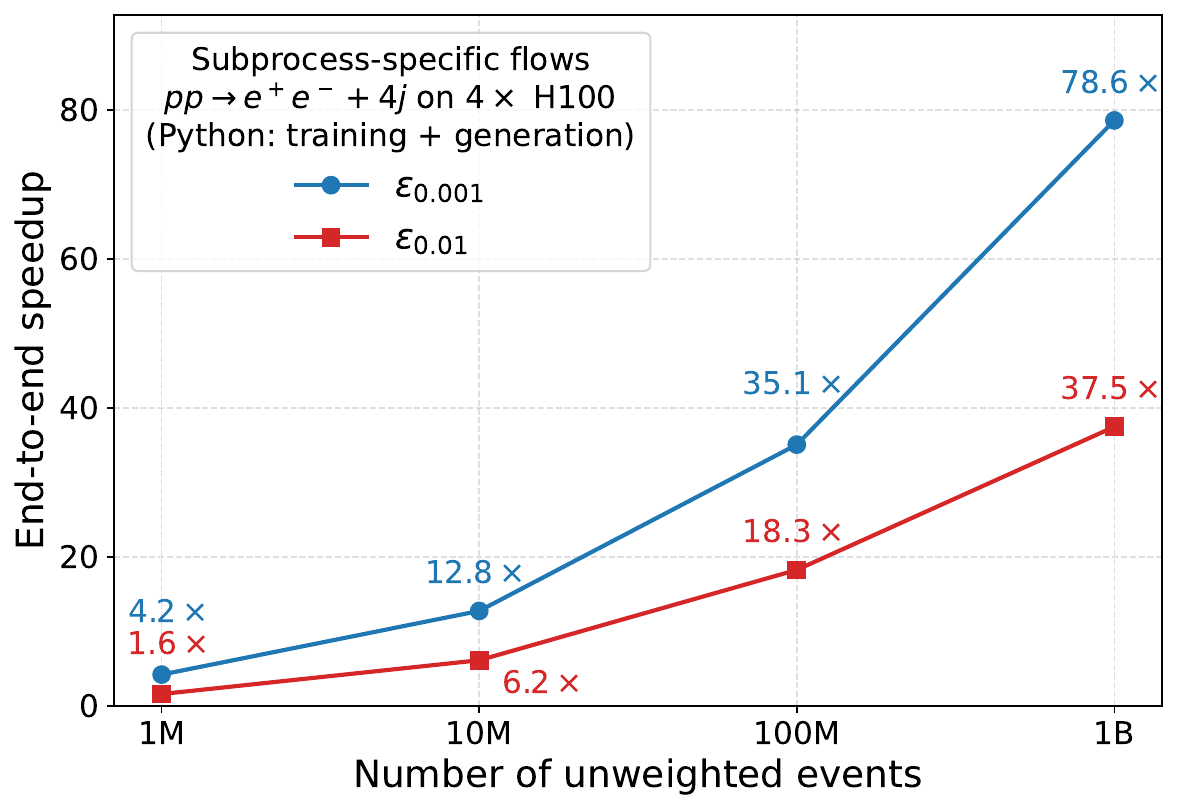} &
        \includegraphics[width=0.48\textwidth]{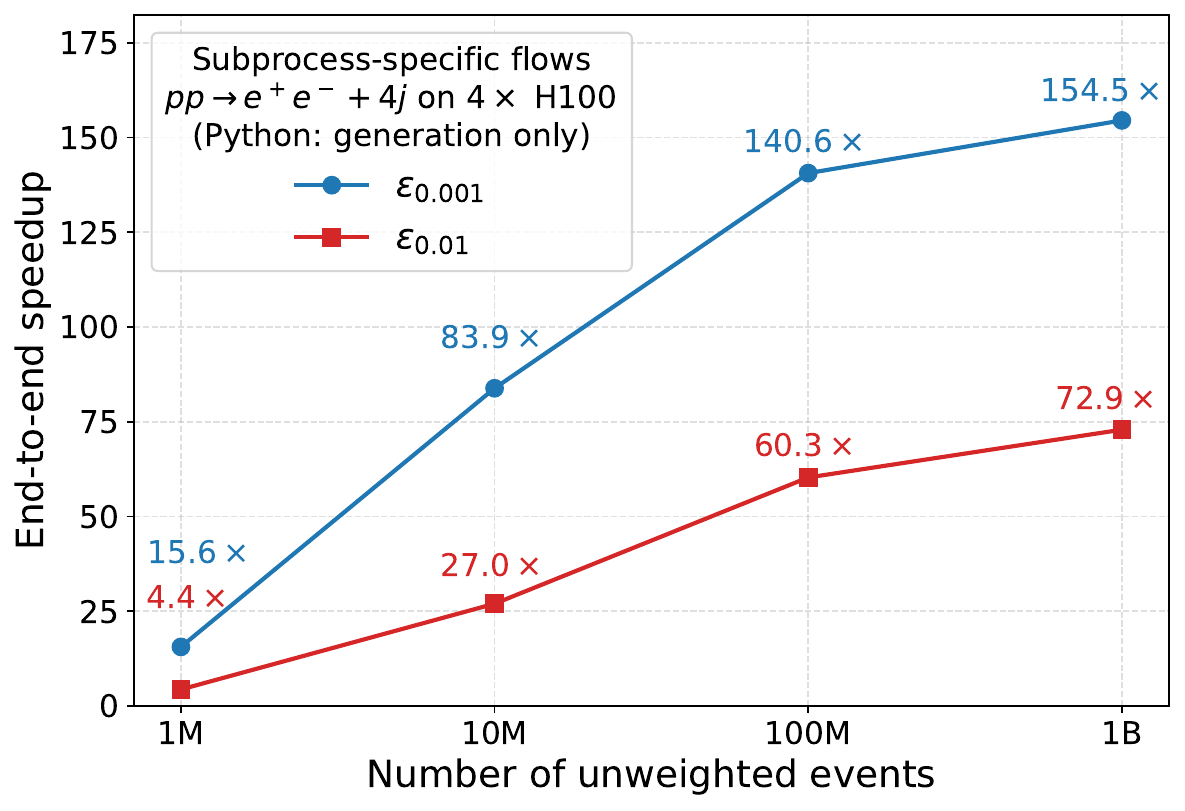}
    \end{tabular}
    \caption{
    End-to-end (left) and generation-only (right) speedups for
    subprocess-specific flow generation in \zjjjj production on
    \(4\times\) H100 GPUs. The left panel uses the measured \pepflow total runtime, including training and direct generation of the requested event sample. The right panel uses the measured generation-only runtime of the final trained subprocess-specific flow checkpoints. The speedup is defined as the \pepvegas runtime divided by the corresponding \pepflow runtime. The \pepvegas runtime is measured at \(10^6\) events and linearly extrapolated to larger target samples, whereas the \pepflow runtimes are measured directly for every target sample size. Results are shown for \(\varepsilon=10^{-3}\) and \(\varepsilon=10^{-2}\).
    }
    \label{fig:z4j-h100-subprocess-specific-speedup}
\end{figure*}

Figure~\ref{fig:z4j-h100-subprocess-specific-speedup} shows the corresponding speedups for subprocess-specific \zjjjj generation on four H100 GPUs.  The left panel gives the end-to-end speedup, where the \pepflow runtime includes both training and event generation.  The speedup is already larger than unity at
\(10^6\) events, reaching \(4.2\times\) for \(\varepsilon=10^{-3}\) and
\(1.6\times\) for \(\varepsilon=10^{-2}\).  It then grows steadily with the requested event sample size, reaching \(78.6\times\) and \(37.5\times\), respectively, at \(10^9\) events.

This growth reflects the checkpoint-selection procedure used for the training-plus-generation benchmark. For each target sample size, the checkpoint is selected by minimizing the estimated total runtime defined above, balancing the training time needed to reach that checkpoint against its expected event-generation efficiency. After a checkpoint has been selected, the requested event sample is generated directly, and the measured generation time is used in the reported total runtime. The directly measured generation times are systematically shorter than the estimates used for checkpoint selection, with the difference increasing toward larger event samples. For example, for \(\varepsilon=10^{-3}\), the measured generation time is approximately \(17\%\) shorter than the estimate at \(10^7\) events, \(34\%\) shorter at \(10^8\) events, and \(53\%\) shorter at \(10^9\) events. A similar trend is observed for \(\varepsilon=10^{-2}\), reflecting the fixed-batch granularity of subprocess-wise generation. The linear estimates rescale the \(10^6\)-event runtime, for which final-batch oversampling is substantial, whereas in the direct larger-sample runs this overhead is incurred only once per subprocess and is amortized over a much larger event sample. Although the estimates are systematically higher than the subsequent direct measurements, they provide a computationally inexpensive way to select a cost-efficient checkpoint without generating the full target sample with every saved checkpoint. They are used only for checkpoint selection and do not enter the reported total runtimes. Larger requested samples can justify later checkpoints: although they require longer training, their improved unweighting efficiency reduces the generation time over a much larger event sample. The end-to-end speedup therefore reflects the balance between training time and generation efficiency implemented by this checkpoint-selection procedure.

The right panel isolates the generation throughput after training by using the final trained checkpoint. In this generation-only comparison, the speedups are substantially larger, starting from \(15.6\times\) and \(4.4\times\) at \(10^6\) events and reaching \(154.5\times\) and \(72.9\times\) at \(10^9\) events for \(\varepsilon=10^{-3}\) and \(\varepsilon=10^{-2}\), respectively. The gap between the two panels quantifies the cost of obtaining the trained proposal, while the generation-only panel shows the asymptotic benefit once the flow has already been trained. Although the same final checkpoint is used for all sample sizes, the generation-only speedup is not constant at finite event counts. This increase with sample size is primarily caused by subprocess-wise generation in fixed-size batches. For small target samples, most subprocesses are assigned fewer accepted events than are typically produced by a single batch. The full batch must nevertheless be sampled, evaluated, and unweighted, while accepted events beyond the subprocess target are discarded. Because this final-batch oversampling is incurred independently for every subprocess, it constitutes a substantial overhead at small sample sizes but is progressively amortized as the requested sample size increases. The resulting speedup therefore approaches the asymptotic throughput ratio only at sufficiently large sample sizes. The same sample-size dependence is observed for all generation-only speedups reported in this work and can be understood in the same way.

Table~\ref{tab:z4j-h100-subprocess-specific-runtime} shows the runtime performance of the subprocess-specific flow setup for \zjjjj production on four H100 GPUs.  The \pepflow workflow is already faster than \pepvegas generation for all event sample sizes shown, even when the training time up to the selected checkpoint is included.  For \(10^6\) events, the total runtime is
reduced from \(1\)h \(8\)min to \(16\)min \(8\)s for
\(\varepsilon=10^{-3}\), and from \(20\)min \(10\)s to \(12\)min \(28\)s for \(\varepsilon=10^{-2}\).  This shows that, for \zjjjj on four H100 GPUs, the flow-based workflow is already beneficial before reaching the very-large-sample regime.

The advantage becomes much larger as the requested event sample increases. For \(10^9\) events, the measured \pepflow total runtime is about \(14.4\) hours for \(\varepsilon=10^{-3}\), compared with an extrapolated \pepvegas runtime of \(47\) days. For \(\varepsilon=10^{-2}\), the corresponding measured \pepflow runtime is about \(9\) hours, compared with an extrapolated \pepvegas runtime of about \(14\) days. The larger speedup at \(\varepsilon=10^{-3}\) reflects the fact that the tighter unweighting threshold makes \pepvegas generation more expensive, while the learned proposal substantially reduces the corresponding event-weight fluctuations.

The generation-only column isolates the throughput of the trained
subprocess-specific flows.  Once the trained proposal is available,
\(10^9\) unweighted \zjjjj events are generated in about \(7.3\) hours for \(\varepsilon=10^{-3}\) and about \(4.6\) hours for
\(\varepsilon=10^{-2}\).  

\begin{table*}[t]
    \centering
    \caption{
    Runtime comparison for generating unweighted \zjjjj events with
    subprocess-specific flows on \(4\times\) H100 GPUs. The \pepflow ``total'' runtime includes the training time up to the checkpoint selected for the target sample size and the directly measured time required to generate that sample. The \pepflow ``gen. only'' runtime is measured by directly generating the requested sample with the final trained subprocess-specific flow checkpoints. The event-generation components of both \pepflow columns are measured by directly generating the requested samples. Blue \pepvegas entries are linear extrapolations of the measured \(10^6\)-event runtime.
    }
    \label{tab:z4j-h100-subprocess-specific-runtime}
    \begin{tabular}{c c c c c}
        \hline
        \(\varepsilon\) & Number of events
        & \pepvegas
        & \pepflow (total)
        & \pepflow (gen. only) \\
        \hline
        \multirow{4}{*}{\(10^{-3}\)}
        & \(1M\) & 1h\;8min\;12s & 16min\;8s & 4min\;22s \\
        & \(10M\) & \textcolor{blue}{11h\;22min} & 53min\;25s & 8min\;8s \\
        & \(100M\) & \textcolor{blue}{4d\;17h\;40min} & 3h\;14min\;22s & 48min\;30s \\
        & \(1B\) & \textcolor{blue}{47d\;8h\;40min} & 14h\;27min\;41s & 7h\;21min\;21s \\
        \hline
        \multirow{4}{*}{\(10^{-2}\)}
        & \(1M\) & 20min\;10s & 12min\;28s & 4min\;36s \\
        & \(10M\) & \textcolor{blue}{3h\;21min\;40s} & 32min\;46s & 7min\;28s \\
        & \(100M\) & \textcolor{blue}{1d\;9h\;36min\;40s} & 1h\;50min\;20s & 33min\;26s \\
        & \(1B\) & \textcolor{blue}{14d\;6min\;40s} & 8h\;57min\;56s & 4h\;36min\;32s \\
        \hline
    \end{tabular}
\end{table*}

\subsection{Grouped Flow Results on Four H100 GPUs}
\label{sec:results-h100-grouped}

\begin{figure*}[t]
    \centering
    \begin{tabular}{cc}
        \includegraphics[width=0.48\textwidth]{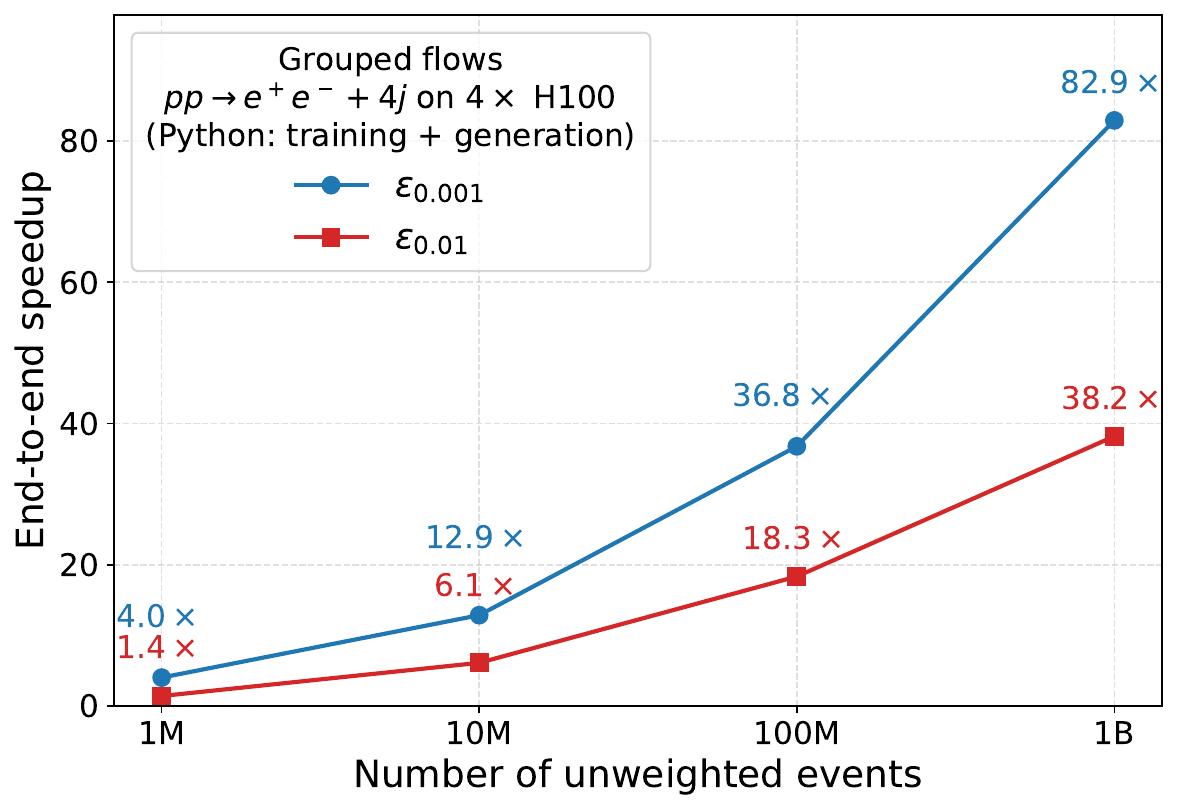} &
        \includegraphics[width=0.48\textwidth]{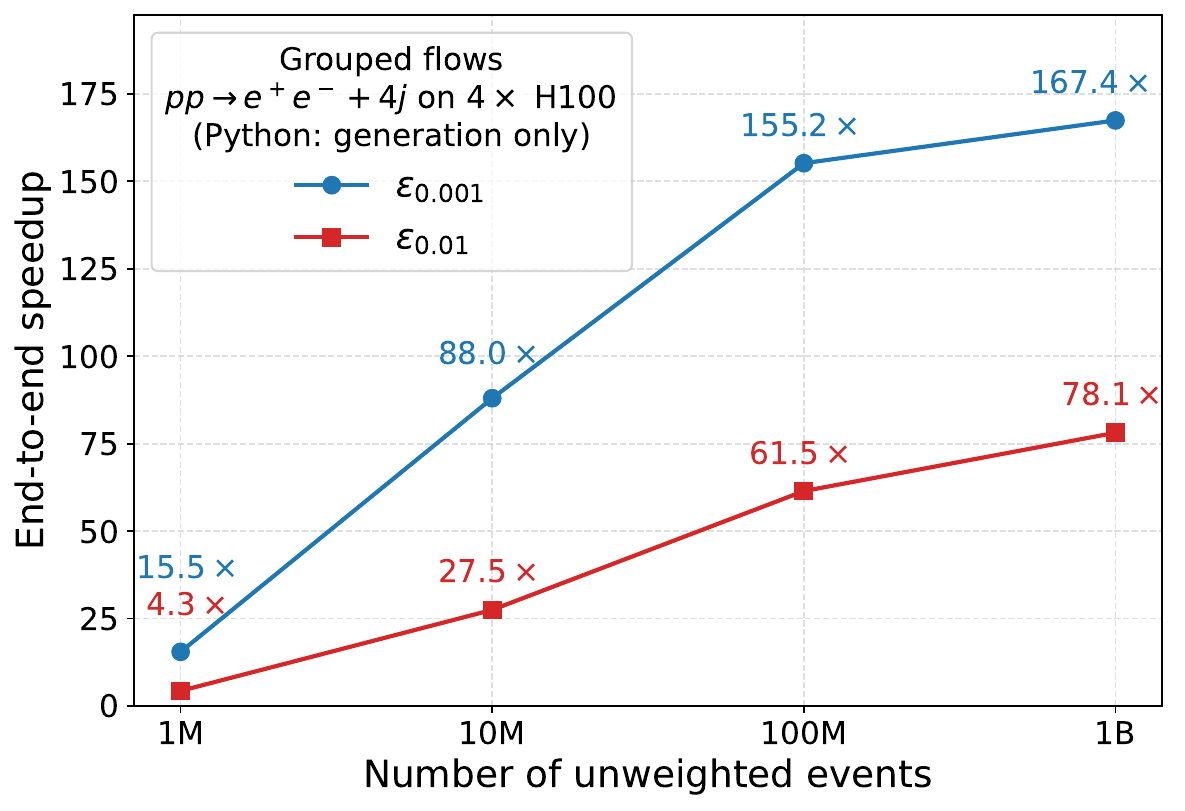}
    \end{tabular}
    \caption{
    End-to-end (left) and generation-only (right) speedups for grouped-flow generation in \zjjjj production on \(4\times\) H100 GPUs. The left panel uses the measured \pepflow total runtime, including training and direct generation of the requested event sample. The right panel uses the measured generation-only runtime of the final trained grouped-flow checkpoints. The speedup is defined as the \pepvegas runtime divided by the corresponding \pepflow grouped-flow runtime. The \pepvegas runtime is measured at \(10^6\) events and linearly extrapolated to larger target samples, whereas the \pepflow runtimes are measured directly for every target sample size. Results are shown for \(\varepsilon=10^{-3}\) and \(\varepsilon=10^{-2}\).
    }
    \label{fig:z4j-h100-grouped-speedup}
\end{figure*}

Figure~\ref{fig:z4j-h100-grouped-speedup} shows the corresponding speedups for the grouped-flow \zjjjj setup on four H100 GPUs. The timing prescription is the same as for the subprocess-specific benchmark: the left panel includes both training and event generation, with the checkpoint selected separately for each target sample size by minimizing the estimated total runtime defined above. After the checkpoint has been selected, the requested event sample is generated directly and the measured generation time is used in the reported total runtime. The speedup therefore reflects the balance between the additional training time needed to reach a given checkpoint and the improved generation efficiency obtained from that checkpoint.

With the training cost included, the grouped-flow workflow gives large end-to-end speedups over \pepvegas generation.  The speedup increases with the requested event sample size, reaching \(82.9\times\) for \(\varepsilon=10^{-3}\) and \(38.2\times\) for \(\varepsilon=10^{-2}\) at \(10^9\) events.  As in the subprocess-specific case, the tighter \(\varepsilon=10^{-3}\) threshold gives the larger speedup, because direct unweighting becomes more expensive while the learned proposal reduces the corresponding event-weight fluctuations.

The right panel isolates the event-generation throughput of the trained grouped-flow model by using the final checkpoint.  In this generation-only comparison, the speedups are substantially larger, reaching \(167.4\times\) for \(\varepsilon=10^{-3}\) and \(78.1\times\) for \(\varepsilon=10^{-2}\) at \(10^9\) events. The generation-only speedups are slightly larger than those obtained with subprocess-specific flows, consistent with the improved average sampling efficiency observed for the grouped \zjjjj model.

Table~\ref{tab:z4j-h100-grouped-runtime} shows the runtime performance of the grouped-flow setup for \zjjjj production on four H100 GPUs.  As in the subprocess-specific case, the grouped-flow workflow is faster than \pepvegas generation for all event sample sizes shown, even when the training time up to the selected checkpoint is included.  For \(10^6\) events, the total runtime is
reduced from \(1\)h \(8\)min to \(16\)min \(54\)s for \(\varepsilon=10^{-3}\), and from \(20\)min \(10\)s to \(14\)min \(8\)s for \(\varepsilon=10^{-2}\).  Thus the grouped-flow model already gives an end-to-end gain at the smallest sample size considered.

The advantage grows rapidly with the requested event sample. For
\(10^9\) events, the measured \pepflow total runtime is about
\(13.7\) hours for \(\varepsilon=10^{-3}\), compared with an extrapolated \pepvegas runtime of \(47\) days. For 
\(\varepsilon=10^{-2}\), the corresponding measured \pepflow runtime is about \(8.8\) hours, compared with an extrapolated \pepvegas runtime of about \(14\) days. These total runtimes are slightly shorter than the corresponding subprocess-specific results in
Table~\ref{tab:z4j-h100-subprocess-specific-runtime}, indicating that the shared conditional flows can achieve comparable or better event-generation efficiency for \zjjjj, despite the additional structure of the grouped parameterization.

The generation-only column isolates the throughput of the trained grouped-flow model.  Once the final trained checkpoint is available, \(10^9\) unweighted \zjjjj events are generated in about \(6.8\) hours for \(\varepsilon=10^{-3}\) and about \(4.3\) hours for
\(\varepsilon=10^{-2}\). These timings are also slightly faster than the
subprocess-specific generation-only results, consistent with the mildly improved average proposal quality observed for the grouped \zjjjj model.

On four H100 GPUs, the remaining benchmark processes, \zjjjjj, \ttjjjj, \jjjj, and \jjjjj, exhibit the same overall qualitative behavior as \zjjjj for both flow setups. For each of these processes, the \pepflow workflow is faster than \pepvegas generation for all target sample sizes and both unweighting thresholds considered. The speedups generally increase with the target sample size, with the generation-only values exceeding the corresponding end-to-end values. Their magnitude is process dependent, with the largest gains generally obtained for the higher-multiplicity benchmarks. Detailed speedup plots and the corresponding runtime tables for the subprocess-specific and grouped-flow setups are provided in
Appendices~\ref{app:additional-h100-subprocess-specific-results}
and~\ref{app:additional-h100-grouped-flow-results}, respectively.

\begin{table*}[t]
    \centering
    \caption{
    Runtime comparison for generating unweighted \zjjjj events with grouped flows on \(4\times\) H100 GPUs. The \pepflow ``total'' runtime includes the training time up to the checkpoint selected for the target sample size and the directly measured time required to generate that sample. The \pepflow ``gen. only'' runtime is measured by directly generating the requested sample with the final trained grouped-flow checkpoints. The event-generation components of both \pepflow columns are measured by directly generating the requested samples. Blue \pepvegas entries are linear extrapolations of the measured \(10^6\)-event runtime.
    }
    \label{tab:z4j-h100-grouped-runtime}
    \begin{tabular}{c c c c c}
        \hline
        \(\varepsilon\) & Number of events
        & \pepvegas
        & \pepflow (total)
        & \pepflow (gen. only) \\
        \hline
        \multirow{4}{*}{\(10^{-3}\)}
        & \(1M\) & 1h\;8min\;12s & 16min\;54s & 4min\;24s \\
        & \(10M\) & \textcolor{blue}{11h\;22min} & 53min & 7min\;45s \\
        & \(100M\) & \textcolor{blue}{4d\;17h\;40min} & 3h\;5min\;28s & 43min\;57s \\
        & \(1B\) & \textcolor{blue}{47d\;8h\;40min} & 13h\;42min\;30s & 6h\;47min\;23s \\
        \hline
        \multirow{4}{*}{\(10^{-2}\)}
        & \(1M\) & 20min\;10s & 14min\;8s & 4min\;39s \\
        & \(10M\) & \textcolor{blue}{3h\;21min\;40s} & 32min\;56s & 7min\;20s \\
        & \(100M\) & \textcolor{blue}{1d\;9h\;36min\;40s} & 1h\;49min\;55s & 32min\;49s \\
        & \(1B\) & \textcolor{blue}{14d\;6min\;40s} & 8h\;47min\;55s & 4h\;18min\;7s \\
        \hline
    \end{tabular}
\end{table*}

\subsection{Relation to Previous \Pepper Flow-Based Studies}
\label{sec:relation-to-previous-pepper-flows}

The results presented above address a different aspect of flow-based event generation from the earlier \Pepper studies in
Refs.~\cite{Bothmann:2025lwg,Bothmann:2026dar}. The first of these studies applied continuous normalizing flows trained with flow matching to representative Drell--Yan and $t \bar t + \text{jets}$ partonic channels and compared their proposal quality with that of coupling-layer flows. The continuous models achieved higher unweighting efficiencies and favorable scaling with increasing jet multiplicity. The subsequent study examined the accompanying tradeoff between proposal quality and model-evaluation cost. Using RegFlow to transfer part of the continuous-flow proposal-quality advantage to fast coupling-layer models, it demonstrated that these improvements can yield substantial parton-level event-generation wall-time gains.

The present work follows a technically independent approach. It uses neither continuous normalizing flows trained with flow matching nor RegFlow. Instead, we train helicity-conditioned rational-quadratic coupling flows directly from \Pepper evaluations. The flows are initialized from the existing \Vegas grids and optimized with a variance-based objective using online training interleaved with off-policy replay-buffer updates. Their explicit invertible structure provides fast sampling and direct density evaluation in the Python generation loop, thereby reducing the per-event generation cost. Together with online training and replay-buffer updates that reuse previous \Pepper evaluations, this design aims to improve proposal quality while controlling both the training and event-generation wall time.

The principal distinction is the unit of generation and the scope of the
workflow. The earlier studies benchmarked learned mappings and their generation performance for representative partonic channels. Here, the learned proposals are deployed across all subprocesses of complete physical processes. We address the resulting full-process requirements through either subprocess-specific flows or group-shared conditional flows, allocate the requested event sample among subprocesses, perform unweighting directly from Python, and write out the standard event samples. The reported end-to-end runtimes include the cost of training up to the selected checkpoint and the subsequent production of the complete event sample. In addition, the generation-only benchmarks directly measure the throughput of the final trained proposals up to \(10^9\) generated events.

Because the benchmark units, model constructions, hardware setups, and timing prescriptions differ, the numerical speedups reported here should not be interpreted as a direct head-to-head comparison with Refs.~\cite{Bothmann:2025lwg,Bothmann:2026dar}. Those studies establish the proposal-quality advantages of flow matching and the corresponding quality--evaluation-cost tradeoff, while the present results demonstrate that independently trained coupling flows can be coordinated across complete processes with many subprocesses and converted into large end-to-end wall-clock gains for large event samples.

\section{Data Availability}

The Python-side normalizing-flow code developed for this work is publicly available~\cite{pepper-flows}. The repository contains the implementations of the subprocess-specific and grouped conditional flow models, together with scripts for training, checkpoint evaluation, and learned-proposal event generation. The \Pyper bindings described in Sec.~\ref{sec:Pepper-workflow} are distributed as part of the public \Pepper code repository~\cite{pepper-repo} and are described in the 
online documentation~\cite{pyper-guide}.

\section{Conclusion}
\label{sec:conclusion}

In this paper, we have developed the first end-to-end Python-controlled event-generation workflow with \Pepper that trains and deploys normalizing-flow proposals across all subprocesses of many-jet processes. The central advance is the coordinated deployment of these proposals across complete subprocess collections and their direct use for production-scale event generation. The Python layer controls flow sampling, proposal-density evaluation, and accept-reject unweighting, while using \Pepper via its new \Pyper interface to evaluate the matrix elements, PDFs, and phase-space factors that define the target density, and to write out standard event samples. Since the learned model changes only the proposal distribution and its density is included in the corrected event weights, any imperfections in the flow training only affects the generation efficiency, but does not bias physics results.

The proposal is implemented using helicity-conditioned rational-quadratic coupling flows initialized from the existing \Vegas grids and optimized to reduce event-weight fluctuations. The training combines online updates with buffer replay. During the online updates, new phase-space points are drawn from the evolving flow proposal and evaluated by \Pepper. Previously evaluated batches are retained in subprocess-specific replay buffers and reused in additional off-policy optimization steps, with the stored proposal densities and helicity-configuration probabilities providing the required importance corrections. This combination allows the proposal to continue exploring through fresh online samples while extracting additional training information from expensive matrix-element evaluations that have already been performed.

We studied two strategies for extending this construction to complete many-jet processes. In the subprocess-specific flow setup, each partonic subprocess is assigned an independent flow, allowing the proposal to specialize to its individual phase-space structure. In the grouped flow setup, subprocesses with related initial- and final-state parton content share a conditional flow and are distinguished through trainable process embeddings. The grouped and subprocess-specific models use the same per-subprocess training
statistics, so their comparison isolates the effect of parameter sharing. For \zjjjj, the grouped flows achieve a slightly higher average unweighting efficiency and a comparable, mildly improved \(\mathrm{ESS}/N\), although their less uniform batch structure leads to a longer training wall time. The results for the additional processes show that the relative benefit of parameter sharing depends on the size and composition of the subprocess collection.

We benchmarked both strategies for \zjjjj, \zjjjjj, \ttjjjj, \jjjj, and \jjjjj production. In particular, we successfully generated \(10^9\) unweighted events for every benchmark process and for both unweighting thresholds on four H100 GPUs. To summarize the achieved production times, we quote the grouped-flow results, which give the shortest end-to-end runtime across the benchmarks. Writing each result as \(T_{\rm total}/T_{\rm gen}\), where \(T_{\rm total}\) includes training up to the checkpoint selected for the end-to-end benchmark and \(T_{\rm gen}\) is the generation-only time using the final trained checkpoint, the runtimes for \(\varepsilon=10^{-3}\) are approximately \(13.7/6.8\) hours for \zjjjj, \(5.6/2.8\) days for \zjjjjj, \(2.9/2.3\) days for \ttjjjj, \(4.8/3.2\) hours for \jjjj, and \(28.7/23.7\) hours for \jjjjj. For \(\varepsilon=10^{-2}\), the corresponding runtimes are approximately \(8.8/4.3\) hours, \(2.8/1.2\) days, \(1.3/1.1\) days, \(3.2/2.4\) hours, and \(14.8/9.8\) hours, respectively.

On four H100 GPUs, the \pepflow workflow is faster than conventional \pepvegas generation for every process, event sample size, and
unweighting threshold considered, already from \(10^6\) requested events. At \(10^9\) events, the largest end-to-end speedup relative to the linearly
extrapolated \pepvegas runtime reaches \(109.7\times\), including the
training cost up to the selected checkpoint, while the largest
generation-only speedup reaches \(248.1\times\). For the representative \zjjjj process, grouped flows reduce the total runtime from a linearly extrapolated \(47\) days of \pepvegas generation to approximately \(13.7\) hours for \(\varepsilon=10^{-3}\), corresponding to an \(83.2\times\) end-to-end speedup. For \(\varepsilon=10^{-2}\), the runtime is reduced from a linearly extrapolated \(14\) days to \(8.8\) hours, corresponding to a \(38.4\times\) speedup. Once the final grouped-flow checkpoint is available, the corresponding generation-only speedups reach \(168.4\times\) and \(79.0\times\), respectively. The subprocess-specific setup gives similarly large gains, and the single-RTX study shows that the acceleration is not restricted to large multi-GPU systems when sufficiently large event samples are required.

These results demonstrate that learned proposals can be deployed across complete processes with many subprocesses and translated into substantial end-to-end wall-clock gains, even after accounting for flow training, cross-section-based subprocess allocation, unweighting, and direct generation of the requested samples. The separation between the end-to-end and generation-only results also highlights the value of reusing trained checkpoints across multiple production campaigns.

A natural next step is to integrate the complete training and generation workflow directly into \Pepper as an automated event-generation mode. A user would then need to specify only a \Pepper-supported process, the desired sample size, the unweighting threshold, and the available computing resources. Based on the number and partonic structure of the contributing subprocesses, \Pepper could automatically choose between subprocess-specific and grouped conditional flows, or use short pilot trainings to estimate which strategy gives the smaller total production time. During training, it could monitor the proposal efficiency together with the measured training and generation throughput and stop automatically once the expected reduction in generation
time no longer compensates for the cost of further optimization. In this way, the checkpoint-selection procedure used in this work could be turned into an online stopping criterion tailored to the requested sample size. Such a self-configuring workflow would hide the technical details of flow construction, subprocess grouping, and checkpoint selection from the user while retaining their performance benefits.
Another natural next step is to extend and study the implementation for event generation at higher perturbative orders,
which is planned as soon as the NLO implementation of \Pepper is available.
Together with extensions to still higher multiplicities and more general event-generation settings, this provides a promising route toward meeting the growing simulation demands of the HL-LHC.

\begin{acknowledgments}
The authors gratefully acknowledge the use of the MUSICA supercomputer. The computational results have been achieved using the Austrian Scientific Computing (ASC) infrastructure and the project ÖAW-MUSICA-2026. The authors thank Karla Maria Tame Narvaez, Taylor Childers, and Max Knobbe for useful discussions. E.B.\ and D.W.\ thank Timo Janßen for helpful discussions clarifying the normalizing-flow setup of Ref.~\cite{Bothmann:2025lwg}. C.L.\ acknowledges and is grateful for the CERN openlab summer student programme and the NextGEN trigger project, during which C.L.'s portion of this work was carried out.
 
\end{acknowledgments}

\appendix

\section{Additional Training Diagnostics for Subprocess-Specific and Grouped Flows}
\label{app:additional-training-diagnostics}

In this appendix, we present the training diagnostics corresponding to
Fig.~\ref{fig:z4j-grouped-training-diagnostics} for the remaining benchmark processes. For each process, the left panel shows the mean unweighting efficiency for \(\varepsilon=10^{-2}\), and the right panel shows the mean \(\mathrm{ESS}/N\), both as functions of the training round.

\begin{figure*}[t]
    \centering
    \begin{tabular}{cc}
        \includegraphics[width=0.48\textwidth]{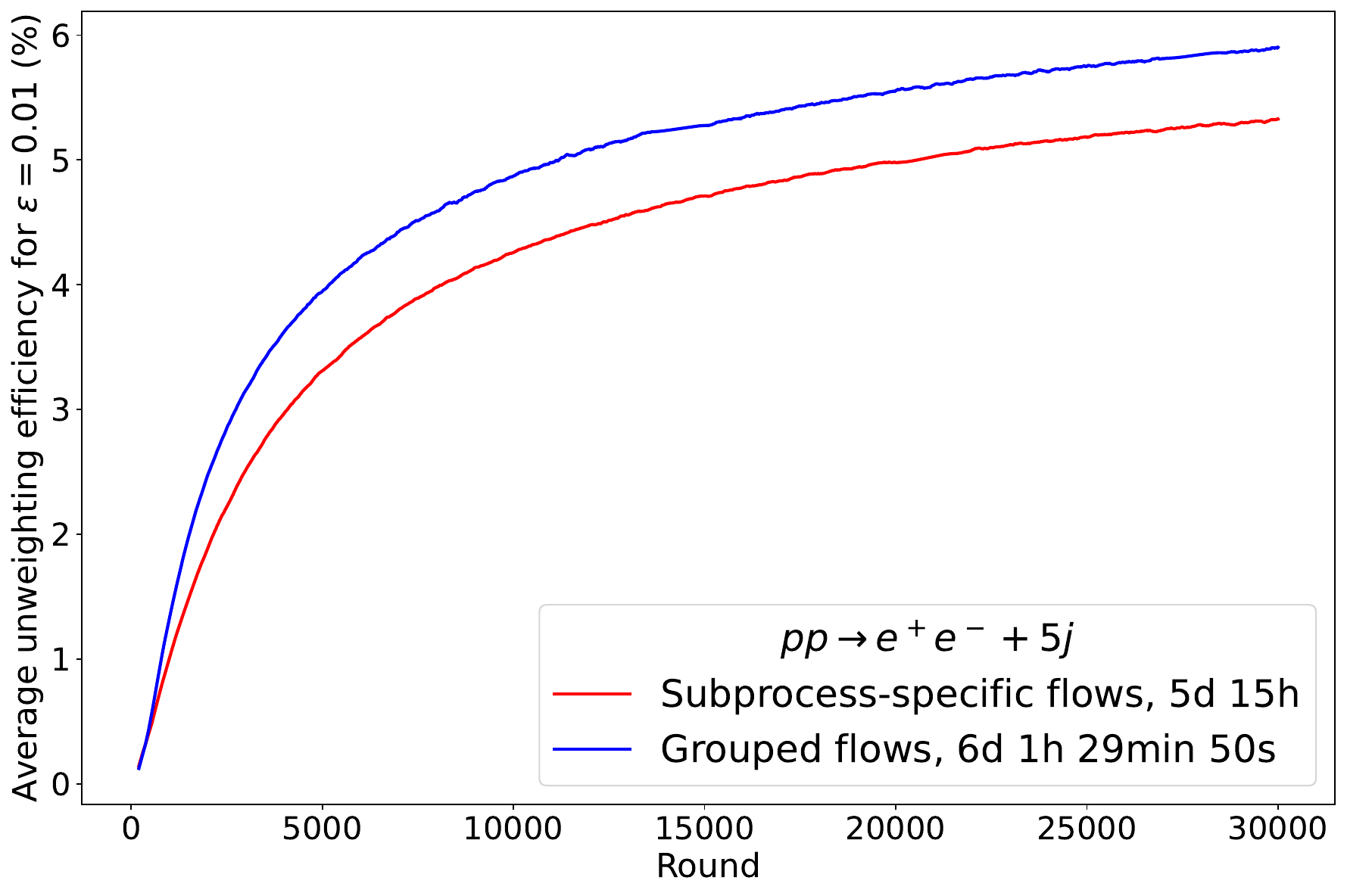} &
        \includegraphics[width=0.48\textwidth]{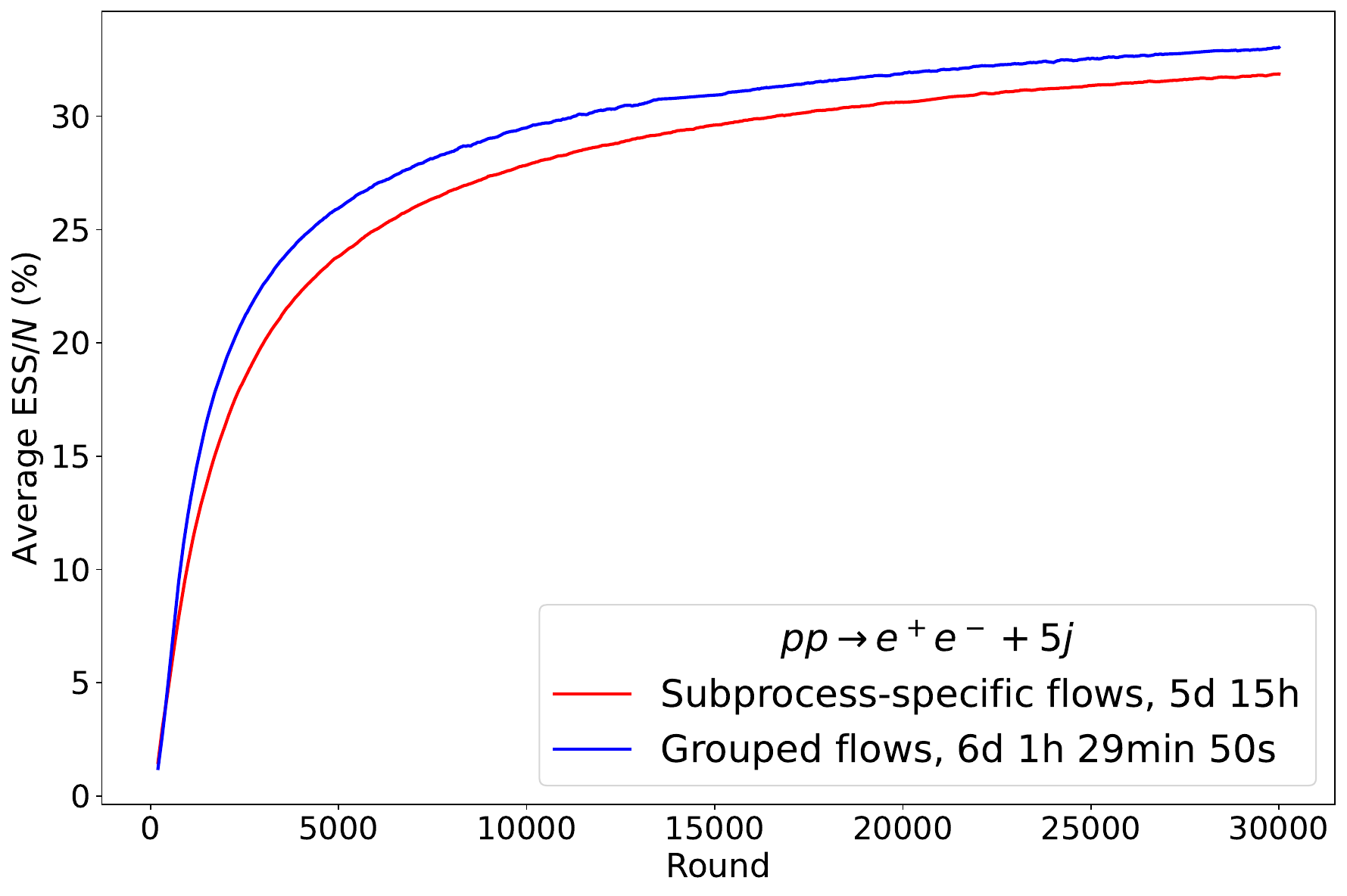}
    \end{tabular}
    \caption{
    Training diagnostics for \zjjjjj production on four H100 GPUs.
    Both quantities are averaged over the \(168\) partonic subprocesses.
    }
    \label{fig:z5j-grouped-training-diagnostics}
\end{figure*}

\begin{figure*}[t]
    \centering
    \begin{tabular}{cc}
        \includegraphics[width=0.48\textwidth]{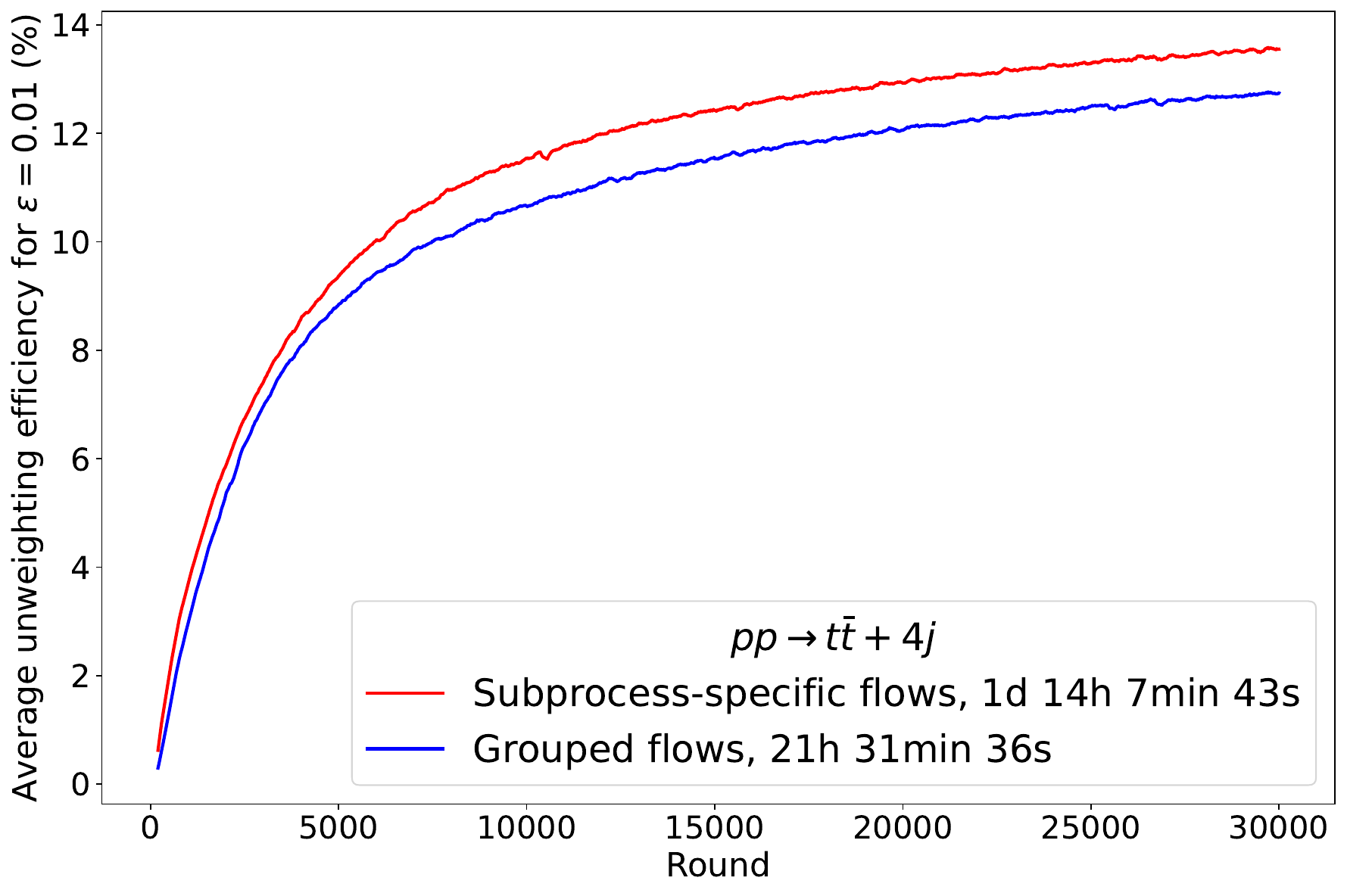} &
        \includegraphics[width=0.48\textwidth]{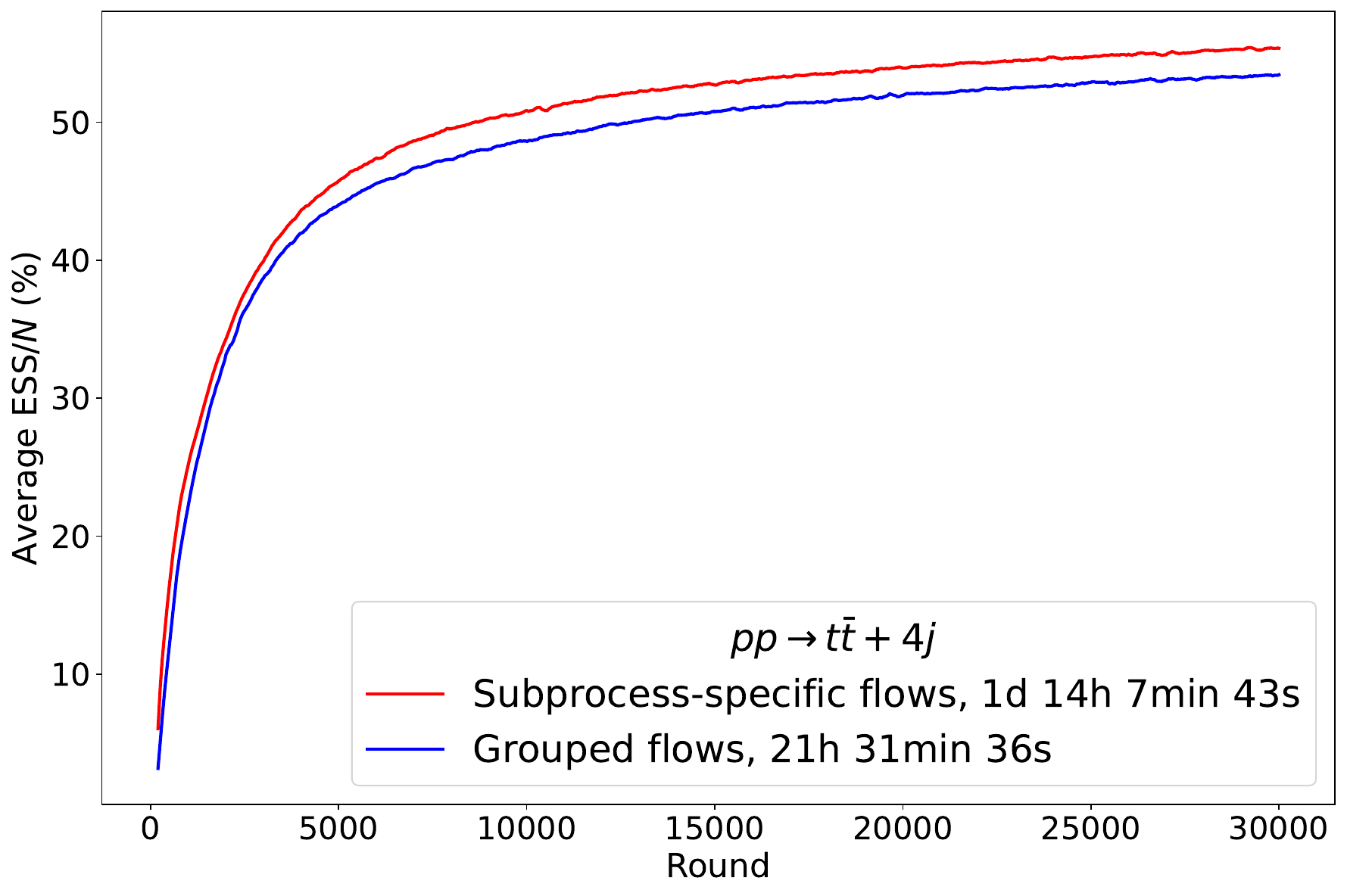}
    \end{tabular}
    \caption{
    Training diagnostics for \ttjjjj production on four H100 GPUs.
    Both quantities are averaged over the \(20\) partonic subprocesses.
    }
    \label{fig:tt4j-grouped-training-diagnostics}
\end{figure*}

\begin{figure*}[t]
    \centering
    \begin{tabular}{cc}
        \includegraphics[width=0.48\textwidth]{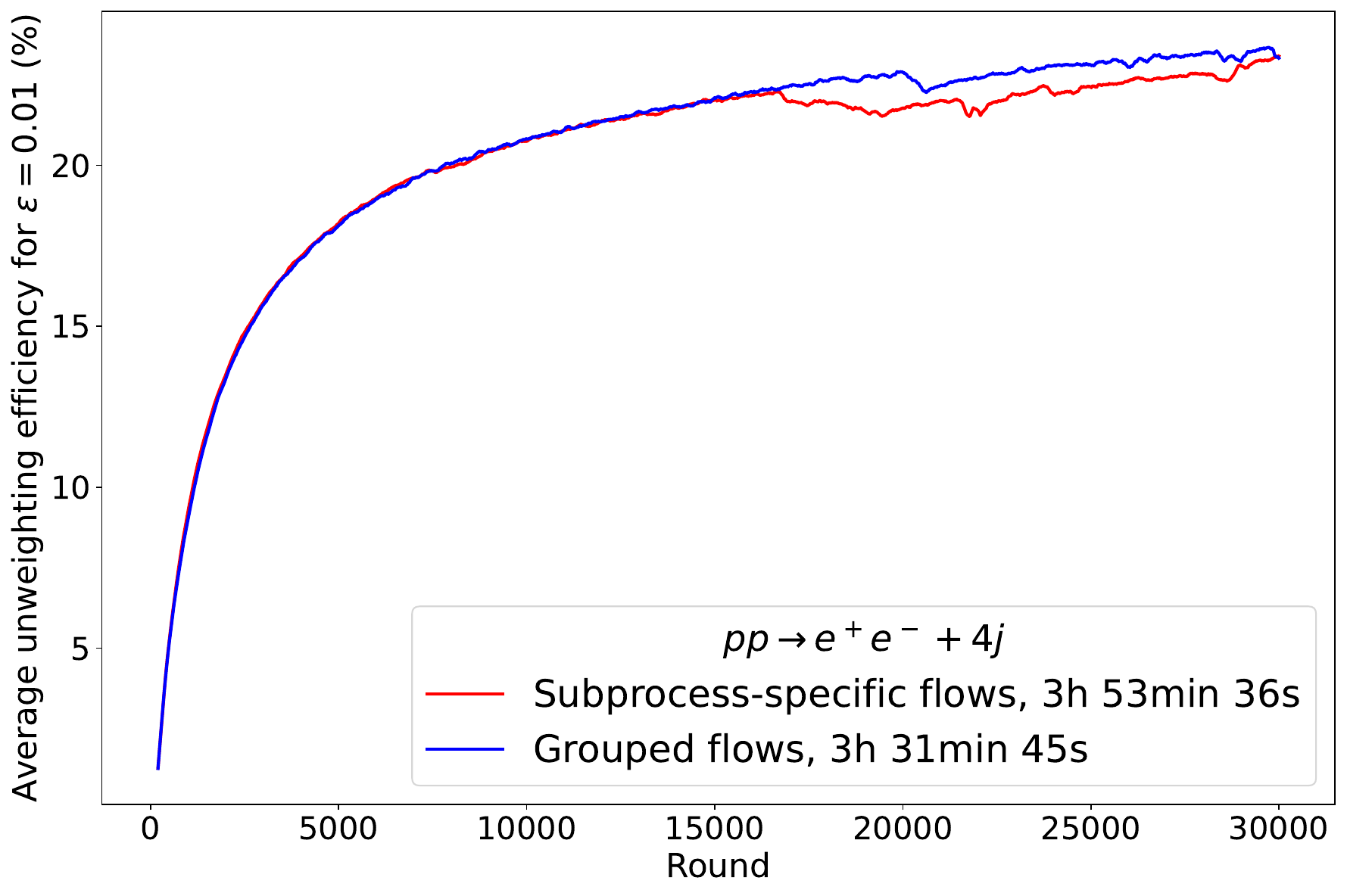} &
        \includegraphics[width=0.48\textwidth]{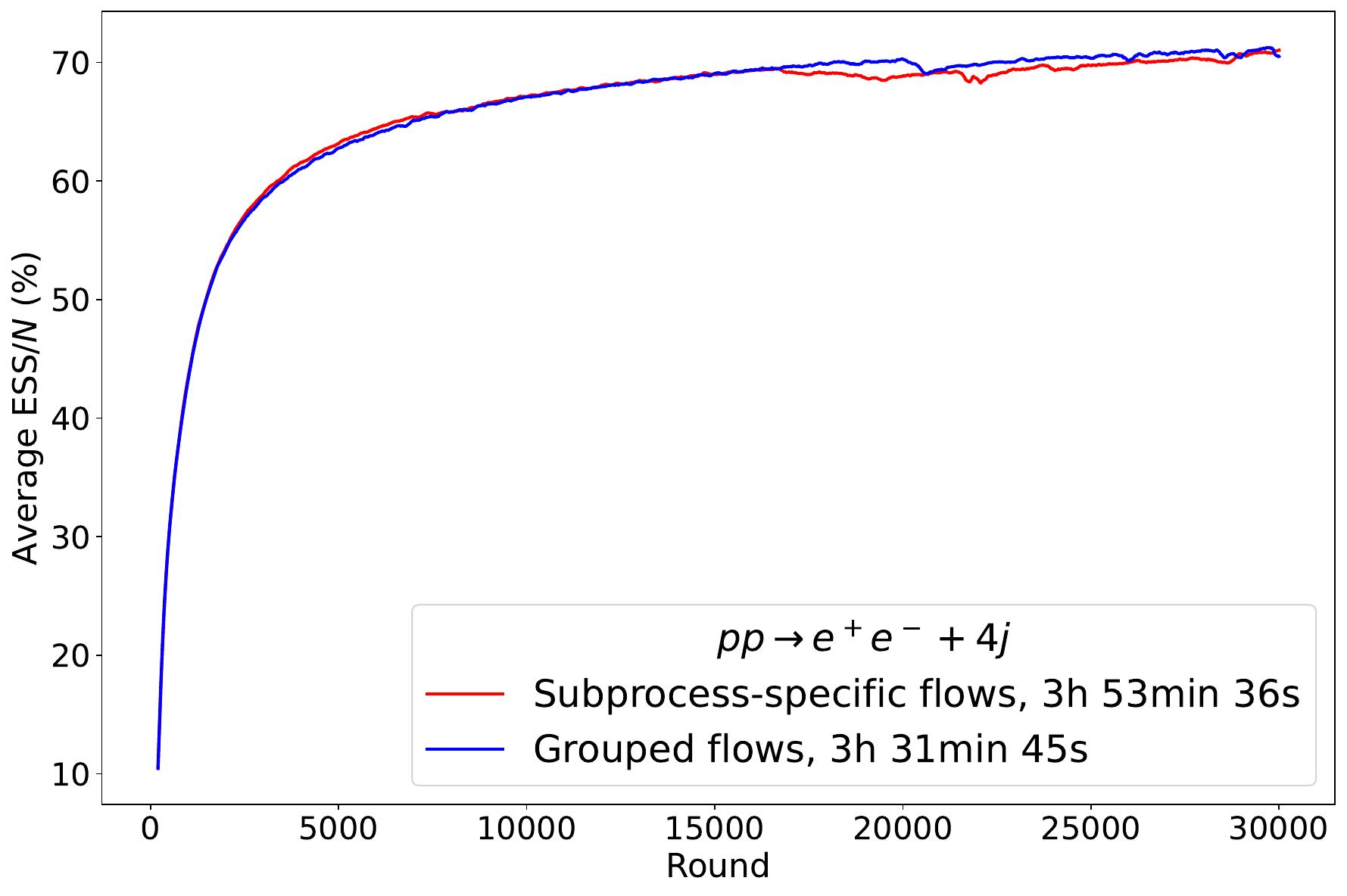}
    \end{tabular}
    \caption{
    Training diagnostics for \jjjj production on four H100 GPUs.
    Both quantities are averaged over the \(20\) partonic subprocesses.
    }
    \label{fig:4j-grouped-training-diagnostics}
\end{figure*}

\begin{figure*}[t]
    \centering
    \begin{tabular}{cc}
        \includegraphics[width=0.48\textwidth]{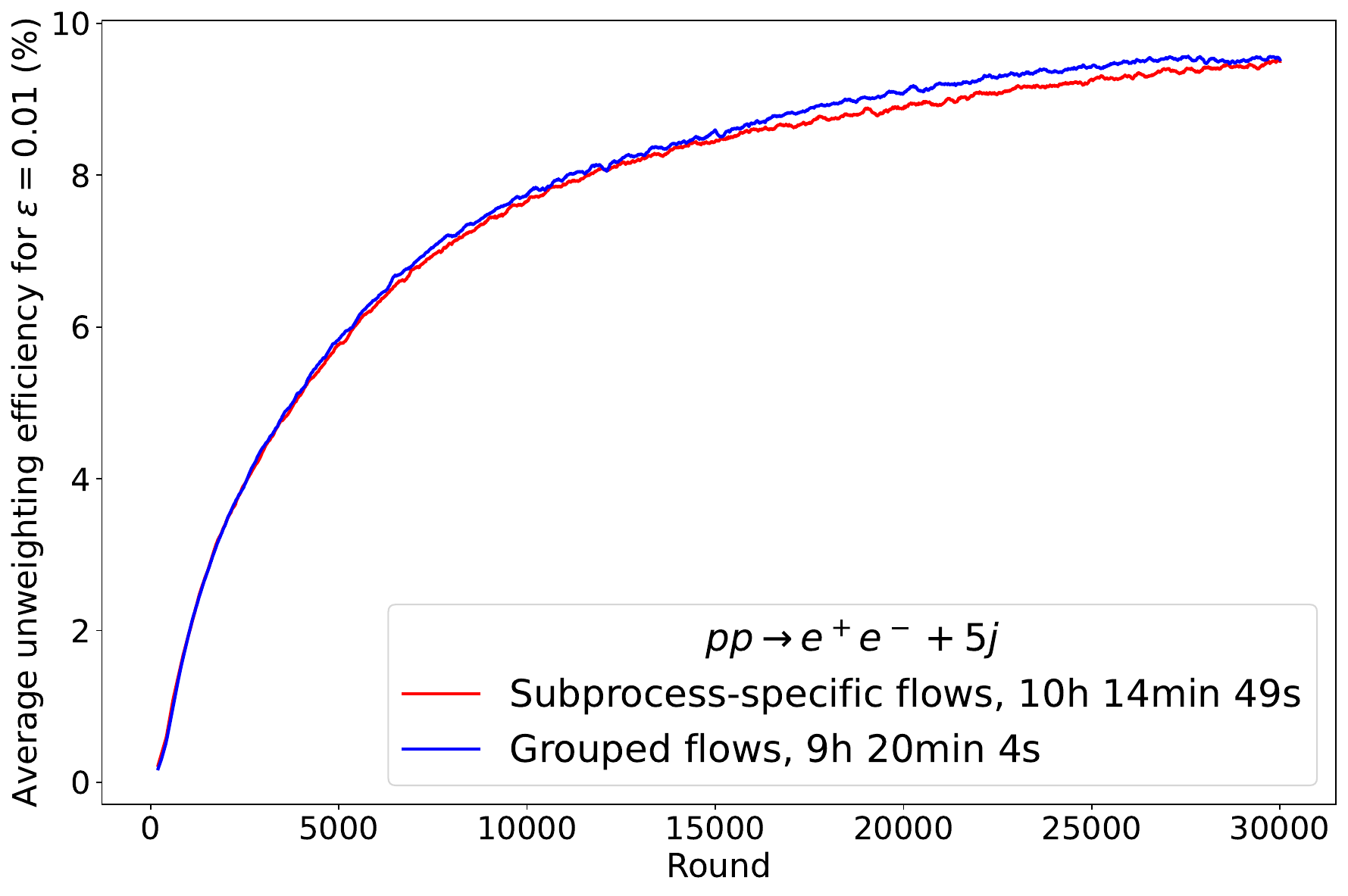} &
        \includegraphics[width=0.48\textwidth]{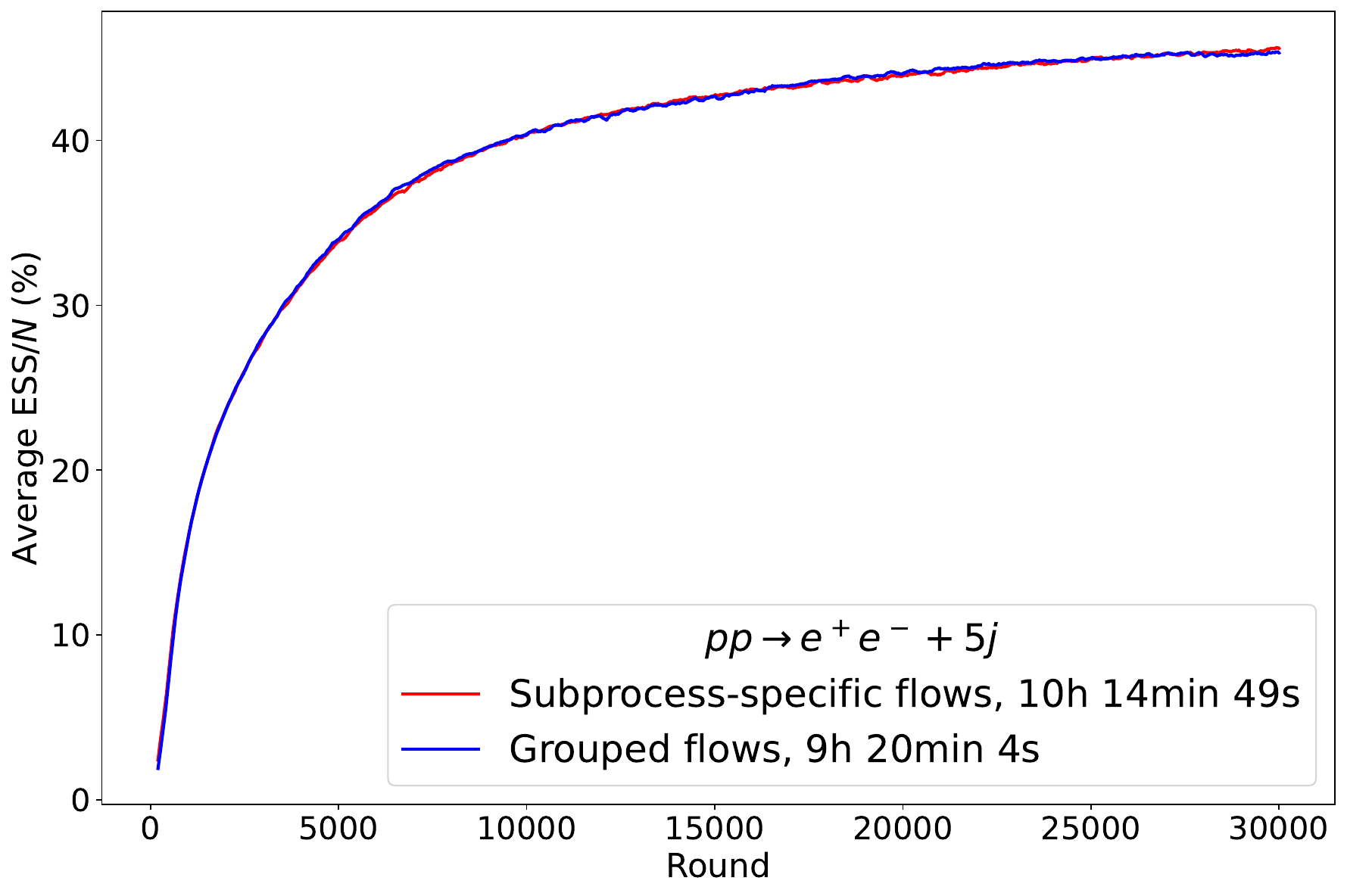}
    \end{tabular}
    \caption{
    Training diagnostics for \jjjjj production on four H100 GPUs.
    Both quantities are averaged over the \(24\) partonic subprocesses.
    }
    \label{fig:5j-grouped-training-diagnostics}
\end{figure*}

Figures~\ref{fig:z5j-grouped-training-diagnostics}--\ref{fig:5j-grouped-training-diagnostics} show that the relative performance of grouped and subprocess-specific flows depends on the subprocess structure. For \zjjjjj, which contains
\(168\) subprocesses, the grouped flow achieves higher mean unweighting efficiency and \(\mathrm{ESS}/N\) during most of the training, following the same pattern as \zjjjj in Fig.~\ref{fig:z4j-grouped-training-diagnostics}. This is consistent with beneficial parameter sharing among related subprocesses. The improvement in proposal quality comes at the cost of a longer training wall time, consistent with the less uniform batch structure of the grouped implementation at fixed subprocess-specific training statistics.

For \jjjj and \jjjjj, which contain \(20\) and \(24\) subprocesses,
respectively, the two approaches give nearly identical unweighting efficiency and \(\mathrm{ESS}/N\), while the grouped setup trains faster. The
\ttjjjj benchmark shows a clearer tradeoff: grouping the \(20\) subprocesses substantially reduces the training time, whereas subprocess-specific flows achieve higher mean unweighting efficiency and \(\mathrm{ESS}/N\).

Overall, grouping improves proposal quality for the large subprocess
collections in \zjjjj and \zjjjjj. For the smaller collections, its main advantage is reduced training time, with proposal quality remaining comparable or becoming modestly lower.

\section{Additional Subprocess-Specific Flow Results on Four H100 GPUs}
\label{app:additional-h100-subprocess-specific-results}

In this appendix, we report additional four-H100 benchmark results for the subprocess-specific flow setup. The main text uses \zjjjj production as the representative process; here we present the corresponding results for \zjjjjj, \ttjjjj, \jjjj, and \jjjjj. The timing definitions follow Sec.~\ref{sec:results}. The left panel of each figure shows the end-to-end speedup, for which the \pepflow runtime includes both training and event generation. The right panel shows the generation-only speedup obtained with the final trained subprocess-specific flow checkpoints. For the end-to-end comparison, the checkpoint is selected by minimizing the estimated training-plus-generation time defined in Sec.~\ref{sec:results}. After the checkpoint has been selected, the requested event sample is generated directly and the measured generation time is used in the reported total runtime. All \pepflow runtimes reported in this appendix are measured directly; none are obtained by extrapolation.

\begin{figure*}[t]
    \centering
    \begin{tabular}{cc}
        \includegraphics[width=0.48\textwidth]{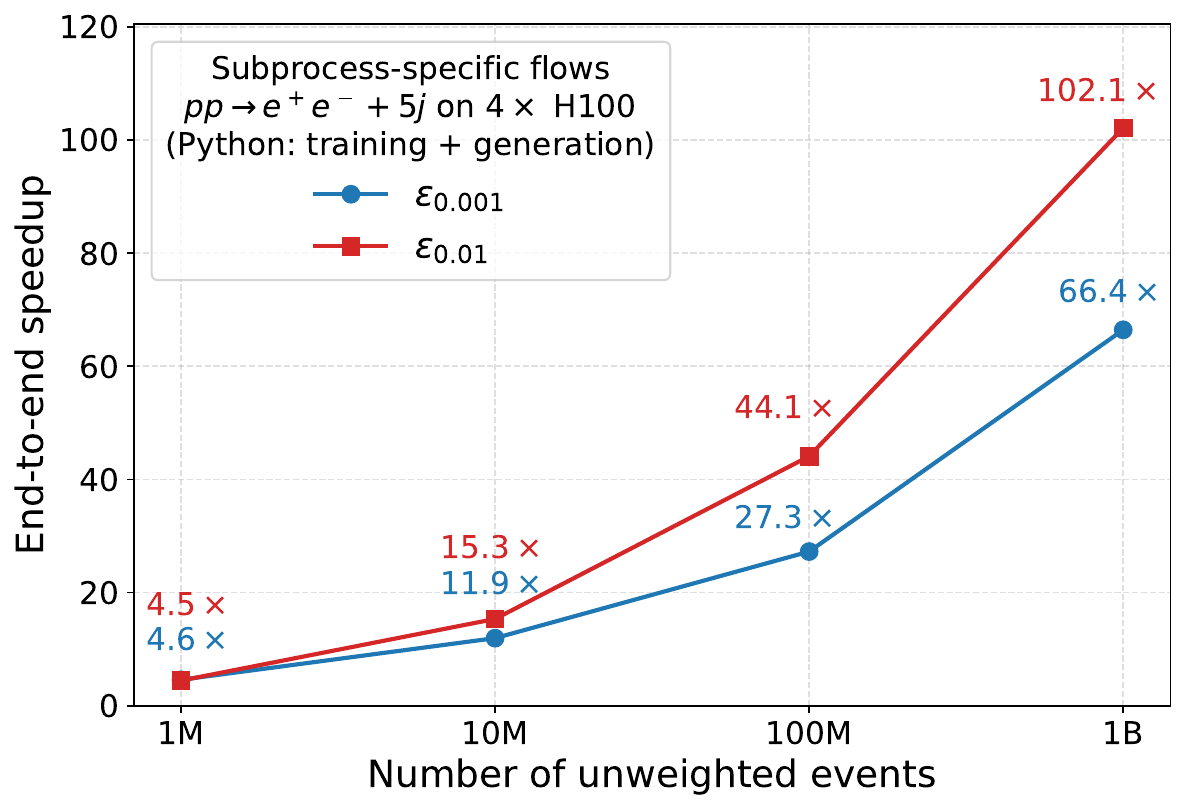} &
        \includegraphics[width=0.48\textwidth]{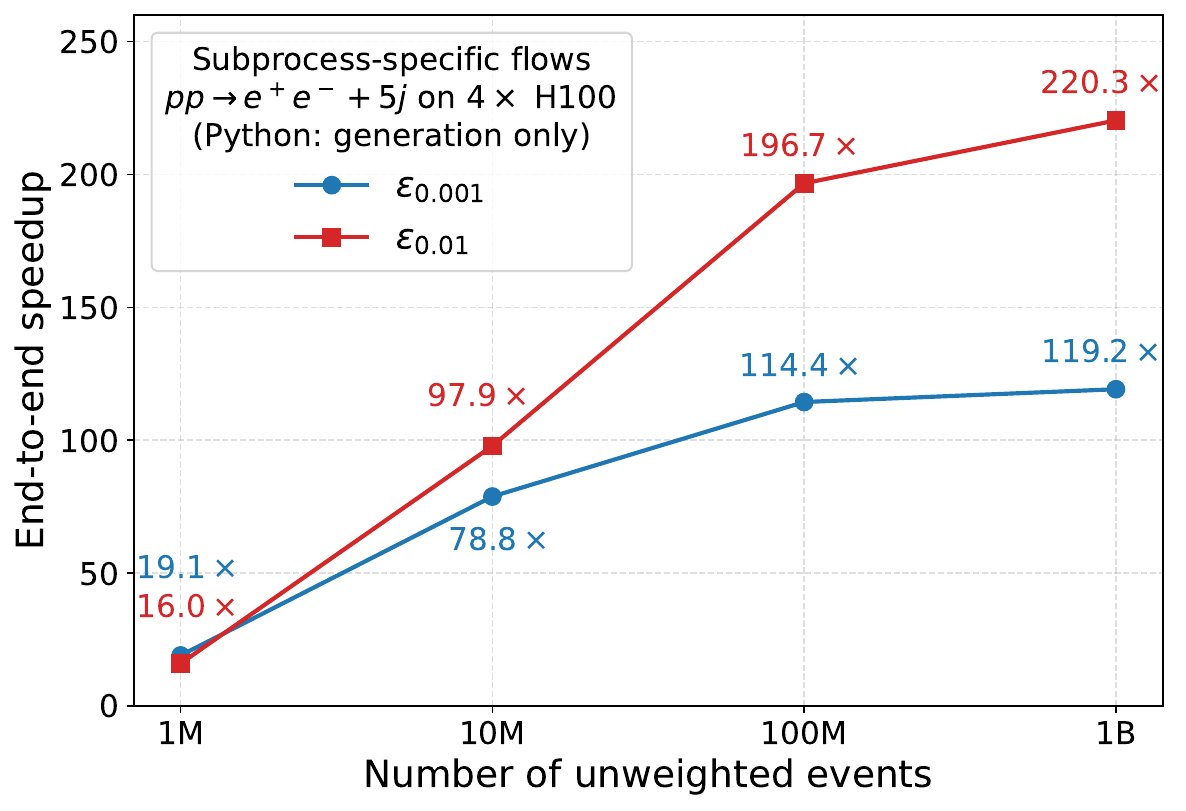}
    \end{tabular}
    \caption{
    End-to-end (left) and generation-only (right) speedups for
    subprocess-specific flow generation in \zjjjjj production on
    \(4\times\) H100 GPUs. The left panel uses the measured \pepflow total runtime, including training and event generation, while the right panel uses the measured generation-only runtime of the final trained subprocess-specific flow checkpoints. The \pepvegas runtime is measured at \(10^6\) events and linearly extrapolated to larger target samples.
    }
    \label{fig:z5j-h100-subprocess-specific-speedup}
\end{figure*}

\begin{figure*}[t]
    \centering
    \begin{tabular}{cc}
        \includegraphics[width=0.48\textwidth]{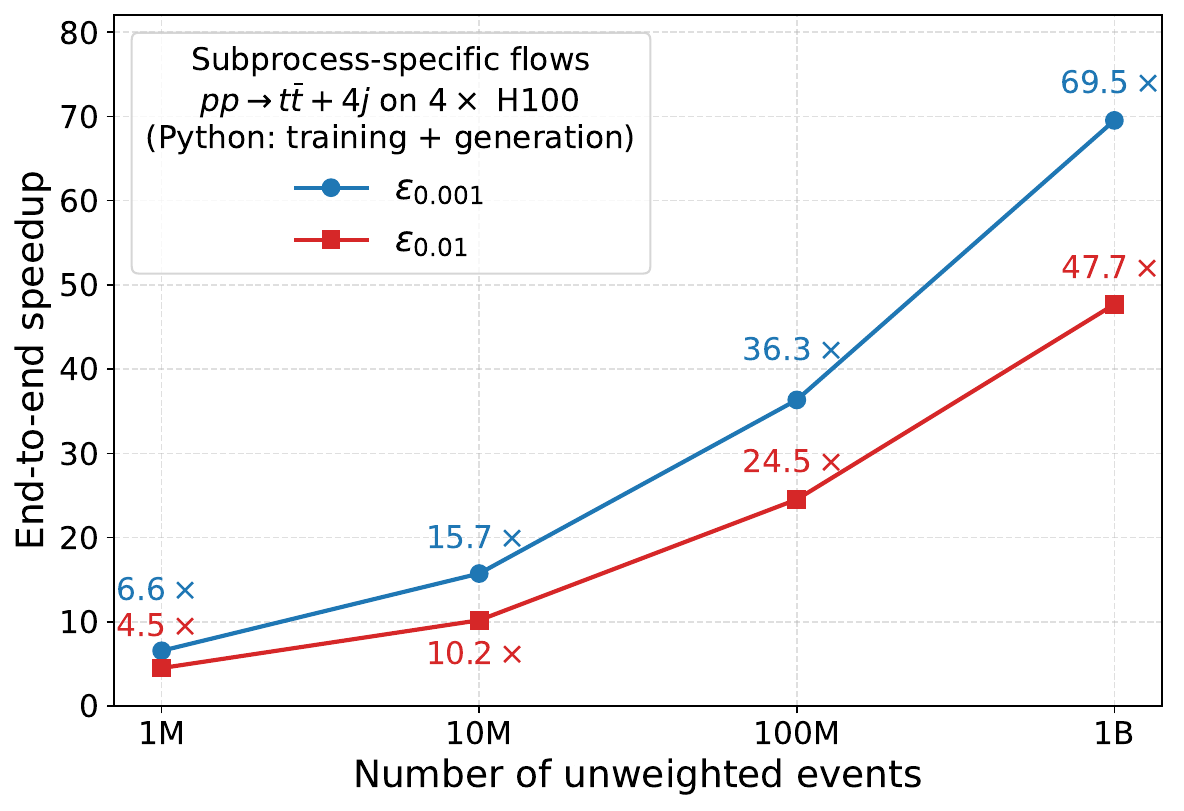} &
        \includegraphics[width=0.48\textwidth]{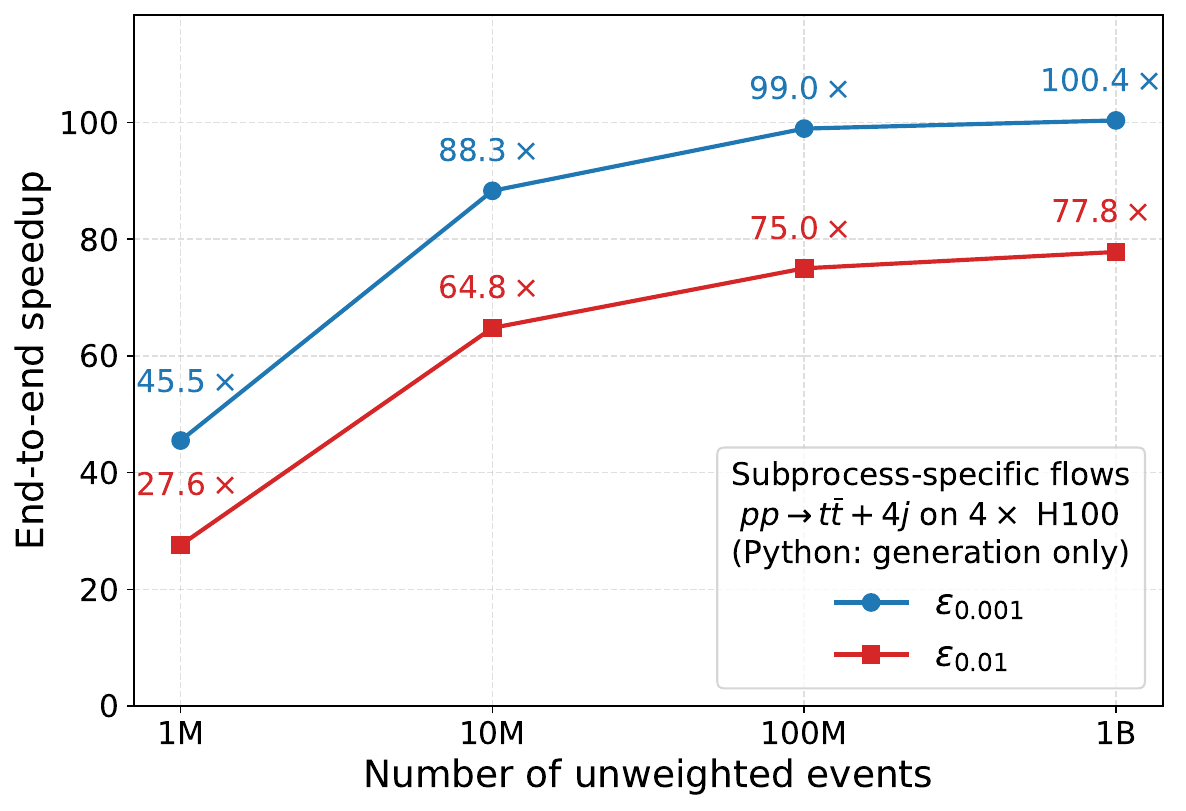}
    \end{tabular}
    \caption{
    End-to-end (left) and generation-only (right) speedups for
    subprocess-specific flow generation in \ttjjjj production on
    \(4\times\) H100 GPUs. The left panel uses the measured \pepflow total runtime, including training and event generation, while the right panel uses the measured generation-only runtime of the final trained subprocess-specific flow checkpoints. The \pepvegas runtime is measured at \(10^6\) events and linearly extrapolated to larger target samples.
    }
    \label{fig:tt4j-h100-subprocess-specific-speedup}
\end{figure*}

\begin{figure*}[t]
    \centering
    \begin{tabular}{cc}
        \includegraphics[width=0.48\textwidth]{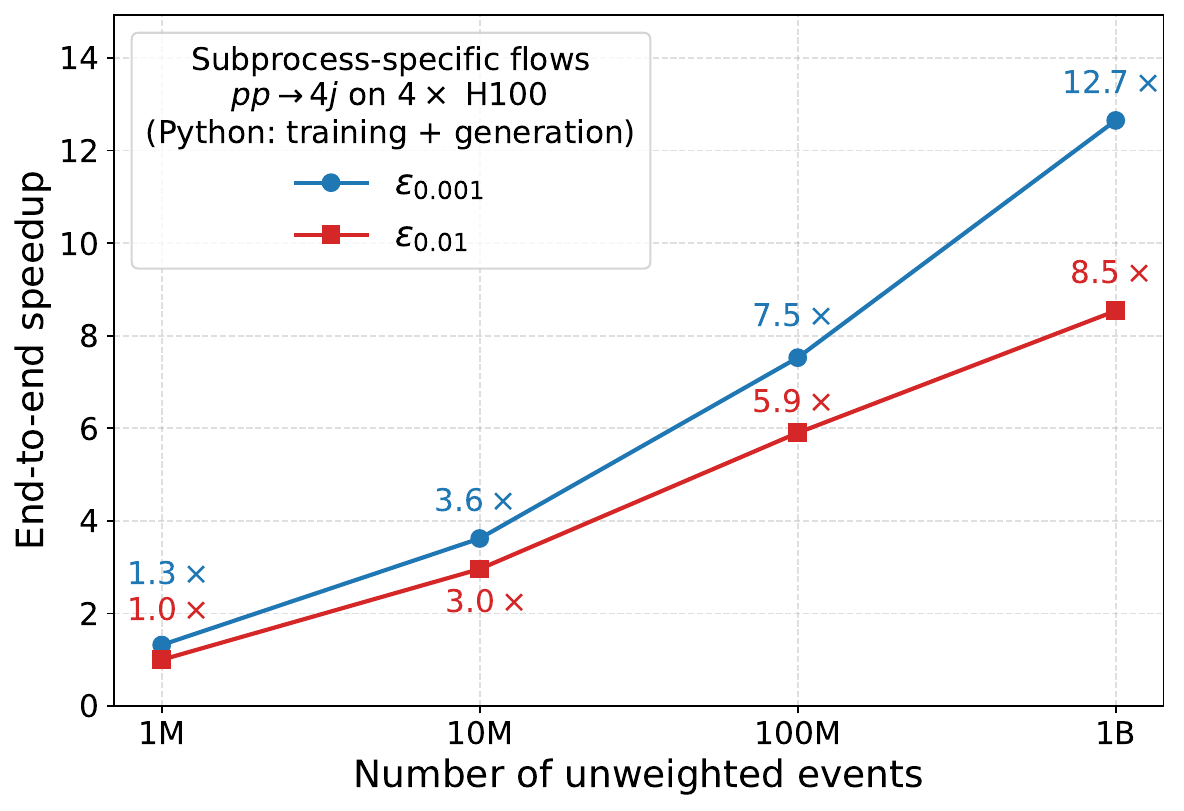} &
        \includegraphics[width=0.48\textwidth]{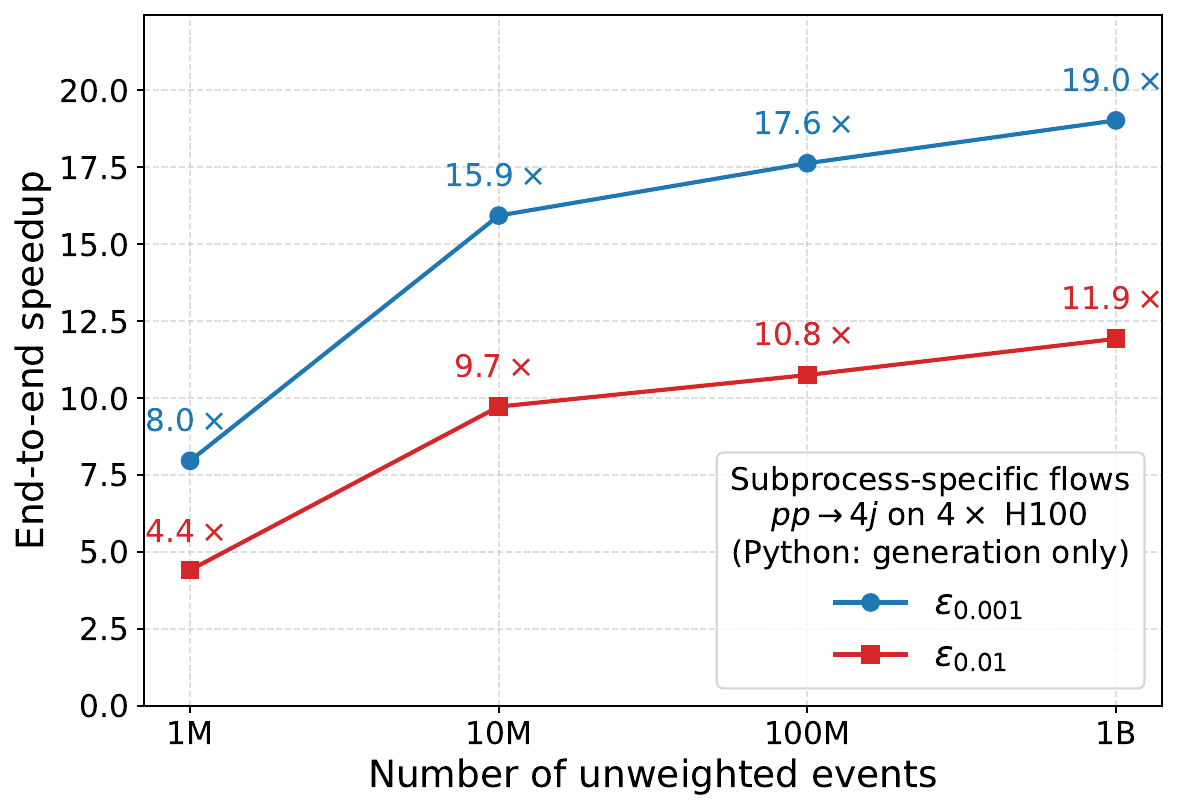}
    \end{tabular}
    \caption{
    End-to-end (left) and generation-only (right) speedups for
    subprocess-specific flow generation in \jjjj production on
    \(4\times\) H100 GPUs. The left panel uses the measured \pepflow total runtime, including training and event generation, while the right panel uses the measured generation-only runtime of the final trained subprocess-specific flow checkpoints. The \pepvegas runtime is measured at \(10^6\) events and linearly extrapolated to larger target samples.
    }
    \label{fig:fourj-h100-subprocess-specific-speedup}
\end{figure*}

\begin{figure*}[t]
    \centering
    \begin{tabular}{cc}
        \includegraphics[width=0.48\textwidth]{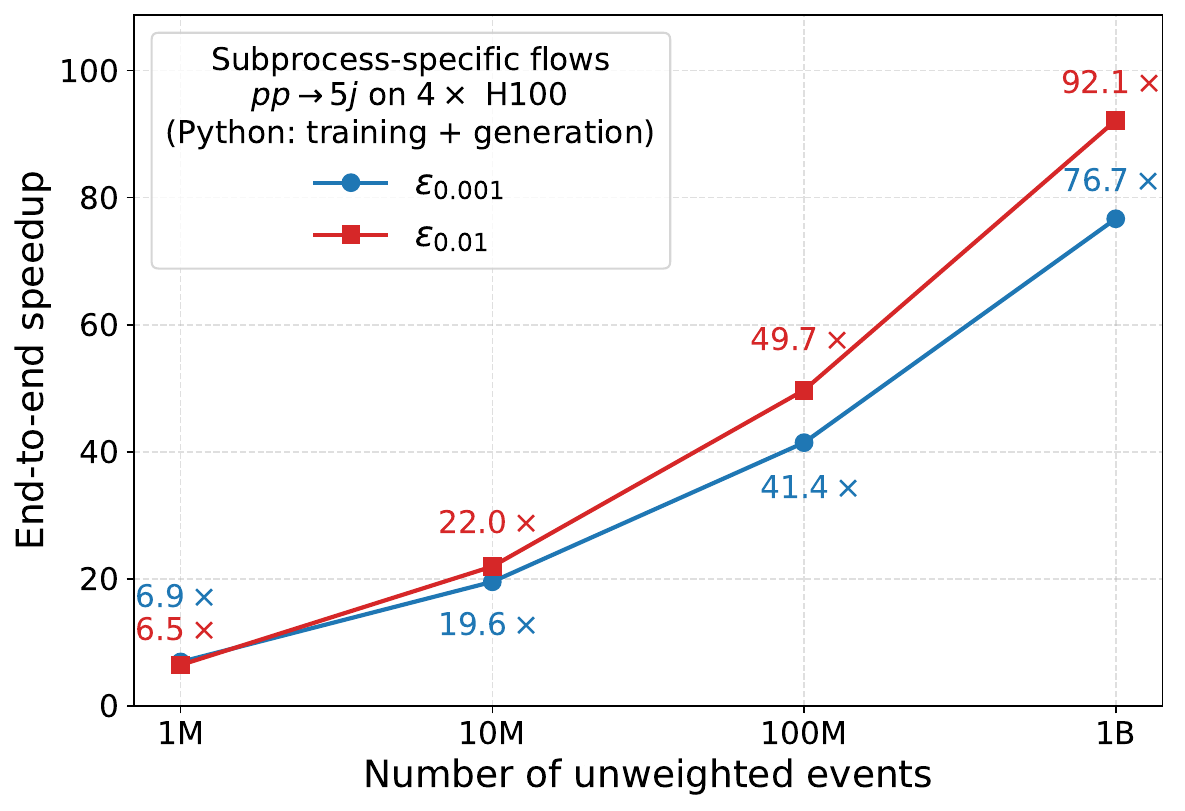} &
        \includegraphics[width=0.48\textwidth]{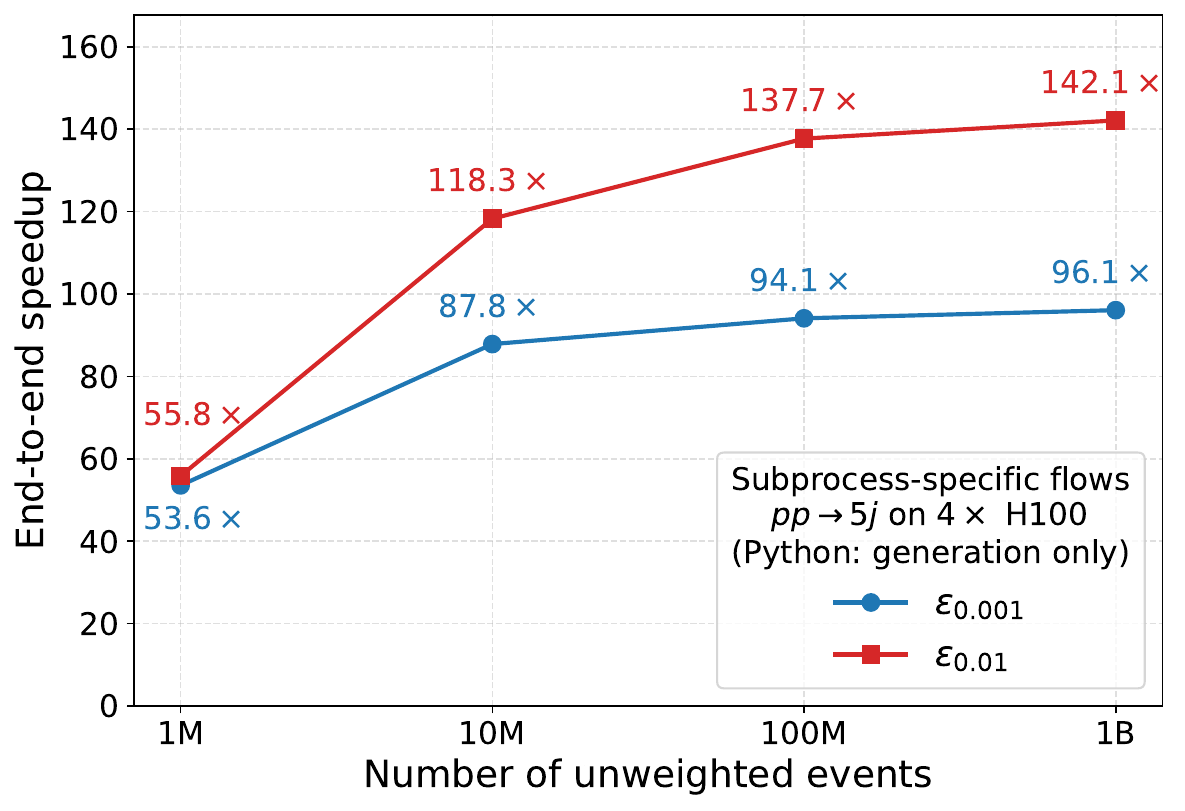}
    \end{tabular}
    \caption{
    End-to-end (left) and generation-only (right) speedups for
    subprocess-specific flow generation in \jjjjj production on
    \(4\times\) H100 GPUs. The left panel uses the measured \pepflow total runtime, including training and event generation, while the right panel uses the measured generation-only runtime of the final trained subprocess-specific flow checkpoints. The \pepvegas runtime is measured at \(10^6\) events and linearly extrapolated to larger target samples.
    }
    \label{fig:fivej-h100-subprocess-specific-speedup}
\end{figure*}

Figures~\ref{fig:z5j-h100-subprocess-specific-speedup}--\ref{fig:fivej-h100-subprocess-specific-speedup} show end-to-end speedups larger than unity for all four additional processes, already at \(10^6\) events. The speedup generally increases with the requested sample size, reflecting the tradeoff between training cost and generation efficiency. Larger target samples can favor later checkpoints whose improved unweighting efficiency compensates for the additional training time.

The largest gains are observed for the higher-multiplicity processes. For \zjjjjj, the end-to-end speedup at \(10^9\) events reaches \(66.4\times\) for \(\varepsilon=10^{-3}\) and \(102.1\times\) for
\(\varepsilon=10^{-2}\), while the corresponding generation-only speedups are \(119.2\times\) and \(220.3\times\). The \jjjjj benchmark shows a similar pattern, with end-to-end speedups of \(76.7\times\) and \(92.1\times\), and generation-only speedups of \(96.1\times\) and
\(142.1\times\).

For \ttjjjj, the end-to-end speedups at \(10^9\) events are
\(69.5\times\) and \(47.7\times\), while the generation-only speedups are \(100.4\times\) and \(77.8\times\). The simpler \jjjj process gives the smallest gains, reaching \(12.7\times\) and \(8.5\times\) end-to-end. Nevertheless, the \pepflow workflow remains faster than \pepvegas generation for every sample size shown.

The absolute runtimes underlying Figs.~\ref{fig:z5j-h100-subprocess-specific-speedup}--\ref{fig:fivej-h100-subprocess-specific-speedup} are listed in Tables~\ref{tab:z5j-h100-subprocess-specific-runtime}--\ref{tab:5j-h100-subprocess-specific-runtime}. The \pepflow total runtime includes training and event generation, while the generation-only runtime isolates event generation with the final trained subprocess-specific flow checkpoints. Both \pepflow columns contain directly measured runtimes. For the \pepvegas workflow, the \(10^6\)-event runtime is measured directly, while the runtimes for \(10^7\), \(10^8\), and \(10^9\) events are obtained by linearly rescaling the measured \(10^6\)-event runtime and are shown in blue.

\begin{table*}[t]
    \centering
    \caption{
    Runtime comparison for generating unweighted \zjjjjj events with
    subprocess-specific flows on \(4\times\) H100 GPUs. The \pepflow total and generation-only runtimes are measured directly. Blue \pepvegas entries are linear extrapolations of the measured \(10^6\)-event runtime.
    }
    \label{tab:z5j-h100-subprocess-specific-runtime}
    \begin{tabular}{c c c c c}
        \hline
        \(\varepsilon\) & Number of events
        & \pepvegas
        & \pepflow (total)
        & \pepflow (gen. only) \\
        \hline
        \multirow{4}{*}{\(10^{-3}\)}
        & \(1M\) & 9h\;38min\;17s & 2h\;5min\;55s & 30min\;21s \\
        & \(10M\) & \textcolor{blue}{4d\;22min\;50s} & 8h\;3min\;56s & 1h\;13min\;21s \\
        & \(100M\) & \textcolor{blue}{40d\;3h\;48min\;20s} & 1d\;11h\;21min\;51s & 8h\;25min\;31s \\
        & \(1B\) & \textcolor{blue}{401d\;14h\;3min\;20s} & 6d\;1h\;2min\;56s & 3d\;8h\;50min\;57s \\
        \hline
        \multirow{4}{*}{\(10^{-2}\)}
        & \(1M\) & 7h\;24min\;24s & 1h\;39min\;49s & 27min\;43s \\
        & \(10M\) & \textcolor{blue}{3d\;2h\;4min} & 4h\;49min\;41s & 45min\;24s \\
        & \(100M\) & \textcolor{blue}{30d\;20h\;40min} & 16h\;47min\;42s & 3h\;45min\;57s \\
        & \(1B\) & \textcolor{blue}{308d\;14h\;40min} & 3d\;32min\;5s & 1d\;9h\;36min\;50s \\
        \hline
    \end{tabular}
\end{table*}

\begin{table*}[t]
    \centering
    \caption{
    Runtime comparison for generating unweighted \ttjjjj events with subprocess-specific flows on \(4\times\) H100 GPUs. The \pepflow total and generation-only runtimes are measured directly. Blue \pepvegas entries are linear extrapolations of the measured \(10^6\)-event runtime.
    }
    \label{tab:tt4j-h100-subprocess-specific-runtime}
    \begin{tabular}{c c c c c}
        \hline
        \(\varepsilon\) & Number of events
        & \pepvegas
        & \pepflow (total)
        & \pepflow (gen. only) \\
        \hline
        \multirow{4}{*}{\(10^{-3}\)}
        & \(1M\) & 4h\;56min\;32s & 45min\;11s & 6min\;31s \\
        & \(10M\) & \textcolor{blue}{2d\;1h\;25min\;20s} & 3h\;8min\;29s & 33min\;35s \\
        & \(100M\) & \textcolor{blue}{20d\;14h\;13min\;20s} & 13h\;35min\;54s & 4h\;59min\;39s \\
        & \(1B\) & \textcolor{blue}{205d\;22h\;13min\;20s} & 2d\;23h\;5min\;12s & 1d\;8h\;29min\;36s \\
        \hline
        \multirow{4}{*}{\(10^{-2}\)}
        & \(1M\) & 1h\;50min\;23s & 24min\;28s & 4min \\
        & \(10M\) & \textcolor{blue}{18h\;23min\;50s} & 1h\;48min\;20s & 17min\;2s \\
        & \(100M\) & \textcolor{blue}{7d\;15h\;58min\;20s} & 7h\;30min\;2s & 2h\;27min\;9s \\
        & \(1B\) & \textcolor{blue}{76d\;15h\;43min\;20s} & 1d\;14h\;34min\;40s & 23h\;38min\;23s \\
        \hline
    \end{tabular}
\end{table*}

\begin{table*}[t]
    \centering
    \caption{
    Runtime comparison for generating unweighted \jjjj events with
    subprocess-specific flows on \(4\times\) H100 GPUs. The \pepflow total and generation-only runtimes are measured directly. Blue \pepvegas entries are linear extrapolations of the measured \(10^6\)-event runtime.
    }
    \label{tab:4j-h100-subprocess-specific-runtime}
    \begin{tabular}{c c c c c}
        \hline
        \(\varepsilon\) & Number of events
        & \pepvegas
        & \pepflow (total)
        & \pepflow (gen. only) \\
        \hline
        \multirow{4}{*}{\(10^{-3}\)}
        & \(1M\) & 3min\;43s & 2min\;49s & 28s \\
        & \(10M\) & \textcolor{blue}{37min\;10s} & 10min\;16s & 2min\;20s \\
        & \(100M\) & \textcolor{blue}{6h\;11min\;40s} & 49min\;23s & 21min\;5s \\
        & \(1B\) & \textcolor{blue}{2d\;13h\;56min\;40s} & 4h\;53min\;46s & 3h\;15min\;25s \\
        \hline
        \multirow{4}{*}{\(10^{-2}\)}
        & \(1M\) & 1min\;46s & 1min\;45s & 23s \\
        & \(10M\) & \textcolor{blue}{17min\;40s} & 5min\;57s & 1min\;48s \\
        & \(100M\) & \textcolor{blue}{2h\;56min\;40s} & 29min\;56s & 16min\;26s \\
        & \(1B\) & \textcolor{blue}{1d\;5h\;26min\;40s} & 3h\;26min\;55s & 2h\;28min\;7s \\
        \hline
    \end{tabular}
\end{table*}

\begin{table*}[t]
    \centering
    \caption{
    Runtime comparison for generating unweighted \jjjjj events with
    subprocess-specific flows on \(4\times\) H100 GPUs. The \pepflow total and generation-only runtimes are measured directly. Blue \pepvegas entries are linear extrapolations of the measured \(10^6\)-event runtime.
    }
    \label{tab:5j-h100-subprocess-specific-runtime}
    \begin{tabular}{c c c c c}
        \hline
        \(\varepsilon\) & Number of events
        & \pepvegas
        & \pepflow (total)
        & \pepflow (gen. only) \\
        \hline
        \multirow{4}{*}{\(10^{-3}\)}
        & \(1M\) & 2h\;18min\;21s & 19min\;55s & 2min\;35s \\
        & \(10M\) & \textcolor{blue}{23h\;3min\;30s} & 1h\;10min\;45s & 15min\;45s \\
        & \(100M\) & \textcolor{blue}{9d\;14h\;35min} & 5h\;33min\;49s & 2h\;27min\;2s \\
        & \(1B\) & \textcolor{blue}{96d\;1h\;50min} & 1d\;6h\;4min\;35s & 1d\;4s \\
        \hline
        \multirow{4}{*}{\(10^{-2}\)}
        & \(1M\) & 1h\;23min\;46s & 12min\;55s & 1min\;28s \\
        & \(10M\) & \textcolor{blue}{13h\;57min\;40s} & 38min\;6s & 7min\;3s \\
        & \(100M\) & \textcolor{blue}{5d\;19h\;36min\;40s} & 2h\;48min\;30s & 1h\;48s \\
        & \(1B\) & \textcolor{blue}{58d\;4h\;6min\;40s} & 15h\;9min\;6s & 9h\;49min\;24s \\
        \hline
    \end{tabular}
\end{table*}

\section{Additional Grouped Flow Results on Four H100 GPUs}
\label{app:additional-h100-grouped-flow-results}

In this appendix, we report additional four-H100 benchmark results for the grouped-flow setup. The main text uses \zjjjj production as the representative process; here we present the corresponding results for \zjjjjj, \ttjjjj, \jjjj, and \jjjjj. The timing definitions follow Sec.~\ref{sec:results}. The left panel of each figure shows the end-to-end speedup, for which the \pepflow runtime includes both training and event generation. The right panel shows the generation-only speedup obtained with the final trained grouped-flow checkpoints. For the end-to-end comparison, the checkpoint is selected by minimizing the estimated training-plus-generation time defined in Sec.~\ref{sec:results}. After the checkpoint has been selected, the requested event sample is generated directly and the measured generation time is used in the reported total runtime. All \pepflow runtimes reported in this appendix are measured directly; none are obtained by extrapolation.

\begin{figure*}[t]
    \centering
    \begin{tabular}{cc}
        \includegraphics[width=0.48\textwidth]{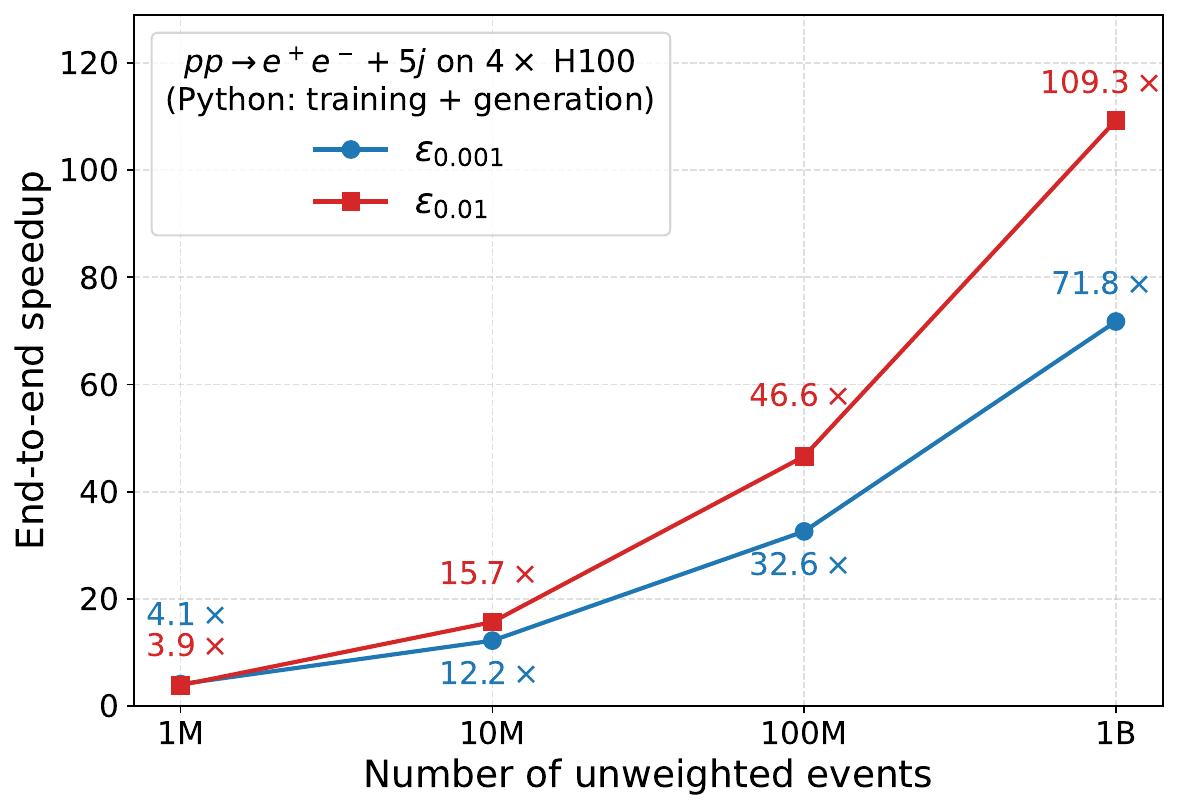} &
        \includegraphics[width=0.48\textwidth]{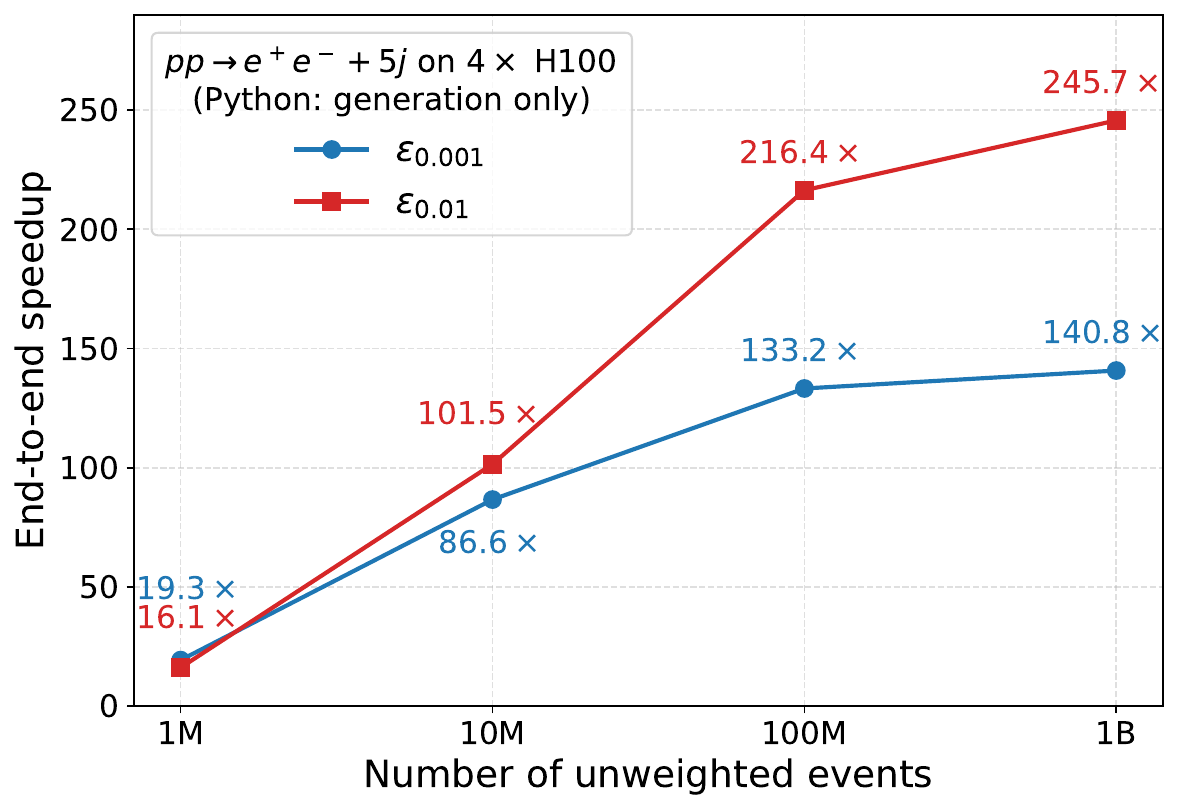}
    \end{tabular}
    \caption{
    End-to-end (left) and generation-only (right) speedups for grouped-flow generation in \zjjjjj production on \(4\times\) H100 GPUs. The left panel uses the directly measured \pepflow total runtime, including training and event generation, while the right panel uses the directly measured generation-only runtime of the final trained grouped-flow checkpoints. The \pepvegas runtime is measured at \(10^6\) events and linearly extrapolated to larger target samples.
    }
    \label{fig:z5j-h100-grouped-speedup}
\end{figure*}

\begin{figure*}[t]
    \centering
    \begin{tabular}{cc}
        \includegraphics[width=0.48\textwidth]{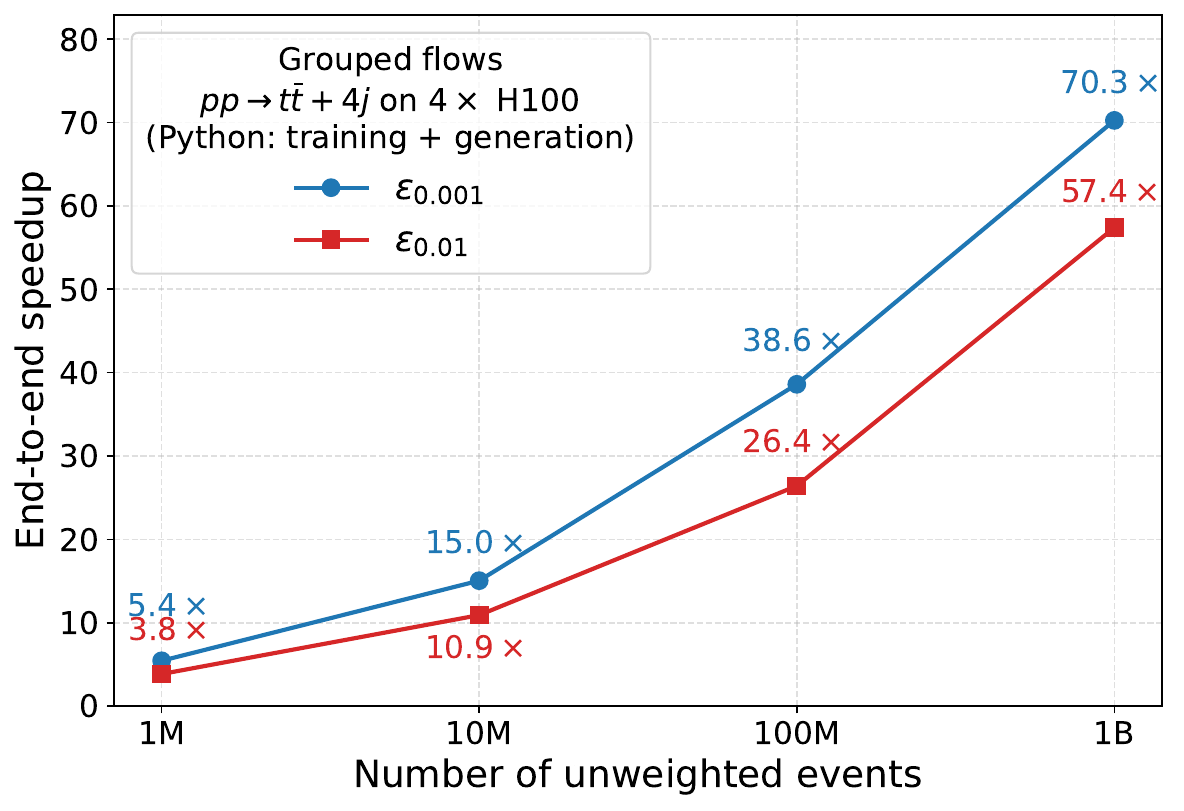} &
        \includegraphics[width=0.48\textwidth]{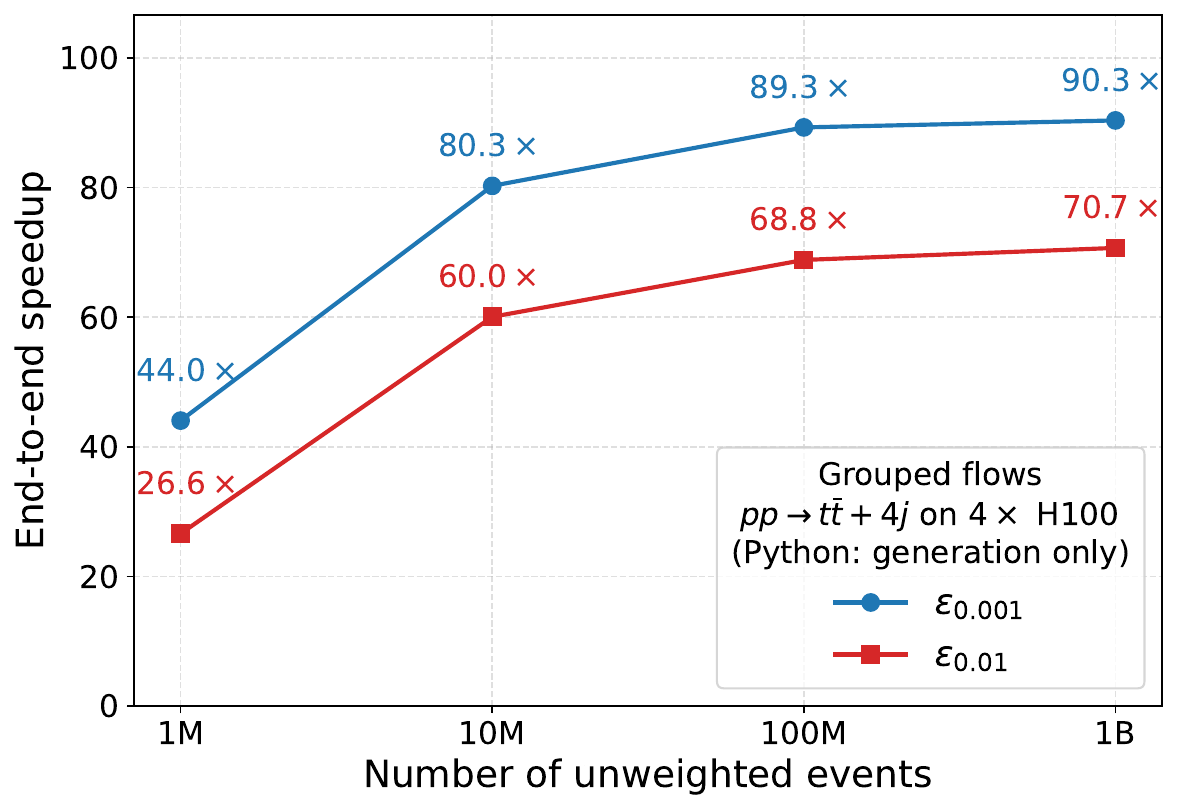}
    \end{tabular}
    \caption{
    End-to-end (left) and generation-only (right) speedups for grouped-flow generation in \ttjjjj production on \(4\times\) H100 GPUs. The left panel uses the directly measured \pepflow total runtime, including training and event generation, while the right panel uses the directly measured generation-only runtime of the final trained grouped-flow checkpoints. The \pepvegas runtime is measured at \(10^6\) events and linearly extrapolated to larger target samples.
    }
    \label{fig:tt4j-h100-grouped-speedup}
\end{figure*}

\begin{figure*}[t]
    \centering
    \begin{tabular}{cc}
        \includegraphics[width=0.48\textwidth]{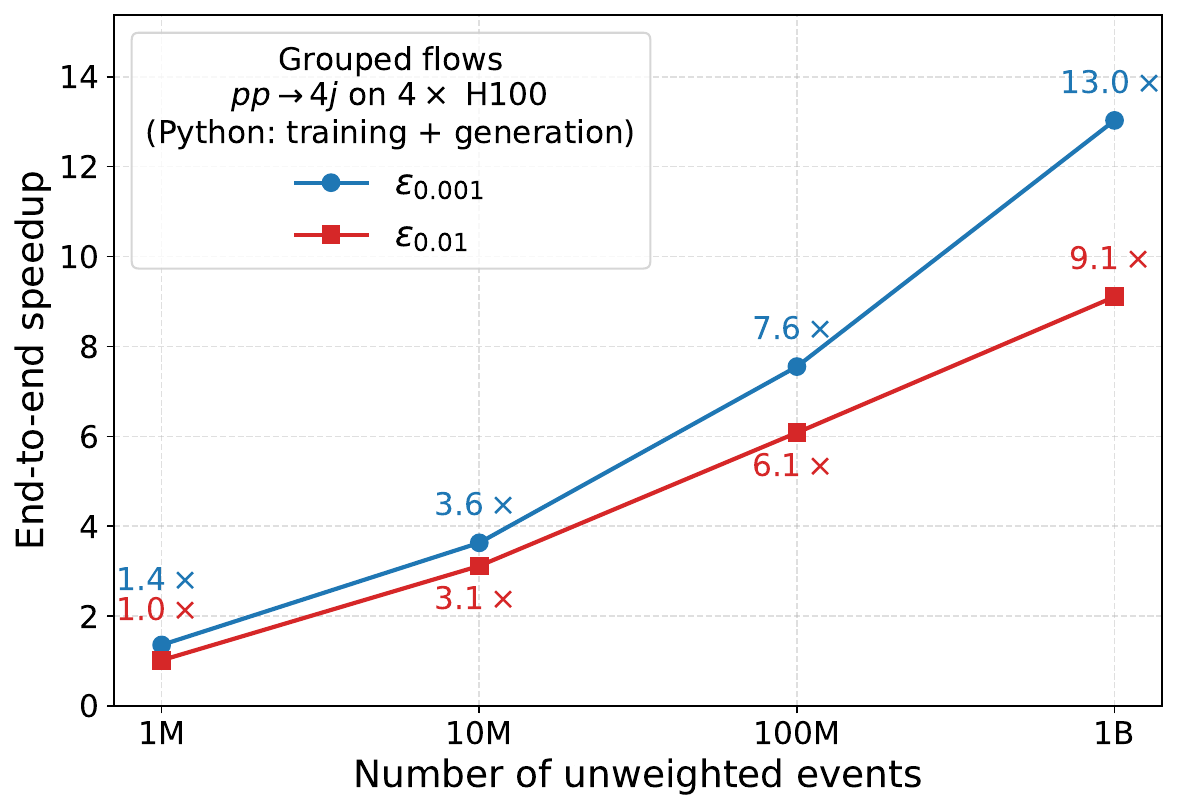} &
        \includegraphics[width=0.48\textwidth]{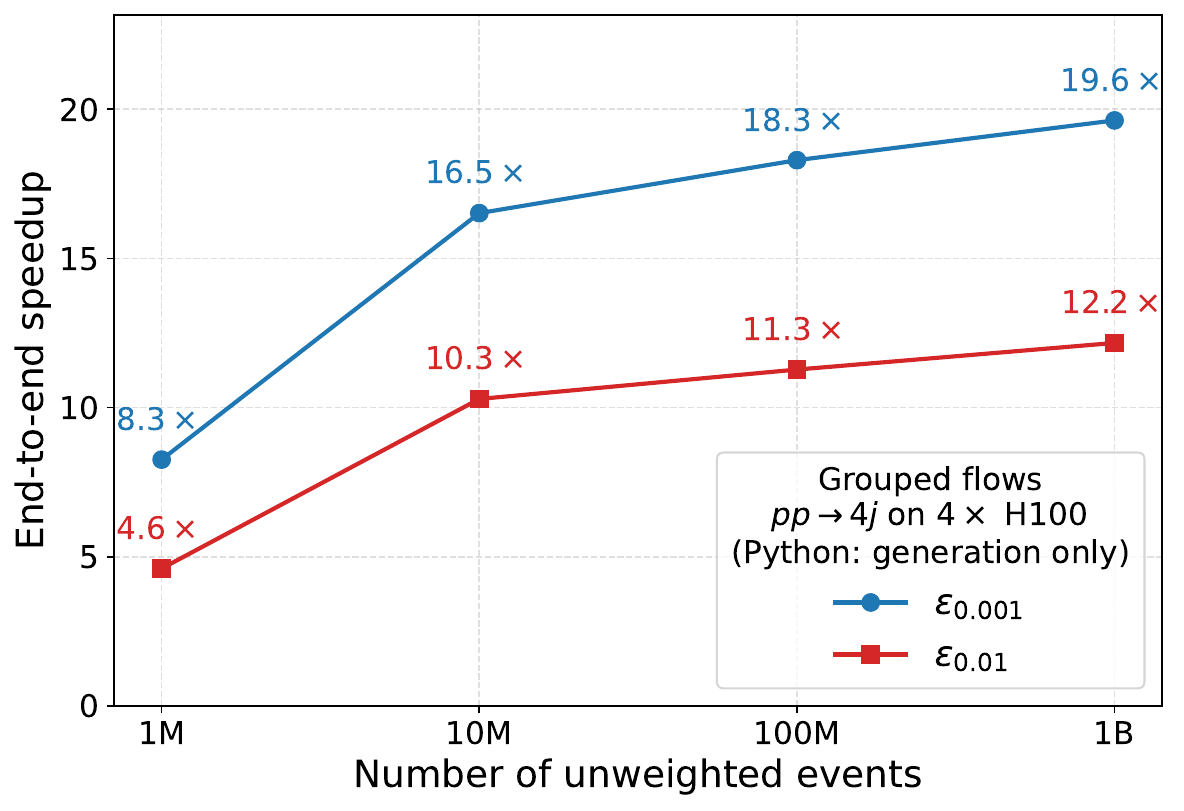}
    \end{tabular}
    \caption{
    End-to-end (left) and generation-only (right) speedups for grouped-flow generation in \jjjj production on \(4\times\) H100 GPUs. The left panel uses the directly measured \pepflow total runtime, including training and event generation, while the right panel uses the directly measured generation-only runtime of the final trained grouped-flow checkpoints. The \pepvegas runtime is measured at \(10^6\) events and linearly extrapolated to larger target samples.
    }
    \label{fig:fourj-h100-grouped-speedup}
\end{figure*}

\begin{figure*}[t]
    \centering
    \begin{tabular}{cc}
        \includegraphics[width=0.48\textwidth]{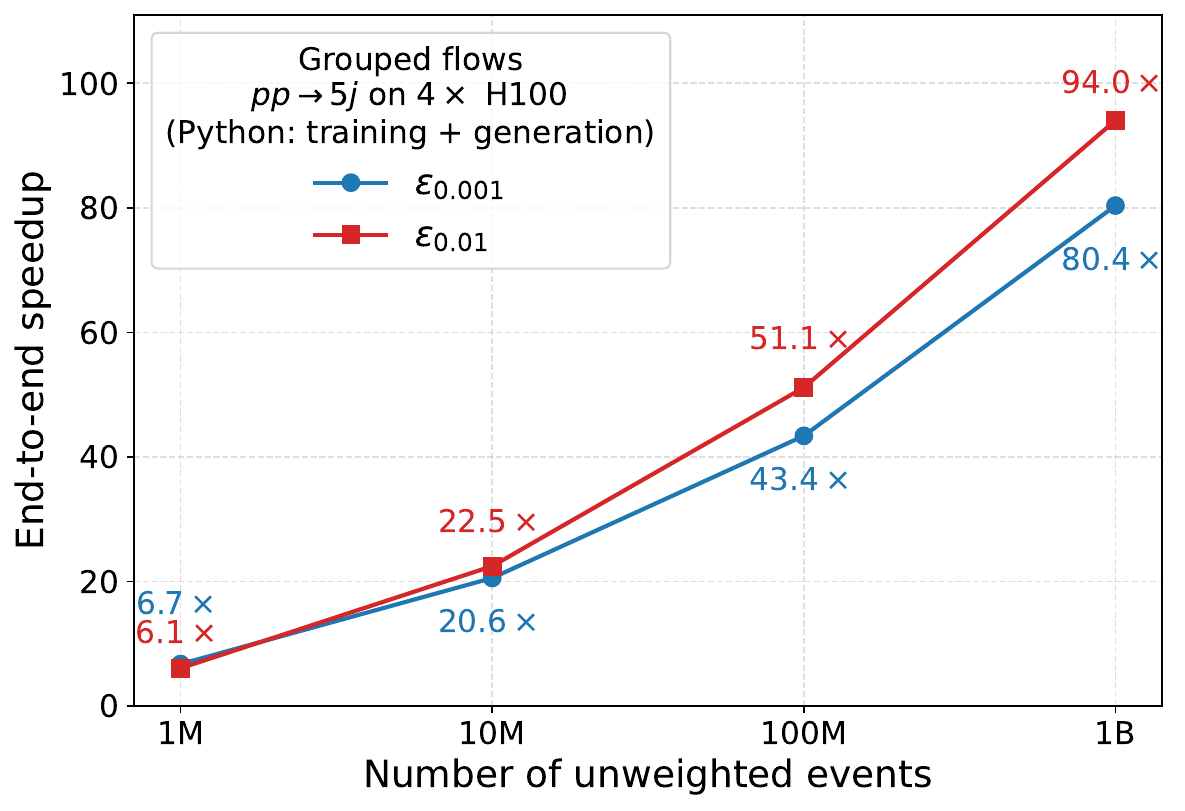} &
        \includegraphics[width=0.48\textwidth]{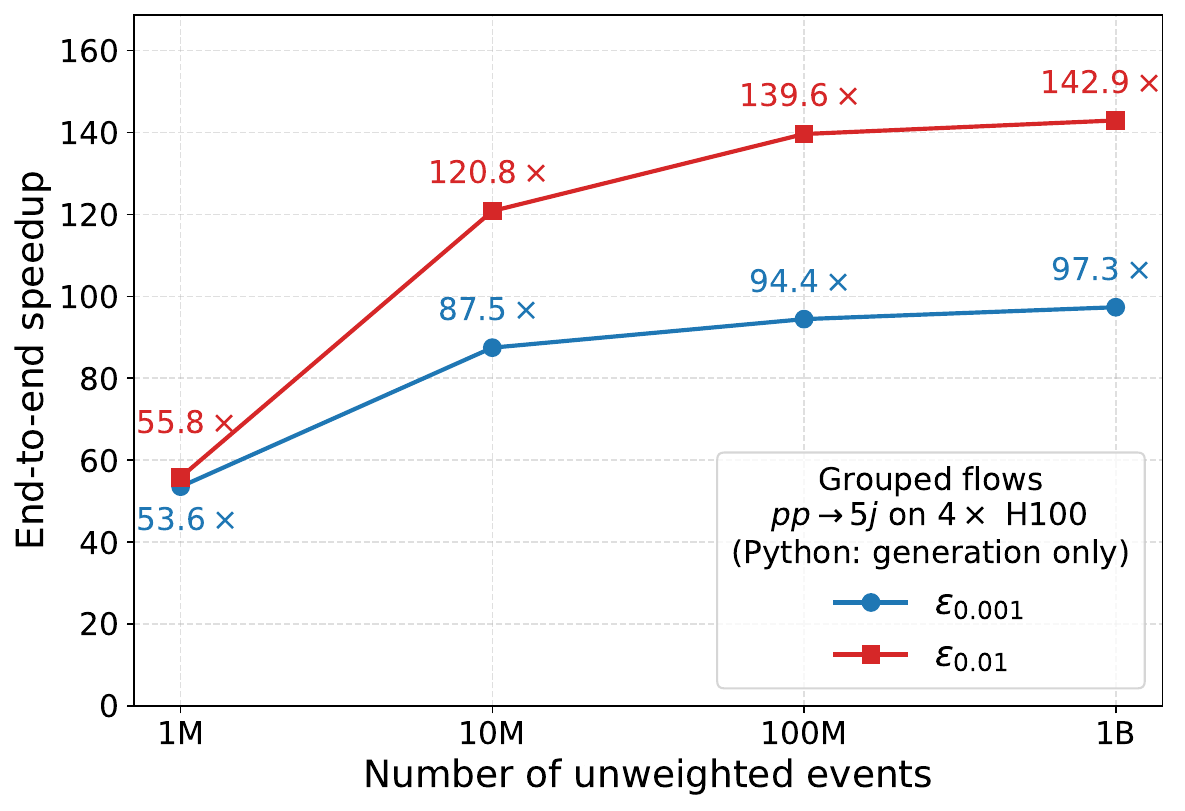}
    \end{tabular}
    \caption{
    End-to-end (left) and generation-only (right) speedups for grouped-flow generation in \jjjjj production on \(4\times\) H100 GPUs. The left panel uses the directly measured \pepflow total runtime, including training and event generation, while the right panel uses the directly measured generation-only runtime of the final trained grouped-flow checkpoints. The \pepvegas runtime is measured at \(10^6\) events and linearly extrapolated to larger target samples.
    }
    \label{fig:fivej-h100-grouped-speedup}
\end{figure*}

Figures~\ref{fig:z5j-h100-grouped-speedup}--\ref{fig:fivej-h100-grouped-speedup} show end-to-end speedups larger than unity for all four additional processes, already at \(10^6\) events. The speedup generally increases with the requested sample size because larger samples can favor later checkpoints whose higher unweighting efficiency compensates for the additional training time. The largest gains are obtained for the higher-multiplicity processes. For \zjjjjj, the end-to-end speedup at \(10^9\) events reaches \(71.8\times\) for \(\varepsilon=10^{-3}\) and \(109.3\times\) for \(\varepsilon=10^{-2}\), while the corresponding generation-only speedups are \(140.8\times\) and \(245.7\times\). The \jjjjj benchmark shows similarly large gains, with end-to-end speedups of \(80.4\times\) and \(94.0\times\), and generation-only speedups of \(97.3\times\) and \(142.9\times\).

For \ttjjjj, the end-to-end speedups at \(10^9\) events are
\(70.3\times\) and \(57.4\times\), while the generation-only speedups are \(90.3\times\) and \(70.7\times\). The simpler \jjjj process gives the smallest gains, reaching \(13.0\times\) and \(9.1\times\) end-to-end. Nevertheless, the grouped-flow workflow remains faster than \pepvegas generation for every sample size shown. Overall, these results demonstrate that parameter sharing within parton-content groups preserves the large runtime gains observed with subprocess-specific flows.

The absolute runtimes underlying Figs.~\ref{fig:z5j-h100-grouped-speedup}--\ref{fig:fivej-h100-grouped-speedup} are listed in Tables~\ref{tab:z5j-h100-grouped-runtime}--\ref{tab:5j-h100-grouped-runtime}. The \pepflow total runtime includes
training and event generation, while the generation-only runtime isolates event generation with the final trained grouped-flow checkpoints. Both \pepflow columns contain directly measured runtimes. For the \pepvegas workflow, the \(10^6\)-event runtime is measured directly, while the runtimes for \(10^7\), \(10^8\), and \(10^9\) events are obtained by linearly rescaling the measured \(10^6\)-event runtime and are shown in blue.

\begin{table*}[t]
    \centering
    \caption{
    Runtime comparison for generating unweighted \zjjjjj events with grouped flows on \(4\times\) H100 GPUs. The \pepflow total and generation-only runtimes are measured directly. Blue \pepvegas entries are linear extrapolations of the measured \(10^6\)-event runtime.
    }
    \label{tab:z5j-h100-grouped-runtime}
    \begin{tabular}{c c c c c}
        \hline
        \(\varepsilon\) & Number of events
        & \pepvegas
        & \pepflow (total)
        & \pepflow (gen. only) \\
        \hline
        \multirow{4}{*}{\(10^{-3}\)}
        & \(1M\) & 9h\;38min\;17s & 2h\;20min\;58s & 29min\;59s \\
        & \(10M\) & \textcolor{blue}{4d\;22min\;50s} & 7h\;53min\;37s & 1h\;6min\;45s \\
        & \(100M\) & \textcolor{blue}{40d\;3h\;48min\;20s} & 1d\;5h\;34min\;20s & 7h\;14min\;3s \\
        & \(1B\) & \textcolor{blue}{401d\;14h\;3min\;20s} & 5d\;14h\;19min\;35s & 2d\;20h\;28min\;31s \\
        \hline
        \multirow{4}{*}{\(10^{-2}\)}
        & \(1M\) & 7h\;24min\;24s & 1h\;54min\;2s & 27min\;32s \\
        & \(10M\) & \textcolor{blue}{3d\;2h\;4min} & 4h\;43min\;27s & 43min\;48s \\
        & \(100M\) & \textcolor{blue}{30d\;20h\;40min} & 15h\;53min\;42s & 3h\;25min\;24s \\
        & \(1B\) & \textcolor{blue}{308d\;14h\;40min} & 2d\;19h\;47min\;3s & 1d\;6h\;8min\;53s \\
        \hline
    \end{tabular}
\end{table*}

\begin{table*}[t]
    \centering
    \caption{
    Runtime comparison for generating unweighted \ttjjjj events with grouped flows on \(4\times\) H100 GPUs. The \pepflow total and generation-only runtimes are measured directly. Blue \pepvegas entries are linear extrapolations of the measured \(10^6\)-event runtime.
    }
    \label{tab:tt4j-h100-grouped-runtime}
    \begin{tabular}{c c c c c}
        \hline
        \(\varepsilon\) & Number of events
        & \pepvegas
        & \pepflow (total)
        & \pepflow (gen. only) \\
        \hline
        \multirow{4}{*}{\(10^{-3}\)}
        & \(1M\) & 4h\;56min\;32s & 54min\;34s & 6min\;44s \\
        & \(10M\) & \textcolor{blue}{2d\;1h\;25min\;20s} & 3h\;17min\;5s & 36min\;57s \\
        & \(100M\) & \textcolor{blue}{20d\;14h\;13min\;20s} & 12h\;48min\;14s & 5h\;32min\;11s \\
        & \(1B\) & \textcolor{blue}{205d\;22h\;13min\;20s} & 2d\;22h\;20min\;50s & 2d\;7h\;42min\;9s \\
        \hline
        \multirow{4}{*}{\(10^{-2}\)}
        & \(1M\) & 1h\;50min\;23s & 28min\;42s & 4min\;9s \\
        & \(10M\) & \textcolor{blue}{18h\;23min\;50s} & 1h\;41min\;9s & 18min\;23s \\
        & \(100M\) & \textcolor{blue}{7d\;15h\;58min\;20s} & 6h\;57min\;55s & 2h\;40min\;23s \\
        & \(1B\) & \textcolor{blue}{76d\;15h\;43min\;20s} & 1d\;8h\;3min\;23s & 1d\;2h\;1min\;48s \\
        \hline
    \end{tabular}
\end{table*}

\begin{table*}[t]
    \centering
    \caption{
    Runtime comparison for generating unweighted \jjjj events with grouped flows on \(4\times\) H100 GPUs. The \pepflow total and generation-only runtimes are measured directly. Blue \pepvegas entries are linear extrapolations of the measured \(10^6\)-event runtime.
    }
    \label{tab:4j-h100-grouped-runtime}
    \begin{tabular}{c c c c c}
        \hline
        \(\varepsilon\) & Number of events
        & \pepvegas
        & \pepflow (total)
        & \pepflow (gen. only) \\
        \hline
        \multirow{4}{*}{\(10^{-3}\)}
        & \(1M\) & 3min\;43s & 2min\;44s & 27s \\
        & \(10M\) & \textcolor{blue}{37min\;10s} & 10min\;34s & 2min\;15s \\
        & \(100M\) & \textcolor{blue}{6h\;11min\;40s} & 49min\;13s & 20min\;19s \\
        & \(1B\) & \textcolor{blue}{2d\;13h\;56min\;40s} & 4h\;45min\;17s & 3h\;9min\;23s \\
        \hline
        \multirow{4}{*}{\(10^{-2}\)}
        & \(1M\) & 1min\;46s & 1min\;44s & 23s \\
        & \(10M\) & \textcolor{blue}{17min\;40s} & 5min\;40s & 1min\;43s \\
        & \(100M\) & \textcolor{blue}{2h\;56min\;40s} & 29min\;3s & 15min\;40s \\
        & \(1B\) & \textcolor{blue}{1d\;5h\;26min\;40s} & 3h\;14min\;3s & 2h\;25min\;11s \\
        \hline
    \end{tabular}
\end{table*}

\begin{table*}[t]
    \centering
    \caption{
    Runtime comparison for generating unweighted \jjjjj events with grouped flows on \(4\times\) H100 GPUs. The \pepflow total and generation-only runtimes are measured directly. Blue \pepvegas entries are linear extrapolations of the measured \(10^6\)-event runtime.
    }
    \label{tab:5j-h100-grouped-runtime}
    \begin{tabular}{c c c c c}
        \hline
        \(\varepsilon\) & Number of events
        & \pepvegas
        & \pepflow (total)
        & \pepflow (gen. only) \\
        \hline
        \multirow{4}{*}{\(10^{-3}\)}
        & \(1M\) & 2h\;18min\;21s & 20min\;30s & 2min\;35s \\
        & \(10M\) & \textcolor{blue}{23h\;3min\;30s} & 1h\;7min\;17s & 15min\;49s \\
        & \(100M\) & \textcolor{blue}{9d\;14h\;35min} & 5h\;18min\;53s & 2h\;26min\;30s \\
        & \(1B\) & \textcolor{blue}{96d\;1h\;50min} & 1d\;4h\;41min\;46s & 23h\;41min\;14s \\
        \hline
        \multirow{4}{*}{\(10^{-2}\)}
        & \(1M\) & 1h\;23min\;46s & 13min\;46s & 1min\;30s \\
        & \(10M\) & \textcolor{blue}{13h\;57min\;40s} & 37min\;17s & 6min\;56s \\
        & \(100M\) & \textcolor{blue}{5d\;19h\;36min\;40s} & 2h\;43min\;46s & 1h \\
        & \(1B\) & \textcolor{blue}{58d\;4h\;6min\;40s} & 14h\;50min\;54s & 9h\;46min\;3s \\
        \hline
    \end{tabular}
\end{table*}

\section{Subprocess-Specific Flow Results on a Single RTX GPU}
\label{app:rtx-results}

In this appendix, we report subprocess-specific flow benchmarks on a single NVIDIA RTX 4000 Ada Generation GPU, complementing the four-H100 results presented in the main text.

For each process, the RTX benchmarks use the same \Pepper cache containing the optimized \Vegas grids as the corresponding four-H100 benchmarks. The \pepvegas reference timing is measured on the same RTX 4000 Ada Generation GPU used for \pepflow generation, with a batch size of \(131072\) events. As in the four-H100 benchmarks, the construction of the shared cache is treated as common preprocessing and is excluded from the reported runtimes.

The RTX benchmark differs from the four-H100 setup in how the training time is obtained. On four H100 GPUs, the subprocess-specific workflow can keep all \(134\) \zjjjj subprocess flows in memory and train them in parallel.  A single NVIDIA RTX 4000 Ada Generation GPU does not have enough memory for this fully parallel setup. The subprocesses are therefore divided into smaller chunks, which are processed sequentially during the flow-density evaluation and backward pass.

For the RTX timing study, we use the full set of saved model checkpoints from the four-H100 training run as candidate model states, but repeat the checkpoint selection using RTX-specific timing information. We apply the same training-plus-generation prescription as defined in Sec.~\ref{sec:computational-setup}. For each saved checkpoint \(c\), we measure
the RTX time \(T_{\rm gen}(c;10^6)\) required to generate \(10^6\) unweighted events and use the same linear rescaling to estimate its generation cost for each target sample size \(N\).

The RTX training contribution \(T_{\rm train}(c)\) is estimated from a separate timing run of the chunked single-GPU training setup. We run this setup for
\(5000\) rounds, measure the average wall time per round, and rescale it to the round corresponding to checkpoint \(c\). This RTX training run is used only to
estimate the training time; the candidate model states remain those saved during the four-H100 training run. The RTX-specific training and generation time estimates are then combined to select a checkpoint separately for each target sample size. Once a checkpoint has been selected, its event-generation runtime is measured by directly generating the requested number of events whenever computationally feasible. Only runtimes for which direct generation is impractical are obtained by extrapolation; these entries are marked in blue in the tables.

\begin{figure*}[t]
    \centering
    \begin{tabular}{ccc}
        \includegraphics[width=0.32\textwidth]{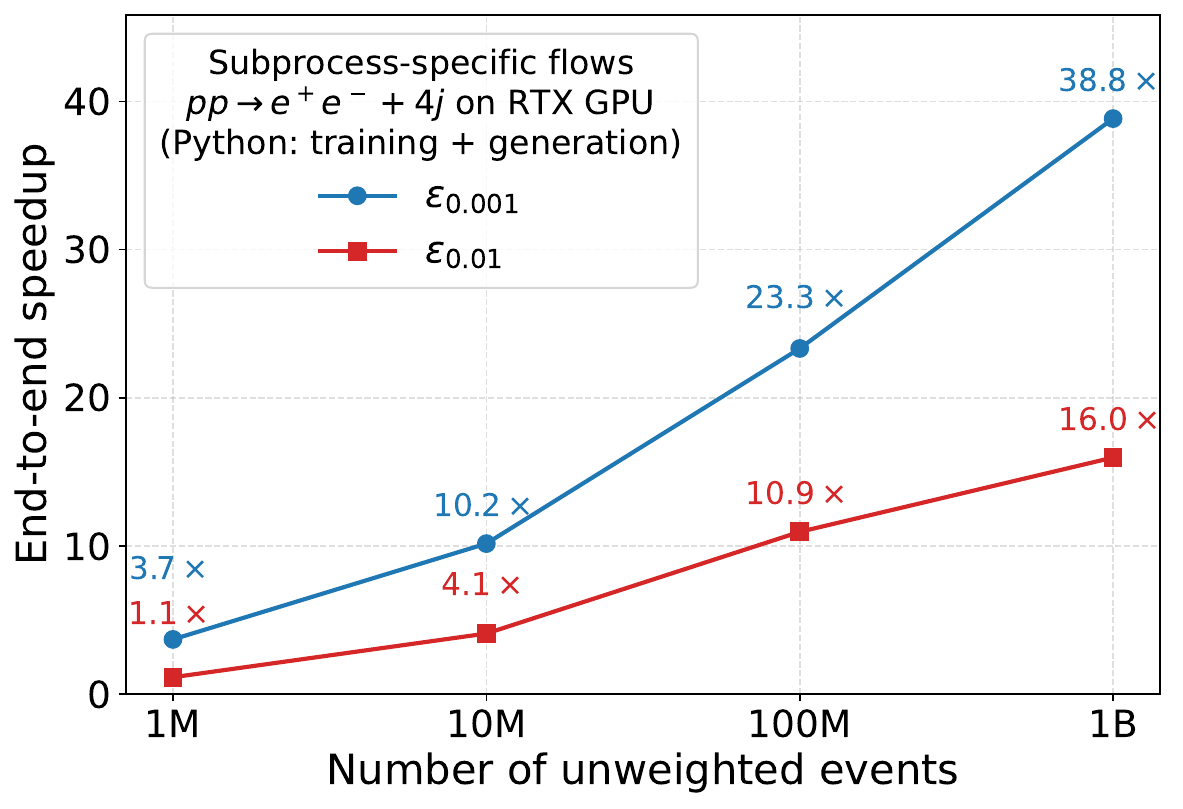} &
        \includegraphics[width=0.32\textwidth]{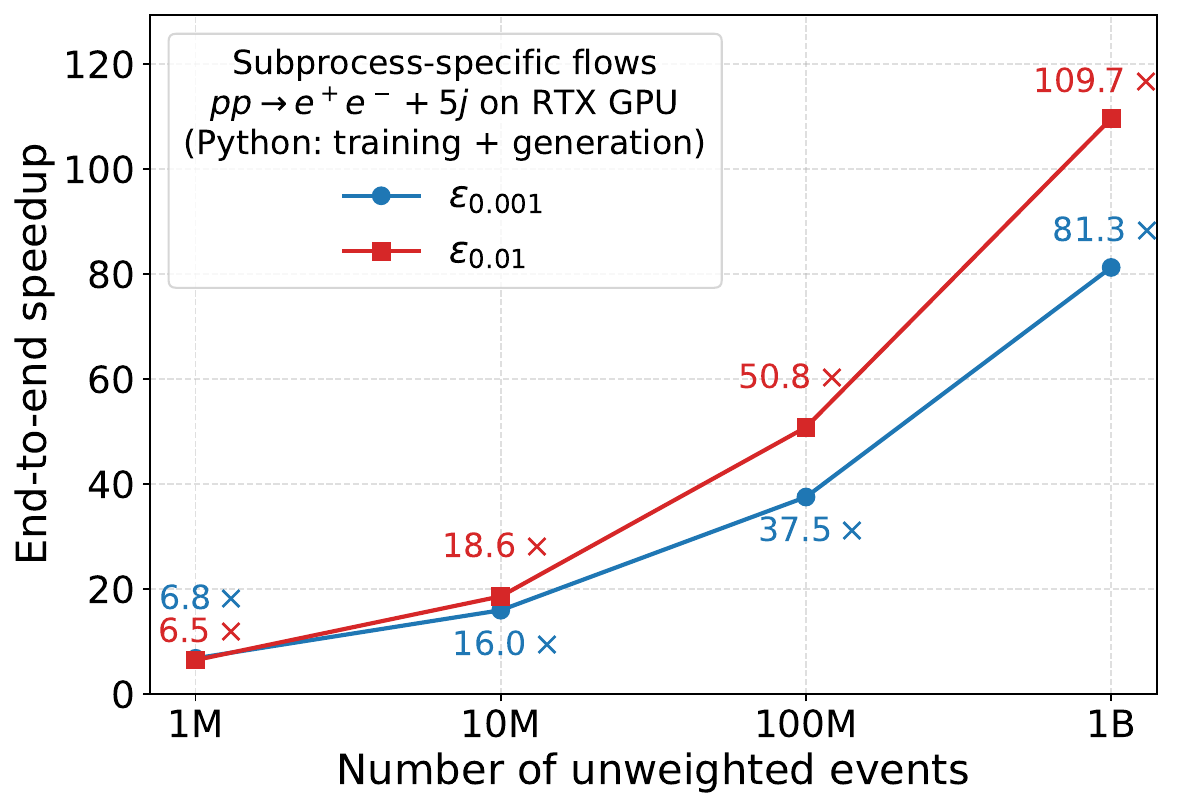} &
        \includegraphics[width=0.32\textwidth]{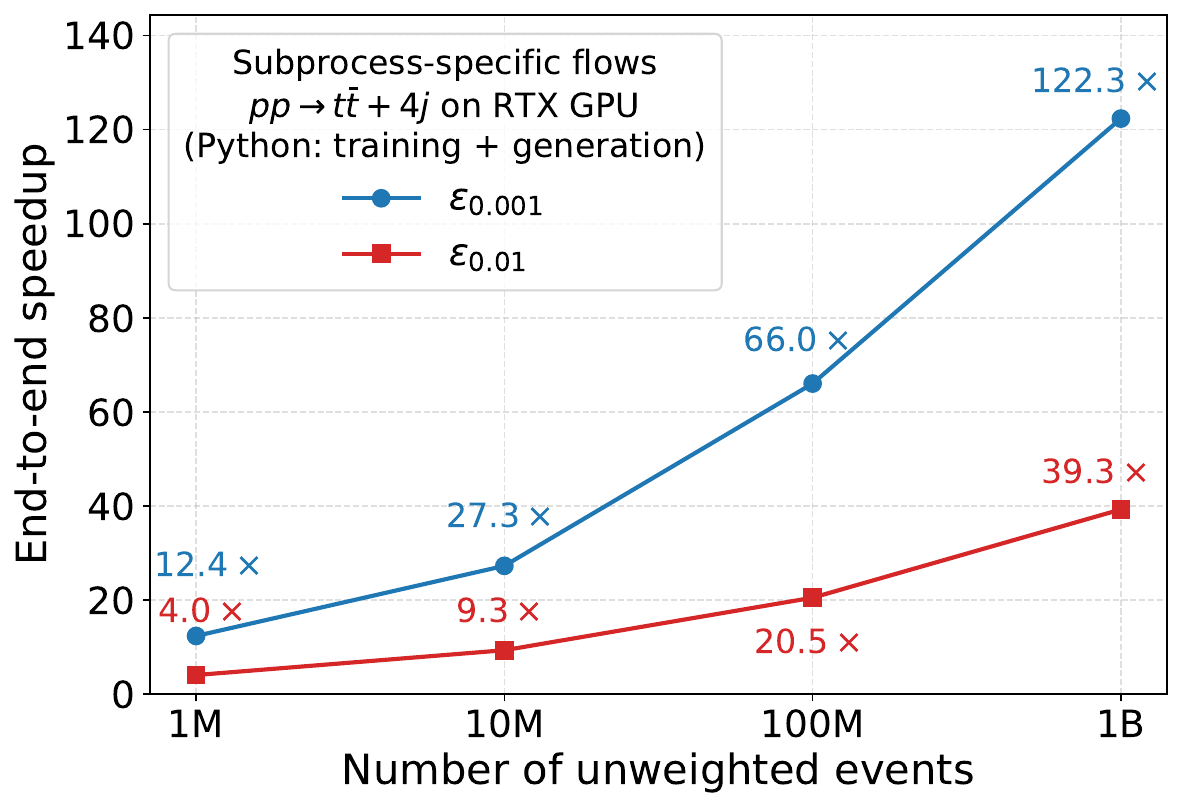} \\
        \includegraphics[width=0.32\textwidth]{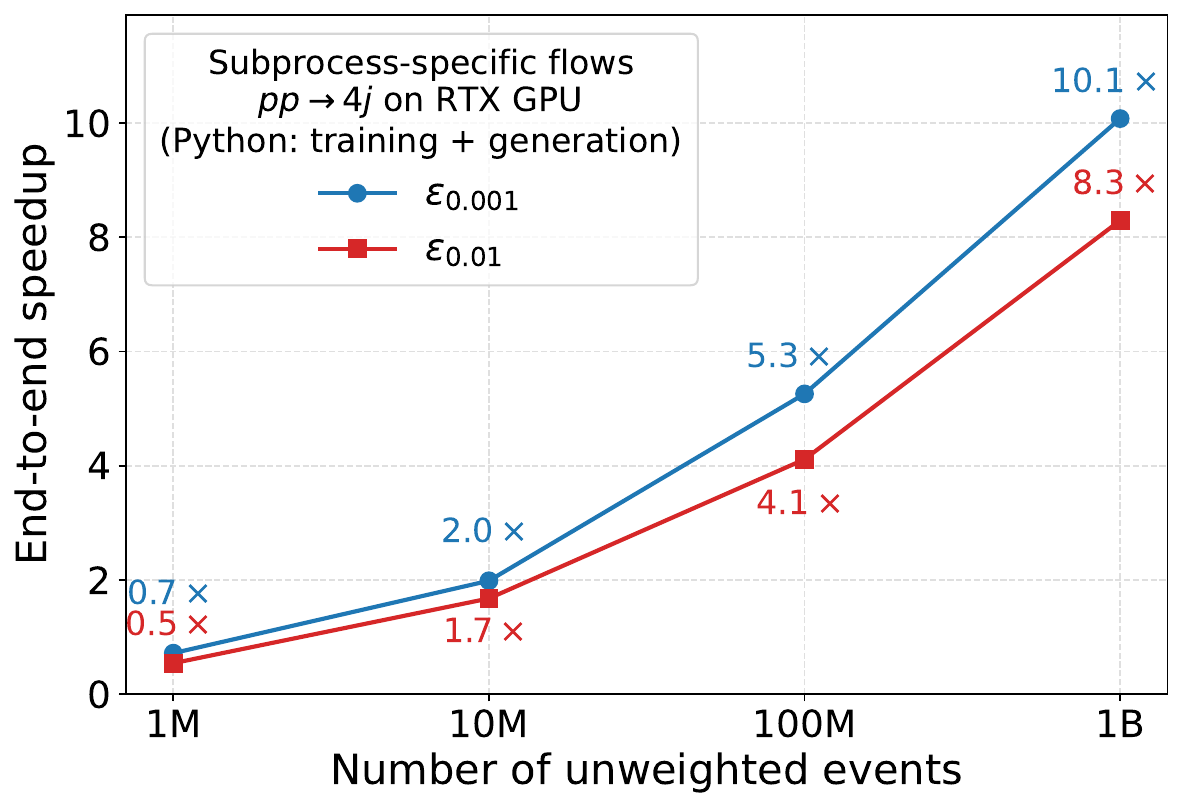} &
        \includegraphics[width=0.32\textwidth]{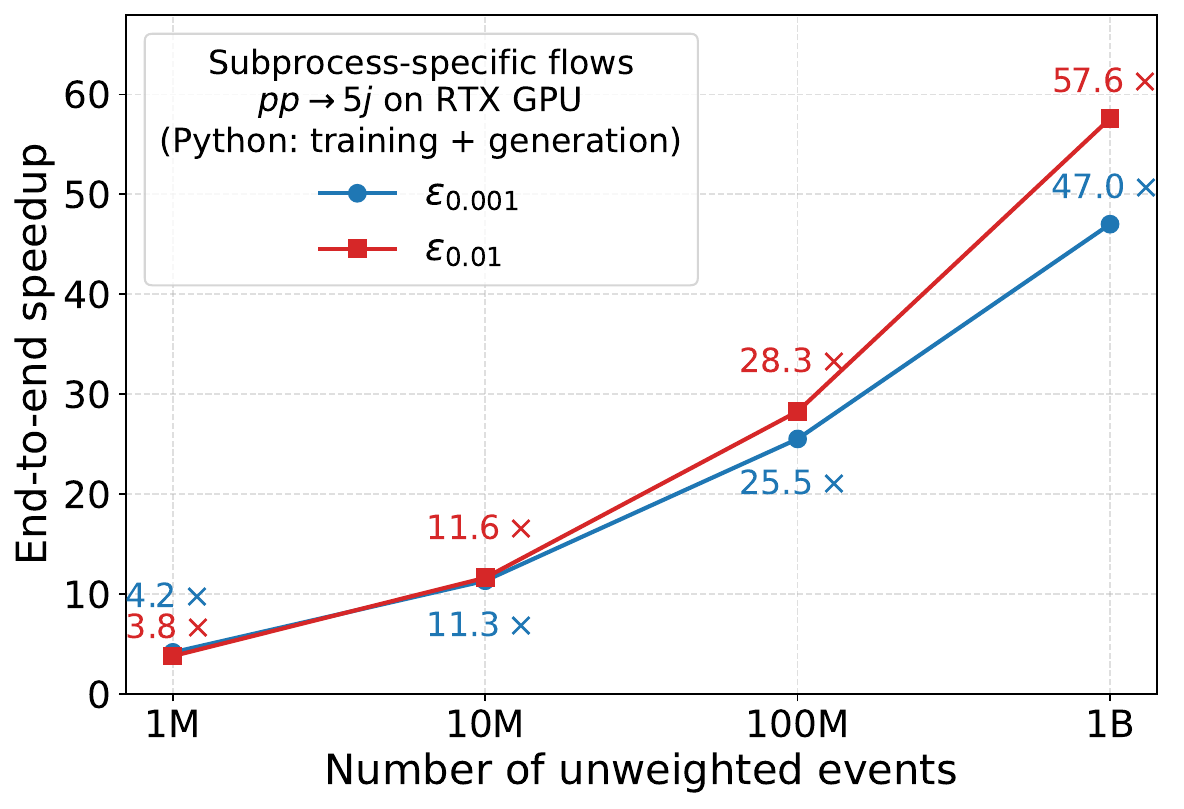} &
        \phantom{\includegraphics[width=0.32\textwidth]{results/resultsA/5j_rtx_speedup.pdf}}
    \end{tabular}
    \caption{
    End-to-end speedup of subprocess-specific flow generation on a single NVIDIA RTX 4000 Ada Generation GPU. The speedup is defined as the \pepvegas runtime divided by the \pepflow runtime, where the latter includes the estimated RTX training contribution and the measured or extrapolated event-generation time of the selected checkpoint. Results are shown for \zjjjj, \zjjjjj, \ttjjjj, \jjjj, and \jjjjj, and for \(\varepsilon=10^{-3}\) and \(\varepsilon=10^{-2}\).
    }
    \label{fig:rtx-speedups}
\end{figure*}

Figure~\ref{fig:rtx-speedups} shows the resulting end-to-end speedups. At \(10^6\) events, the training contribution limits the gain, and \pepvegas generation remains faster for the \jjjj benchmark at both unweighting thresholds. As the requested sample size increases, later checkpoints can become optimal because their higher unweighting efficiency compensates for the additional training time. The \pepflow workflow already becomes faster for \jjjj production at \(10^7\) events.

At \(10^9\) events, the \jjjj speedup reaches \(10.1\times\) and
\(8.3\times\) for \(\varepsilon=10^{-3}\) and \(\varepsilon=10^{-2}\), respectively. The gains are substantially larger for the higher-multiplicity processes. For \zjjjj, the corresponding speedups are \(38.8\times\) and \(16.0\times\), while for \zjjjjj they reach \(81.3\times\) and \(109.7\times\). The \ttjjjj and \jjjjj
benchmarks show the same qualitative trend. These results demonstrate that the \pepflow workflow remains effective on a single workstation GPU, with the largest gains obtained for expensive processes and large target samples.

The absolute runtimes underlying Fig.~\ref{fig:rtx-speedups} are listed in Tables~\ref{tab:z4j-rtx-runtime}--\ref{tab:5j-rtx-runtime}. The \pepflow runtime includes the estimated RTX training contribution and the measured or extrapolated event-generation time of the selected checkpoint. Results are shown for both unweighting thresholds and for target sample sizes from \(10^6\) to \(10^9\) events. For the \pepvegas workflow, the \(10^6\)-event runtime is measured directly, while the runtimes for \(10^7\), \(10^8\), and \(10^9\) events are obtained by linearly rescaling the measured \(10^6\)-event generation time. For the \pepflow workflow, the generation time is measured directly whenever computationally feasible. Blue entries indicate target sample sizes that were not generated directly; in these cases, the generation time is estimated by linearly rescaling the runtime of the largest smaller sample generated directly with the selected checkpoint. The estimated training contribution is then added to this extrapolated generation time to obtain the reported total runtime.

\begin{table*}[t]
    \centering
    \caption{
    Estimated runtime comparison for generating unweighted \zjjjj events with subprocess-specific flows on a single NVIDIA RTX 4000 Ada Generation GPU. The \pepflow runtime includes the estimated RTX training contribution and event generation with the selected checkpoint. Blue entries are extrapolated estimates for target sample sizes that were not generated directly.
    }
    \label{tab:z4j-rtx-runtime}
    \begin{tabular}{c c c c}
        \hline
        \(\varepsilon\) & Number of events
        & \pepvegas
        & \pepflow (total)\\
        \hline
        \multirow{4}{*}{\(10^{-3}\)}
        & \(1M\) & 5h\;46min\;45s & 1h\;33min\;59s \\
        & \(10M\) & \textcolor{blue}{2d\;9h\;47min\;30s} & 5h\;41min\;26s \\
        & \(100M\) & \textcolor{blue}{24d\;1h\;55min} & 1d\;46min\;42s \\
        & \(1B\) & \textcolor{blue}{240d\;19h\;10min} & \textcolor{blue}{6d\;4h\;48min\;28s} \\
        \hline
        \multirow{4}{*}{\(10^{-2}\)}
        & \(1M\) & 1h\;38min\;25s & 1h\;26min\;36s \\
        & \(10M\) & \textcolor{blue}{16h\;24min\;10s} & 4h\;55s \\
        & \(100M\) & \textcolor{blue}{6d\;20h\;1min\;40s} & 14h\;59min \\
        & \(1B\) & \textcolor{blue}{68d\;8h\;16min\;40s} & \textcolor{blue}{4d\;6h\;45min\;12s} \\
        \hline
    \end{tabular}
\end{table*}

\begin{table*}[t]
    \centering
    \caption{
    Estimated runtime comparison for generating unweighted \zjjjjj events with subprocess-specific flows on a single NVIDIA RTX 4000 Ada Generation GPU. The \pepflow runtime includes the estimated RTX training contribution and event generation with the selected checkpoint. Blue entries are extrapolated estimates for target sample sizes that were not generated directly.
    }
    \label{tab:z5j-rtx-runtime}
    \begin{tabular}{c c c c}
        \hline
        \(\varepsilon\) & Number of events
        & \pepvegas
        & \pepflow (total)\\
        \hline
        \multirow{4}{*}{\(10^{-3}\)}
        & \(1M\) & 2d\;21h\;51min\;7s & 10h\;14min\;47s \\
        & \(10M\) & \textcolor{blue}{29d\;2h\;31min\;10s} & 1d\;19h\;42min\;55s \\
        & \(100M\) & \textcolor{blue}{291d\;1h\;11min\;40s} & \textcolor{blue}{7d\;18h\;2min\;35s} \\
        & \(1B\) & \textcolor{blue}{2910d\;11h\;56min\;40s} & \textcolor{blue}{35d\;19h\;16min\;59s} \\
        \hline
        \multirow{4}{*}{\(10^{-2}\)}
        & \(1M\) & 2d\;3h\;59min\;21s & 8h\;1min\;27s \\
        & \(10M\) & \textcolor{blue}{21d\;15h\;53min\;30s} & 1d\;3h\;54min\;34s \\
        & \(100M\) & \textcolor{blue}{216d\;14h\;55min} & \textcolor{blue}{4d\;6h\;15min\;24s} \\
        & \(1B\) & \textcolor{blue}{2166d\;5h\;10min} & \textcolor{blue}{19d\;18h\;7min\;28s} \\
        \hline
    \end{tabular}
\end{table*}

\begin{table*}[t]
    \centering
    \caption{
    Estimated runtime comparison for generating unweighted \ttjjjj events with subprocess-specific flows on a single NVIDIA RTX 4000 Ada Generation GPU. The \pepflow runtime includes the estimated RTX training contribution and event generation with the selected checkpoint. Blue entries are extrapolated estimates for target sample sizes that were not generated directly.
    }
    \label{tab:tt4j-rtx-runtime}
    \begin{tabular}{c c c c}
        \hline
        \(\varepsilon\) & Number of events
        & \pepvegas
        & \pepflow (total)\\
        \hline
        \multirow{4}{*}{\(10^{-3}\)}
        & \(1M\) & 34h\;24min\;23s & 2h\;47min\;6s \\
        & \(10M\) & \textcolor{blue}{14d\;8h\;3min\;50s} & 12h\;37min\;16s \\
        & \(100M\) & \textcolor{blue}{143d\;8h\;38min\;20s} & 2d\;4h\;7min\;5s \\
        & \(1B\) & \textcolor{blue}{1433d\;14h\;23min\;20s} & \textcolor{blue}{11d\;17h\;13min\;47s} \\
        \hline
        \multirow{4}{*}{\(10^{-2}\)}
        & \(1M\) & 6h\;1min\;50s & 1h\;29min\;23s \\
        & \(10M\) & \textcolor{blue}{2d\;12h\;18min\;20s} & 6h\;28min\;14s \\
        & \(100M\) & \textcolor{blue}{25d\;3h\;3min\;20s} & 1d\;5h\;26min \\
        & \(1B\) & \textcolor{blue}{251d\;6h\;33min\;20s} & \textcolor{blue}{6d\;9h\;38min\;16s} \\
        \hline
    \end{tabular}
\end{table*}

\begin{table*}[t]
    \centering
    \caption{
    Estimated runtime comparison for generating unweighted \jjjj events with subprocess-specific flows on a single NVIDIA RTX 4000 Ada Generation GPU. The \pepflow runtime includes the estimated RTX training contribution and event generation with the selected checkpoint. Blue entries are extrapolated estimates for target sample sizes that were not generated directly.
    }
    \label{tab:4j-rtx-runtime}
    \begin{tabular}{c c c c}
        \hline
        \(\varepsilon\) & Number of events
        & \pepvegas
        & \pepflow (total)\\
        \hline
        \multirow{4}{*}{\(10^{-3}\)}
        & \(1M\) & 19min\;14s & 26min\;50s \\
        & \(10M\) & \textcolor{blue}{3h\;12min\;20s} & 1h\;36min\;55s \\
        & \(100M\) & \textcolor{blue}{1d\;8h\;3min\;20s} & 6h\;5min\;51s \\
        & \(1B\) & \textcolor{blue}{13d\;8h\;33min\;20s} & \textcolor{blue}{1d\;7h\;48min\;26s} \\
        \hline
        \multirow{4}{*}{\(10^{-2}\)}
        & \(1M\) & 8min\;39s & 16min \\
        & \(10M\) & \textcolor{blue}{1h\;26min\;30s} & 51min\;39s \\
        & \(100M\) & \textcolor{blue}{14h\;25min} & 3h\;30min\;25s \\
        & \(1B\) & \textcolor{blue}{6d\;10min} & \textcolor{blue}{17h\;22min\;57s} \\
        \hline
    \end{tabular}
\end{table*}

\begin{table*}[t]
    \centering
    \caption{
    Estimated runtime comparison for generating unweighted \jjjjj events with subprocess-specific flows on a single NVIDIA RTX 4000 Ada Generation GPU. The \pepflow runtime includes the estimated RTX training contribution and event generation with the selected checkpoint. Blue entries are extrapolated estimates for target sample sizes that were not generated directly.
    }
    \label{tab:5j-rtx-runtime}
    \begin{tabular}{c c c c}
        \hline
        \(\varepsilon\) & Number of events
        & \pepvegas
        & \pepflow (total)\\
        \hline
        \multirow{4}{*}{\(10^{-3}\)}
        & \(1M\) & 8h\;10min\;1s & 1h\;57min\;23s \\
        & \(10M\) & \textcolor{blue}{3d\;9h\;40min\;10s} & 7h\;12min\;14s \\
        & \(100M\) & \textcolor{blue}{34d\;41min\;40s} & 1d\;7h\;59min\;44s \\
        & \(1B\) & \textcolor{blue}{340d\;6h\;56min\;40s} & \textcolor{blue}{7d\;5h\;42min\;52s} \\
        \hline
        \multirow{4}{*}{\(10^{-2}\)}
        & \(1M\) & 4h\;43min\;12s & 1h\;14min\;29s \\
        & \(10M\) & \textcolor{blue}{1d\;23h\;12min} & 4h\;3min\;24s \\
        & \(100M\) & \textcolor{blue}{19d\;16h} & 16h\;41min\;31s \\
        & \(1B\) & \textcolor{blue}{196d\;16h} & \textcolor{blue}{3d\;9h\;58min\;43s} \\
        \hline
    \end{tabular}
\end{table*}

\bibliography{refs}

\end{document}